\documentclass[nonacm,acmsmall,screen]{acmart}
\AtBeginDocument{%
  }

\usepackage{amsmath,amsfonts, wasysym}
\usepackage{algorithmic,algorithm}
\usepackage{subcaption}
\usepackage{textcomp}
\usepackage{stfloats}
\usepackage{url}
\usepackage{verbatim}
\usepackage{graphicx}
\usepackage{pifont}
 
\usepackage{bbding}
\usepackage{colortbl}
\usepackage{enumitem}
\usepackage{hyperref}
\usepackage{makecell}

\definecolor{mygreen}{HTML}{82B366}
\definecolor{myred}{HTML}{B85450}
\newcommand{\cmark}{\textcolor{mygreen}{\ding{51}}} 
\newcommand{\xmark}{\textcolor{myred}{\ding{55}}}   

\newcommand{\ccpp}{C/C\raisebox{0.1ex}{\small++}}

\newcommand{\numStudies}{263 }
\newcommand{\numBinaryStudies}{195 }
\newcommand{\numMultiClassStudies}{61 }
\newcommand{\numMultiLabelStudies}{23 }
\newcommand{\numVulSpecificStudies}{20 }
\newcommand{\numOwnDatasets}{87 }

\setcopyright{acmlicensed}
\copyrightyear{2025}
\acmYear{2025}
\acmJournal{TOSEM}
\acmDOI{XXXXXXX.XXXXXXX}

\begin{document}
\title{A Systematic Literature Review on Detecting Software Vulnerabilities with Large Language Models}

\author{Sabrina Kaniewski}
\email{sabrina.kaniewski@hs-esslingen.de}
\orcid{0009-0004-3966-9681}
\affiliation{%
  \institution{Esslingen University}
  \city{Esslingen}
  \country{Germany}}

\author{Fabian Schmidt}
\email{fabian.schmidt@hs-esslingen.de}
\orcid{0000-0003-3958-8932}
\affiliation{%
  \institution{Institute for Intelligent Systems, Esslingen University}
  \city{Esslingen}
  \country{Germany}}

\author{Markus Enzweiler}
\email{markus.enzweiler@hs-esslingen.de}
\orcid{0000-0001-9211-9882}
\affiliation{%
  \institution{Institute for Intelligent Systems, Esslingen University}
  \city{Esslingen}
  \country{Germany}}

\author{Michael Menth}
\email{menth@uni-tuebingen.de}
\orcid{0000-0002-3216-1015}
\affiliation{%
  \institution{Chair of Communication Networks, University of Tübingen}
  \city{Tübingen}
  \country{Germany}}

\author{Tobias Heer}
\email{tobias.heer@hs-esslingen.de}
\orcid{0000-0003-3119-252X}
\affiliation{%
  \institution{Esslingen University}
  \city{Esslingen}
  \country{Germany}}

\renewcommand{\shortauthors}{Kaniewski et al.}

\begin{abstract}
The increasing adoption of Large Language Models (LLMs) in software engineering has sparked interest in their use for software vulnerability detection. However, the rapid development of this field has resulted in a fragmented research landscape, with diverse studies that are difficult to compare due to differences in, e.g., system designs and dataset usage. This fragmentation makes it difficult to obtain a clear overview of the state-of-the-art or compare and categorize studies meaningfully. In this work, we present a comprehensive systematic literature review (SLR) of LLM-based software vulnerability detection. We analyze 263 studies published between January 2020 and November 2025, categorizing them by task formulation, input representation, system architecture, and techniques. Further, we analyze the datasets used, including their characteristics, vulnerability coverage, and diversity. We present a fine-grained taxonomy of vulnerability detection approaches, identify key limitations, and outline actionable future research opportunities. By providing a structured overview of the field, this review improves transparency and serves as a practical guide for researchers and practitioners aiming to conduct more comparable and reproducible research. We publicly release all artifacts and maintain a living repository of LLM-based software vulnerability detection studies at \url{https://github.com/hs-esslingen-it-security/Awesome-LLM4SVD}.
\end{abstract}



\keywords{Software Vulnerability Detection, Vulnerability Detection, Software Engineering, Vulnerability Datasets, Benchmark Datasets, LLM, SLR, Survey}


\maketitle


\section{Introduction}
Vulnerability detection plays a critical role in the software development life cycle, identifying security vulnerabilities before they can be exploited in deployed systems.
The growing complexity of modern software, combined with an ever-increasing threat landscape, has led to a surge in reported vulnerabilities.
The Common Vulnerabilities and Exposures (CVE) records~\cite{CVECommonVulnerabilities} provide unique identifiers for publicly known software vulnerabilities. Over 40,000 CVEs have been published in 2024 alone, with over 12,000 reported in the first quarter of 2025~\cite{CVEMetrics}. 
Complementing CVEs, the Common Weakness Enumeration (CWE)~\cite{CWECommonWeakness} offers a hierarchical classification of vulnerability types, serving as a higher-level abstraction for understanding underlying vulnerability patterns.

Vulnerability assessment pipelines as found in industrial development workflows typically consist of multiple steps: detection (identifying whether code contains a vulnerability), localization (pinpointing the exact location in the code), severity estimation (evaluating potential impact), and repair (patching the code to mitigate the vulnerability), often initiated reactively when a new CVE is disclosed. 
Despite advances in tooling, e.g., static application security testing (SAST) tools, these processes remain largely manual or only partially automated due to the dependency on expert-defined rules (cf. Taghavi and Feyzi~\cite{taghavi2025large} for a comparison between LLMs and established software vulnerability detection methods). Further, these tools suffer from high false positive rates~\cite{croft2023data}.
As such, existing workflows are insufficient to meet the scale of remediation demands.

At the same time, the adoption of Large Language Models (LLMs)\footnote{Due to the lack of a formal definition for LLMs, in this work, we use the term \textit{LLM} broadly to encompass both large-scale language models (e.g., GPT-4) and smaller pre-trained models, such as BERT and CodeBERT~\cite{zhaoCodingPTMsHowFind2024}. We differentiate architectures and scales of LLMs later on in the body of this work.} for software engineering tasks has grown rapidly since the introduction of BERT~\cite{BERT} in 2018. 
LLMs are deep neural network models, typically based on Transformer architectures~\cite{vaswani2017attention}, that are pre-trained on massive corpora of natural language. 
LLMs have demonstrated impressive capabilities in understanding and generating code. 
Industry forecasts further emphasize the transformative potential of LLMs in code generation~\cite{amodei2024cfr}.

However, while LLMs hold promise in code generation, they also introduce new risks. 
Generated code often lacks awareness of existing libraries or internal codebases, which may lead to redundancies and the creation of syntactically different but semantically equivalent code. 
This complicates the detection of vulnerabilities, especially when they are subtle or semantically nuanced. Additionally, LLMs are prone to hallucinations and the generation of insecure code~\cite{hammondpearceAsleepKeyboardAssessing2022,liuNoNeedLift2023,liuExploringEvaluatingHallucinations2024,fuSecurityWeaknessesCopilot2024}.
Development teams may be unaware of vulnerabilities introduced by generated code, whether in their codebase, third-party components, or libraries.

The combination of LLM-generated code and the increasing volume of vulnerabilities calls for automated, scalable, and reliable vulnerability assessment pipelines.
LLMs' ability to generalize across programming languages and representations, as well as their high flexibility through prompt adaptation, make them a promising approach for software vulnerability assessment.
Therefore, recent research has begun to explore the use of LLMs.
In a prior study~\cite{kaniewski2024vulnerability}, we surveyed early work on LLM-based vulnerability handling, including detection, localization, and repair, and highlighted open challenges with regard to scalability, data diversity, and model reliability.
We continued to extensively document the progress in this field, with a focus on vulnerability detection studies as a prerequisite for all subsequent remediation steps.
As the number of studies on LLM-based vulnerability detection continues to grow, we observe a wide variety of system architectures, adaptation techniques, and evaluation methodologies. 
This heterogeneity leads to two pressing challenges:
First, there is \textbf{currently no comprehensive survey that maps the landscape of LLM-based software vulnerability detection methods}, their system designs, and dataset usage. As a result, \textbf{researchers face difficulties identifying trends, gaps, or best practices} in this rapidly evolving area.
Second, studies frequently adopt \textbf{custom datasets, metrics, and data splits}, making it \textbf{difficult to assess progress, reproduce results, or compare models fairly}.
To address these challenges, we contribute this systematic literature review (SLR) of LLM-based software vulnerability detection studies. 
Specifically, our contributions are as follows:
\begin{itemize}
\item We conduct a \textbf{SLR} of studies covering LLM-based software vulnerability detection in the time from January~2020 to November~2025, covering \textbf{\numStudies studies}. 
\item We present a \textbf{comprehensive taxonomy} that categorizes existing approaches based on task formulation (classification and generation), input representation (code and auxiliary information), system architecture (LLM usage), adaptation and orchestration technique, as well as dataset usage. This taxonomy allows for \textbf{meaningful comparison} of studies.
\item We perform an \textbf{in-depth analysis of commonly used and selected datasets}, investigating class balance, CWE coverage as well as diversity, and limitations. Based on this analysis, we provide actionable insights and best practices for future dataset selection and evaluation design to improve cross-study comparability.
\item We discuss \textbf{overarching limitations and challenges} of current LLM-based vulnerability detection approaches and outline \textbf{future research opportunities}.
\item To support the community, we provide the artifacts and maintain a regularly updated list of reviewed studies at \url{https://github.com/hs-esslingen-it-security/Awesome-LLM4SVD} (\textbf{Awesome-LLM4SVD})~\cite{replicationpackage}.
\end{itemize}

The remainder of this paper is structured as follows:
We review related surveys in Section~\ref{sec:related_work}, and describe the SLR methodology in Section~\ref{sec:SLR_methodology}.
Section~\ref{sec:VD_taxonomy} presents the comprehensive taxonomy for LLM-based vulnerability detection.
We extend this taxonomy and analyze datasets used in Section~\ref{sec:datasets}.
We discuss current limitations and future research opportunities in Section~\ref{sec:discussion}.
Finally, we address limitations of this SLR in Section~\ref{sec:limitations} and conclude this review in Section~\ref{sec:conclusion}.



\section{Related Work}
\label{sec:related_work}
With the growing adoption of LLMs across various domains, numerous works have emerged to explore their applications. 
In this section, we discuss related works, covering related SLRs and surveys that intersect most closely with the topic of LLM-based software vulnerability detection, as well as works focusing on vulnerability datasets.

\subsection{Related SLRs and Surveys}
In Table~\ref{tab:related_surveys}, we provide an overview of related surveys, with particular attention to their coverage of the main aspects of this SLR, i.e., focus on LLM-based studies, input representations, adaptation techniques, analysis of datasets, and a comprehensive taxonomy.
In the following, we discuss these surveys and how this SLR differs in coverage from previous works.

Several surveys analyze the application of LLMs in the \textbf{domain of cyber security} \cite{xu2024large,ferrag2024generative}, addressing vulnerability detection as one of several sub-tasks.
Xu et al.~\cite{xu2024large}, for example, cover the domains of software and system security, network security, information and content security, hardware security, and blockchain security. The authors focus on what LLMs have been employed to support these security tasks, and discuss domain-specific adaptation techniques, i.e., fine-tuning, prompt engineering, and external augmentation. Further, they address dataset collection and pre-processing techniques.
Ferrag et al.~\cite{ferrag2024generative} focus on LLMs employed for cyber security and their optimization strategies, briefly discussing advanced training and fine-tuning techniques. The authors also provide an overview of software vulnerability datasets used for fine-tuning.

Surveys in the \textbf{domain of software engineering}~\cite{hou2023large,zhang2023survey,fuAIDevSecOpsLandscape2025} categorize LLM usage by phases of the software development life cycle.
Vulnerability detection is typically part of software testing~\cite{zhang2023survey} or software quality assurance~\cite{hou2023large}.
Fu et al.~\cite{fuAIDevSecOpsLandscape2025} focus on AI for DevSecOps, covering software vulnerability detection and classification. Due to the broader scope of AI, the survey covers fewer studies applying LLM. 
Zhang et al.~\cite{zhang2023survey} focus on the LLM perspective, providing a comprehensive overview of code LLMs, discussing pre-training tasks, and downstream applications.
Hou et al.~\cite{hou2023large} discuss techniques used to optimize LLMs, including prompt engineering and fine-tuning techniques. The authors also provide an overview of input forms used for diverse software engineering tasks (i.e., token- and tree/graph-based) and highlight data collection, selection, and pre-processing processes.
While these surveys provide overviews of the application of LLMs in broader domains, we focus on insights into the state-of-the-art adaptation and orchestration techniques, input representations, and datasets that are specific to software vulnerability detection.

\begin{table}[tb]
\caption{Related surveys intersecting with LLM-based software vulnerability detection.  
Extent to which each topic is covered in the corresponding survey: \CIRCLE\, extensive analysis; \LEFTcircle\, overview or analysis focused on specific aspects; \Circle\, not covered; n/s not specified.}
\label{tab:related_surveys}
\fontsize{8pt}{8pt}\selectfont
\setlength{\tabcolsep}{3pt}
\renewcommand{\arraystretch}{1.25}
\resizebox{\textwidth}{!}{%
    \begin{tabular}{@{}lrlcccccccc@{}}
        \toprule
        & & & & & & \multicolumn{5}{c}{Coverage} \\ \cmidrule{7-11}
        \textbf{Reference} & \textbf{Published} & \textbf{Focus} & \textbf{Time Frame} & \makecell[b]{\textbf{\#Studies} \\ (all | SVD)} & \rotatebox[origin=l]{90}{\textbf{SLR}} & \rotatebox[origin=l]{90}{\textbf{LLM Focus}} & 
        \rotatebox[origin=l]{90}{\makecell[l]{\textbf{Input} \\ \textbf{Repres.}}} & 
        \rotatebox[origin=l]{90}{\makecell[l]{\textbf{Adaptation} \\ \textbf{Techniques}}} & 
        \rotatebox[origin=l]{90}{\makecell[l]{\textbf{Dataset} \\ \textbf{Analysis}}} & 
        \rotatebox[origin=l]{90}{\textbf{Taxonomy}} \\
        \midrule
        \rowcolor{gray!08} \multicolumn{11}{l}{\hspace{-0.5em}\textbf{Cyber Security}} \\ 
        Xu et al. \cite{xu2024large} & Sep 2025 & Cyber security & 2020-2025 & 185 | 22 & \ding{51} & \CIRCLE & \Circle & \CIRCLE & \LEFTcircle\;\, & \Circle \\
        Ferrag et al. \cite{ferrag2024generative} & May 2025 & Cyber security & n/s & n/s | 9 & \ding{55} & \CIRCLE & \Circle & \LEFTcircle & \LEFTcircle\;\, & \Circle \\ \midrule

        \rowcolor{gray!08} \multicolumn{11}{l}{\hspace{-0.5em}\textbf{Software Engineering}} \\ 
        Zhang et al. \cite{zhang2023survey} & Sep 2024 & Software engineering & 2020-2024 & 947 | 71 & \ding{51} & \CIRCLE & \Circle & \Circle & \Circle\;\, & \Circle \\
        Hou et al. \cite{hou2023large} & Dec 2024 & Software engineering & 2017-2024 & 395 | 18 & \ding{51} & \CIRCLE & \CIRCLE & \CIRCLE & \LEFTcircle\;\, & \Circle \\ 
        Fu et al. \cite{fuAIDevSecOpsLandscape2025} & Apr 2025 & DevSecOps & 2017-2023 & 99 | 9 & \ding{51} & \Circle & \Circle & \Circle & \LEFTcircle* & \Circle \\ \midrule

        \rowcolor{gray!08} \multicolumn{11}{l}{\hspace{-0.5em}\textbf{Vulnerability Detection}} \\ 

        Shiri Harzevili et al. \cite{shiri2024systematic} & Nov 2024 & Autom. software vulnerability detection & 2011-2024 & 138 | 7 & \ding{51} & \Circle & \CIRCLE & \Circle & \LEFTcircle\;\, & \Circle \\
        Shereen et al. \cite{shereen2024sok} & Dec 2024 & Autom. vulnerability detection & 2006-2024 & 79 | 10 & \ding{55} & \Circle & \LEFTcircle & \Circle & \LEFTcircle* & \Circle \\
        Shimmi et al. \cite{shimmi2025aibasedsoftwarevulnerabilitydetection} & Jun 2025 & Software vulnerability detection & 2018-2023 & 98 | 13 & \ding{51} & \Circle & \CIRCLE & \Circle & \LEFTcircle* & \CIRCLE \\

        Taghavi Far and Feyzi \cite{taghavi2025large} & Feb 2025 & Software vulnerability detection & n/s & n/s | n/s & \ding{51} & \CIRCLE & \Circle & \LEFTcircle & \LEFTcircle\;\, & \Circle \\
        
        Basic and Giaretta \cite{basic2024large} & Apr 2025 & Code security & n/s & n/s | n/s & \ding{51} & \CIRCLE & \Circle & \LEFTcircle & \Circle\;\, & \Circle \\
        Zhou et al. \cite{zhou2024large} & May 2025 & Vulnerability detection and repair & 2018-2024 & 58 | 40 & \ding{51} & \CIRCLE & \Circle & \CIRCLE & \Circle\;\, & \Circle \\
        Germano et al. \cite{germanoSystematicReviewDetection2025} & Nov 2025 & Code vulnerability detection, repair, expl. & 2018-2025 & 208 | 190 & \ding{51} & \CIRCLE & \Circle & \Circle & \LEFTcircle* & \Circle \\ 
        Sheng et al. \cite{sheng2025large} & Sep 2025 & Software security, vulnerability detection & 2017-2025 & 58 | 58 & \ding{55} & \CIRCLE & \LEFTcircle & \CIRCLE  & \LEFTcircle\;\, & \Circle \\ 
        \bottomrule
        & & & & & & & & & \multicolumn{2}{r}{*List only} \\
        \midrule
        \textbf{This SLR} & Dec 2025 & Software vulnerability detection & 2020-2025 & \numStudies & \ding{51} & \CIRCLE & \CIRCLE & \CIRCLE & \CIRCLE\;\, & \CIRCLE \\
        \bottomrule
    \end{tabular}
    }%
\end{table}

Shiri Harzevili et al.~\cite{shiri2024systematic}, Shereen et al.~\cite{shereen2024sok}, and Shimmi et al.~\cite{shimmi2025aibasedsoftwarevulnerabilitydetection} focus on the application of machine learning and deep learning techniques for \textbf{vulnerability detection}, discussing seven, ten, and thirteen studies applying LLMs, respectively.
These studies discuss machine learning and deep learning specific input (tree, graph, token) ~\cite{shiri2024systematic,shimmi2025aibasedsoftwarevulnerabilitydetection} or program representations (source, binary, intermediate)~\cite{shereen2024sok} and provide an overview of common datasets.
Shimmi et al.~\cite{shimmi2025aibasedsoftwarevulnerabilitydetection} also provide a taxonomy of methods, feature representations, and embedding techniques.
In contrast to these surveys, we focus explicitly on LLM-based studies, discussing and categorizing LLM-specific input representations and adaptation techniques, as well as challenges with existing datasets and dataset usage.
Taghavi Far and Feyz~\cite{taghavi2025large} consider a broader scope of software vulnerability detection, including, e.g., phishing detection, threat detection, and patch recommendation. The authors compare the application of LLMs to traditional program analysis techniques, discuss various fine-tuning techniques, and present selected pre-training and fine-tuning datasets.
The surveys by Basic and Giaretta~\cite{basic2024large}, Zhou et al.~\cite{zhou2024large}, and  Germano et al.~\cite{germanoSystematicReviewDetection2025} are closely aligned with software vulnerability detection, extending to code security and vulnerability repair.
Specifically, Basic and Giaretta~\cite{basic2024large} focus on vulnerabilities introduced by LLM-generated code as well as LLMs for detecting and fixing vulnerabilities (in a unified process), discussing various prompting strategies.
Zhou et al.~\cite{zhou2024large} discuss LLMs for vulnerability detection and repair, focusing on early fine-tuning, prompt engineering (zero-shot, few-shot), and retrieval augmentation techniques.
Germano et al.~\cite{germanoSystematicReviewDetection2025} survey studies for vulnerability detection, repair, and explanation with LLMs, discussing the different models and programming languages applied as well as the distribution of datasets.
Most closely to this SLR is the survey by Sheng et al.~\cite{sheng2025large}, covering vulnerability detection techniques and insights. The authors discuss code pre-processing techniques, cover prompt engineering as well as fine-tuning techniques, and provide an overview of languages and granularities in benchmarks and datasets.
In contrast, this SLR further offers a fine-grained taxonomy of prompting and fine-tuning techniques, incorporating learning paradigms and orchestration strategies. Furthermore, we introduce a temporal analysis of dataset usage and critically evaluate the vulnerability diversity and coverage of common benchmarks. Finally, we deepen the discussion on input representations and the role of auxiliary information in both prompting and tuning contexts, as well as classification and generation tasks.

Overall, this SLR extends the field through a dedicated and in-depth overview of \numStudies studies on LLM-based software vulnerability detection.
The further this research field evolves, the more patterns and divergences can be identified. 
By systematically classifying studies using a comprehensive taxonomy, this SLR reveals focused insights that remain hidden in broader surveys, including how vulnerabilities are represented, which (new) techniques have been adapted and are effective for detection, and how study design influences applicability and comparability.

\subsection{Vulnerability Datasets}

A further distinguishing aspect of this SLR from previous surveys is its in-depth analysis of vulnerability detection datasets. 
High-quality, realistic, and diverse datasets play a crucial role in learning-based software vulnerability detection, prompting studies to analyze, classify, and evaluate their characteristics, quality, and impact on model performance.
Discussed surveys have touched upon vulnerability detection datasets: 
Xu et al.~\cite{xu2024large} and Hou et al.~\cite{hou2023large} categorize dataset collection, i.e., open-source, collected, constructed, or industrial, as well as dataset type, i.e., code-based, text-based, graph-based, software repository-based, or combined. 
Shiri Harzevili et al.~\cite{shiri2024systematic} categorize dataset sources into benchmark, hybrid, open source software, and repository. Further, they differentiate code, text, numerical, and hybrid dataset types.
Shereen et al.~\cite{shereen2024sok} compare frequently used datasets with respect to properties such as language, size, annotation, and granularity.
Similarly, other discussed surveys \cite{ferrag2024generative,fuAIDevSecOpsLandscape2025,shimmi2025aibasedsoftwarevulnerabilitydetection,taghavi2025large,germanoSystematicReviewDetection2025,sheng2025large} list used datasets with respect to, e.g., size, programming language, granularity, or source.
In contrast to previous surveys that primarily list datasets, we present a taxonomy and comprehensive analysis. 
We categorize datasets along type, granularity, source, and labeling methodology, and investigate additional quality dimensions, such as class balance, CWE diversity, and distribution, offering an in-depth analysis of represented vulnerabilities.
Further, we add a time component and discuss emerging trends and use cases for datasets in LLM-based software vulnerability detection.

Other related studies focus explicitly on vulnerability detection datasets:
Hu et al.~\cite{hu2025assessingadvancingbenchmarksevaluating} analyze benchmarks for evaluating LLM-based solutions, focusing on construction methods, programming languages, and metrics.
Lin et al.~\cite{lin2022vulnerability} examine dataset construction methodologies, comparing datasets across six dimensions, such as granularity, vulnerability type, and labeling approach. 
Guo et al.~\cite{guo2024comprehensive} analyze selected datasets in terms of vulnerability distributions and types.
Jain et al.~\cite{jain2023code} provide a code-centric evaluation of commonly used \ccpp datasets for deep learning-based vulnerability detection.
Moreover, the works \cite{guo2023investigation,croft2023data,risse2024top,yadav2024r} provide in-depth analyses of quality issues in selected vulnerability datasets, such as data imbalance, low vulnerability coverage, biased vulnerability distribution, high duplication rates, and label noise.
Despite this growing body of work, a targeted and comprehensive overview of vulnerability detection datasets commonly used for LLM-based software vulnerability detection approaches, along with their limitations and suitability, remains missing. 
This survey addresses this gap, offering a taxonomy and systematic analysis of datasets, their CWE coverage, diversity, and usage, thereby facilitating better comparability and benchmarking in future research.


\section{Methodology}
\label{sec:SLR_methodology}
We followed established guidelines by Kitchenham et al.~\cite{kitchenhamGuidelinesPerformingSystematic2007a,petersenGuidelinesConductingSystematic2015} for conducting SLRs in software engineering, i.e., defining research questions, conducting the search, defining selection criteria, and analyzing the selected studies.
In the following, we outline the research focus of this SLR as well as the three phases of literature search, study selection, and study analysis, as visualized in Figure~\ref{fig:study_selection}.

\subsection{Research Focus}

In this SLR, we address the following research questions \textbf{RQ1–RQ5}: 
\begin{enumerate}[label={\textbf{RQ\arabic*}}]
    \item \textbf{How is the vulnerability detection task formulated and approached in LLM-based systems?} 
    Traditional deep-learning-based approaches predominantly frame the detection task as binary classification (vulnerable vs. non-vulnerable)~\cite{shereen2024sok}. While this formulation simplifies evaluation, it falls short in reflecting real-world software engineering workflows, where developers additionally require information on, e.g., vulnerability types and root causes.  
    As LLMs are capable of producing structured reasoning steps and conceptual explanations, understanding whether LLM-based approaches move beyond this limited formulation is essential.
    We investigate how existing LLM-based approaches formulate the task, i.e., do they maintain the binary framing or do they extend or reformulate it, e.g., as multi-label vulnerability type prediction or vulnerability report generation (cf. Section~\ref{sec:task_formulation}).
    
    \item \textbf{How is input, particularly the semantics of vulnerabilities, represented and encoded for LLMs?} 
    Existing machine learning and deep learning approaches convert code into suitable representations to capture semantics, for example, by constructing graph structures for GNNs or vectorizing code embeddings for CNNs~\cite{shiri2024systematic, shereen2024sok}. For LLMs with their world knowledge, it remains unclear whether code alone is sufficient to capture specific vulnerability semantics. Investigating how studies represent input and whether they incorporate auxiliary information is crucial to understanding what information LLMs can leverage for effective detection (cf. Section~\ref{sec:input_representation}).
    
    \item \textbf{What are the predominant techniques for adapting LLMs to the task of software vulnerability detection?}
    LLMs can be adapted via prompting, fine-tuning, or pre-training. Each technique has different implications for data requirements, computational costs, and practical deployment. Since these constraints affect the feasibility of integrating LLM-based methods into software engineering pipelines, this question investigates which adaptation techniques are used and how well they align with practical constraints (cf. Section~\ref{sec:adaptation_techniques}).
    
    \item \textbf{What datasets are used to train, tune, or evaluate LLM-based vulnerability detection approaches, and how does dataset selection impact applicability?}
    Dataset selection influences the types and variety of vulnerabilities under detection. In terms of applicability, synthetic or oversimplified datasets lead to models with poor generalization. Therefore, understanding the characteristics (such as covered vulnerability types, size, and label quality) of the datasets used in current research is essential to enable reliable benchmarking and assess model generalization to real-world applications (cf. Section~\ref{sec:datasets}).

    \item \textbf{How does dataset selection impact comparability across studies?}
    Dataset selection influences whether results are comparable across studies. Inconsistent evaluation setups as well as not adhering to open science practices make it impossible to compare study results. Identifying commonly used datasets as part of evaluation setups enables researchers to select datasets to improve comparability across studies (cf. Section~\ref{sec:datasets_trends_comparability}).
\end{enumerate}

A joint investigation of \textbf{RQ1}-\textbf{RQ5} is crucial to obtain a holistic understanding of LLM-based vulnerability detection research, from input representations to system design and evaluation, and to identify key levers to enhance robustness, practicability, and comparability.

\subsection{Literature Search}
\label{sec:search_string}


To identify relevant studies, we conducted a systematic literature search across three major databases: IEEE Xplore, ACM Digital Library, and arXiv. 
We selected IEEE Xplore and the ACM Digital Library, as they host peer-reviewed publications from conferences and journals.
We further include arXiv for preprints to capture ongoing research that is still in the submission process, given the rapidly evolving nature of this field.

The search string construction followed a two-step process.
First, we manually identified a set of relevant primary studies in our prior publication~\cite{kaniewski2024vulnerability}.
We used these studies as a starting point to derive vulnerability detection-related terms. 
Second, to ensure systematic coverage of LLM-related terminology, we incorporated keywords used in related SLRs~\cite{hou2023large, zhang2023survey}. 
Specifically, we considered the keywords \textit{LLM}, \textit{Large Language Model}, \textit{Pre-trained}, \textit{Pre-training}, \textit{BERT}, \textit{Codex}, \textit{GPT}, \textit{T5}, \textit{ChatGPT}, and \textit{GPT-*}. We iteratively refined the search string by adding observed major model families and used wildcard extensions to ensure systematic coverage. The final search string is as follows: \\

\setlength{\fboxsep}{10pt}
\noindent
\fcolorbox{black}{gray!08}{%
\parbox{\dimexpr\linewidth - 2\fboxsep - 2\fboxrule\relax}{%
\textbf{\textit{Search String}}: \textit{(LLM OR LLMs OR Large Language Model OR Large Language Models OR pre-train* OR GPT* OR ChatGPT OR T5* OR CodeT5* OR Llama* OR CodeLlama* OR Codex OR BERT OR CodeBERT OR GraphCodeBERT OR Bard OR Claude OR Gemma* OR DeepSeek* OR Gemini* OR Mistral* OR Mixtral* OR Phi* or Qwen* OR CodeQwen* OR WizardCoder*) AND ((vulnerability OR vulnerabilities OR CVE OR CWE) AND (detect* OR identif* OR classif* OR analy* OR discover* OR assess*) AND (software OR program OR code))}
}}

\subsection{Study Selection}

\begin{figure*}[t!]
    \centering 
    \includegraphics[width=\textwidth]{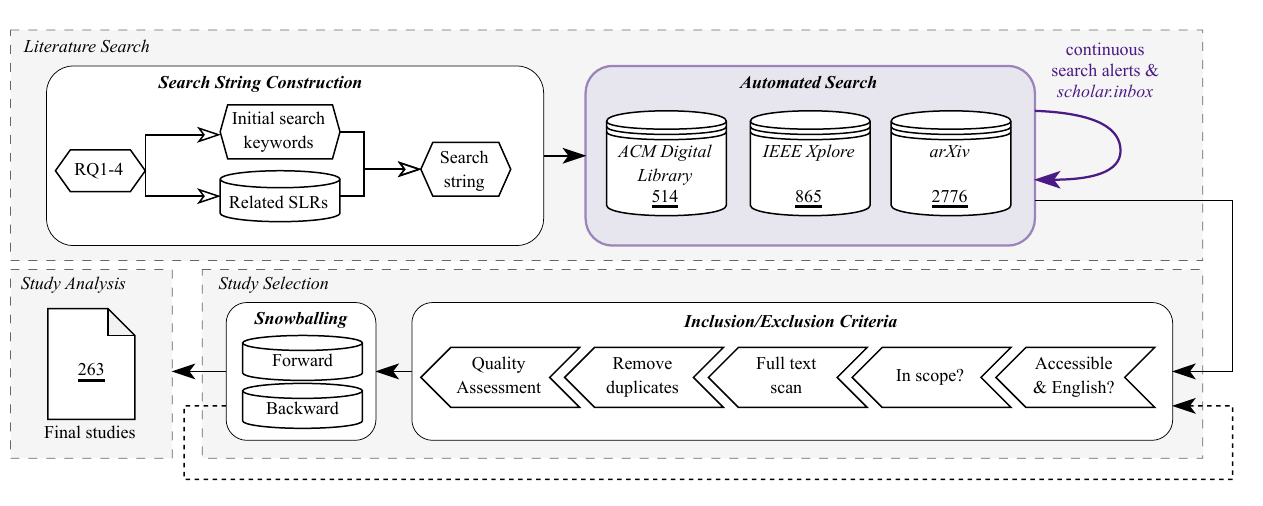}
    \Description{todo.}
    \caption{Literature search and study selection pipeline. We use the constructed search string, cf. Section~\ref{sec:search_string}, for the automated search in the selected databases. Further, we continuously add to the selected studies via search alerts and scholar.inbox notifications. Crawled studies go through the pipeline of inclusion and exclusion criteria as well as snowballing. Study statistics as of December 13\textsuperscript{th} 2025.}
    \label{fig:study_selection}
\end{figure*}

Figure~\ref{fig:study_selection} illustrates the study selection pipeline used in this SLR.
Using the final search string, we ran an automated search across the selected databases to identify studies whose titles or abstracts matched the specified keywords.
We carried out the initial search in December 2024.
After the initial crawl, we continuously integrated newly published material through database-specific search alerts and \textit{scholar.inbox}~\cite{flicke2025scholarinboxpersonalizedpaper}, ensuring that further relevant studies were considered and included.

Following this automated search, we applied a selection process to filter out irrelevant studies based on predefined inclusion~(\cmark) and exclusion~(\xmark) criteria, cf. Table~\ref{tab:inclusion_exclusion}.
We retrieved each accessible study written in English and first filtered by title and abstract before performing a full text screen to assess if it is in scope.
Specifically, we focus on studies that apply LLMs to actively detect software vulnerabilities, rather than applying LLMs to work with outputs of other detection tools, e.g., SAST tools. Further, we specifically focus on software vulnerability detection rather than detection in binary code, smart contracts, network logs, or detection of malware.
We further exclude any duplicates or minor variations of multi- or extended-version papers as well as studies classified as reviews, surveys, theses, abstracts, posters, demonstrations, or tutorials.

Given the inclusion of preprints from arXiv, we conducted a manual quality assessment to determine study eligibility. 
We assessed whether the studies provided a clear contribution, adequately described the proposed workflow and implementation, outlined the evaluation setup (including datasets, baselines, and metrics), and presented coherent findings that support their central claims.

After filtering, we performed backward and forward snowballing. 
We manually reviewed the reference lists of the selected studies (backward snowballing) and used citation tracking via Google Scholar to identify additional papers citing the included studies (forward snowballing).
Studies retrieved through snowballing also pass through the pipeline of inclusion and exclusion criteria.

Finally, at the time of writing this SLR, we cover \numStudies studies. Further studies added through the continuous search will be made available as part of the maintained repository, cf. ~\cite{replicationpackage}.

\begin{table}[tb]
\centering
\caption{Inclusion and exclusion criteria applied in this SLR.}
\label{tab:inclusion_exclusion}
\fontsize{8pt}{8pt}\selectfont
\renewcommand{\arraystretch}{1.25} 
\resizebox{\textwidth}{!}{%
    \begin{tabular}{@{}p{13.5cm}@{}}
        \toprule
        \textbf{Inclusion Criteria} \\ \midrule
    \cmark \, Studies with title or abstract matching the specified search string. \\
    \cmark \, Studies that use and evaluate LLMs for software vulnerability detection, introducing novel techniques or systems. \\[0.75em]
        \textbf{Exclusion Criteria} \\ \midrule
    \xmark \, Studies not written in English. \\
    \xmark \, Studies not accessible in full text. \\
    \xmark \, Studies classified as reviews, surveys, theses, abstracts, posters, demonstrations, or tutorials. \\
    \xmark \, Studies that focus on domains outside source code vulnerability detection, such as binary code, smart contracts, network analysis, or malware detection. \\
    \xmark \, Studies that do not use LLMs to actively detect vulnerabilities but solely to work on outputs produced by other tools. \\
    \xmark \, Studies that mention LLMs only in passing or use them merely as baselines. \\
    \xmark \, Studies identified as duplicates or minor variations of multi- or extended-version papers from the same authors. \\
        \bottomrule
    \end{tabular}
}%
\end{table}


\subsection{Study Analysis}
Following the selection process, we identified \numStudies papers that met the selection criteria. 
During the full-text review of each selected study, we extracted key data points, recording how the vulnerability detection task was defined, which LLMs were used, and what adaptation techniques were employed.
Additionally, we extracted information on input representation formats, datasets used for fine-tuning and evaluation, targeted vulnerability types (e.g., specific CWEs), and evaluation metrics.
This structured analysis enabled us to derive a comprehensive taxonomy of LLM-based software vulnerability detection approaches, address the research questions, and develop a deeper understanding of current trends, open challenges, and emerging directions in LLM-based software vulnerability detection.

\begin{figure*}[t!]
    \centering
    \includegraphics[width=0.97\textwidth]{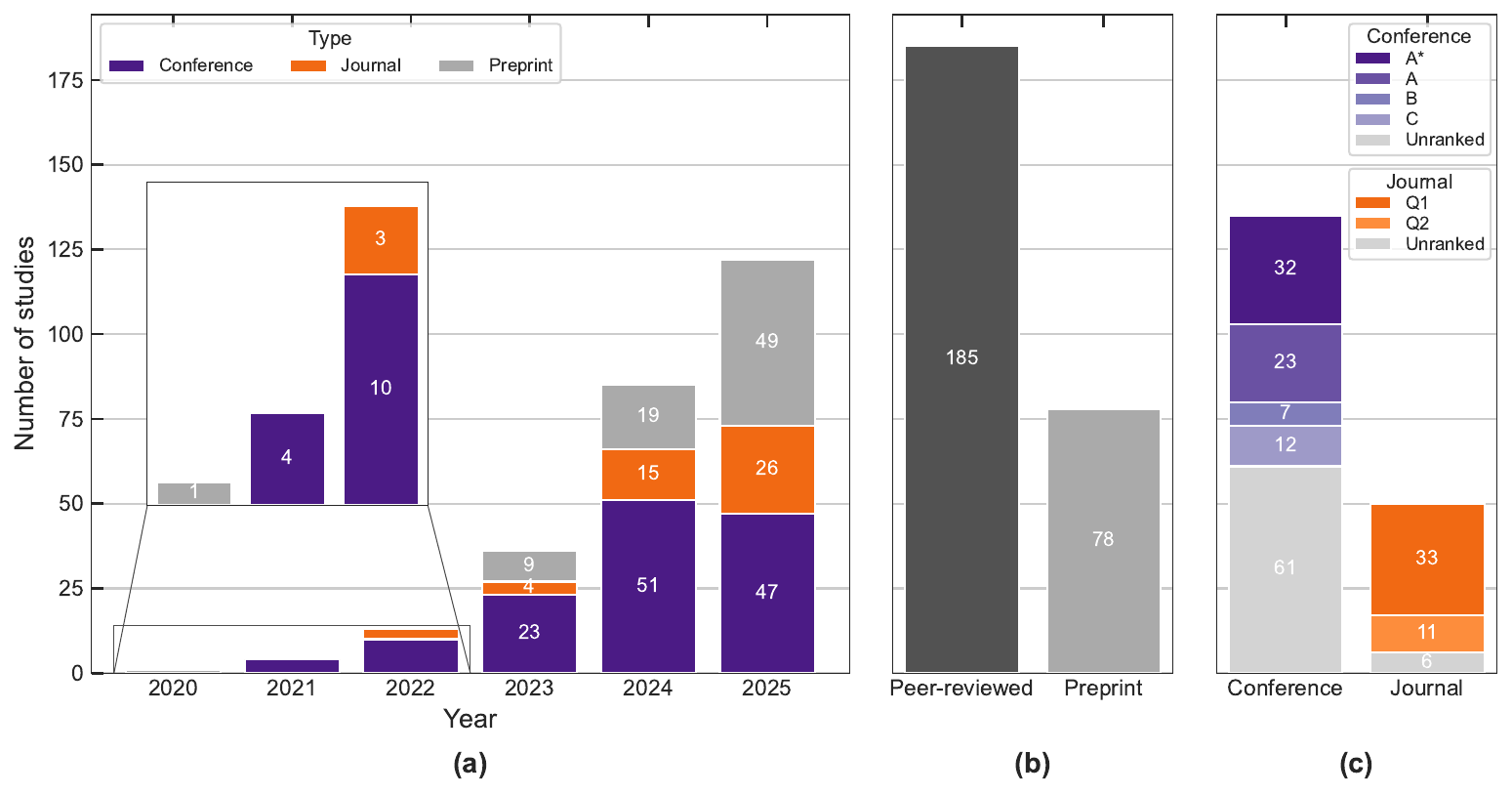}
    \caption{Publication characteristics of the included studies: 
        \textbf{(a)} distribution of studies per year, 
        \textbf{(b)} relation of peer-reviewed vs. preprint (arXiv) studies, and 
        \textbf{(c)} distribution of conference and journal papers across venue tiers, using CORE for conference rankings and SJR for journal rankings.}
    \label{fig:study_analysis}
    \Description{Bar plots displaying (a) the distribution of studies per year - number of studies is rising continuously; (b) relation of peer-reviewed to preprint studies - preprints account for less than half of peer-reviewed studies; (c) break-up of peer-reviewed studies to venue rankings, see Table~\ref{tab:study_venues}.}
\end{figure*}

The earliest included study dates back to 2020 (cf. Fig~\ref{fig:study_analysis} (a)). 
Since then, we can observe a rapid increase in LLM-based vulnerability detection studies.
Notably, 78 (29.7\%) of the selected studies were sourced from arXiv, cf. Fig~\ref{fig:study_analysis} (b), reflecting both the fast-paced development of this field and the prevalence of recent works still undergoing peer review. 
For the peer-reviewed studies, Fig~\ref{fig:study_analysis} (c) visualizes the distribution of the selected studies across venue tiers. 
In addition, Table~\ref{tab:study_venues} lists the respective A*/A conferences as well as Q1 journals with their study counts.

\begin{table}[bt!]
    \centering
    \caption{Selected high-rank venues with their study counts. Venues are associated with the software engineering (\textsuperscript{$\diamondsuit$}), security (\textsuperscript{\textdagger}), and AI domain (\textsuperscript{\sun}), respectively. Venues marked with \textsuperscript{\textopenbullet} have a broader scope, e.g., systems or networks.}
    \label{tab:study_venues}
    \fontsize{8pt}{8pt}\selectfont
    \begin{tabular}{l p{11.8cm}} 
        \toprule
        \textbf{Type} & \textbf{Venue (Count)} \\ 
        \midrule
        \textbf{Conf. A*} & ICSE\textsuperscript{$\diamondsuit$}~(13), EMNLP\textsuperscript{\sun}~(4), USENIX Security\textsuperscript{\textdagger}~(3), ASE\textsuperscript{$\diamondsuit$}~(3), ESEC/FSE\textsuperscript{$\diamondsuit$}~(2), INFOCOM\textsuperscript{\textopenbullet}~(2), ACL\textsuperscript{\textopenbullet}~(2), NDSS\textsuperscript{\textdagger}~(1), SP\textsuperscript{\textdagger}~(1), ICLR\textsuperscript{\sun}~(1) \\ 
        \midrule
        \textbf{Conf. A} & ISSRE\textsuperscript{$\diamondsuit$}~(3), ESORICS\textsuperscript{\textdagger}~(3), ACSAC\textsuperscript{\textdagger}~(2), ISSTA\textsuperscript{$\diamondsuit$}~(2), SANER\textsuperscript{$\diamondsuit$}~(2), 
        MSR\textsuperscript{$\diamondsuit$}~(2), DSN\textsuperscript{\textopenbullet}~(1), EASE\textsuperscript{$\diamondsuit$}~(1), EuroS\&P\textsuperscript{\textdagger}~(1), ASIACCS\textsuperscript{\textdagger}~(1), ICPC\textsuperscript{$\diamondsuit$}~(1), ICSME\textsuperscript{$\diamondsuit$}~(1), ICST\textsuperscript{$\diamondsuit$}~(1), NAACL\textsuperscript{\textopenbullet}~(1), RAID\textsuperscript{\textdagger}~(1) \\ 
        \midrule
        \textbf{Jour. Q1} & TSE\textsuperscript{$\diamondsuit$}~(7), IEEE Access\textsuperscript{\textopenbullet}~(4), COSE\textsuperscript{\textdagger}~(4), EMSE\textsuperscript{$\diamondsuit$}~(4), TOSEM\textsuperscript{$\diamondsuit$}~(3), JISA\textsuperscript{\textdagger}~(3), JSS\textsuperscript{$\diamondsuit$}~(3), IST\textsuperscript{$\diamondsuit$}~(1), Info. Fusion\textsuperscript{\textopenbullet}~(1), Cluster Computing\textsuperscript{\textopenbullet}~(1), TDSC\textsuperscript{\textdagger}~(1), Neural Comput. Appl.\textsuperscript{\textopenbullet}~(1) \\ 
        \bottomrule
    \end{tabular}
\end{table}


\begin{figure*}[b!]
    \centering 
    \includegraphics[width=\linewidth]{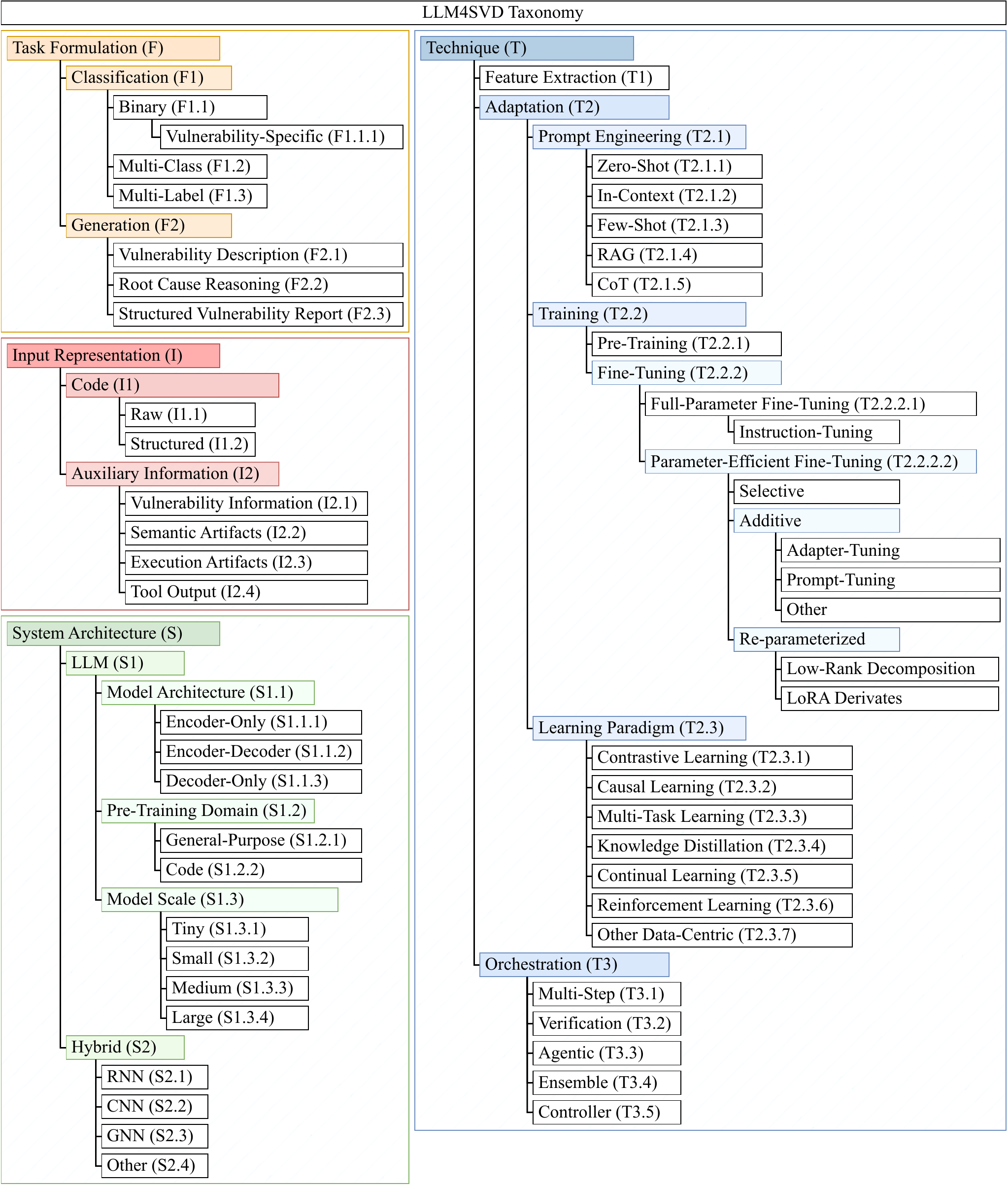}
    \Description{Tree structure showing categories and attribute values for LLM-based vulnerability detection studies.}
    \caption{Taxonomy of LLM-based vulnerability detection studies with numbering in parentheses. We omit individual numbering in lower-level nodes for readability. A study may be associated with multiple values (i.e., white boxes) per category.}
    \label{fig:taxonomy_all}
\end{figure*} 

\section{Vulnerability Detection with LLMs}
\label{sec:VD_taxonomy}
Recent research has adopted a variety of techniques for detecting software vulnerabilities using LLMs. 
In this chapter, we present a comprehensive taxonomy that categorizes the different ways LLMs are applied to vulnerability detection. 
Specifically, we examine task formulations, input representations, system architectures, and techniques. 
The overall taxonomy is illustrated in Figure~\ref{fig:taxonomy_all}. 
We discuss each category on the basis of the surveyed studies in the following subsections.


\subsection{Task Formulation (F)}
\label{sec:task_formulation}
Software vulnerability detection can be framed in multiple ways, depending on how the task is formulated and what objectives are prioritized beyond detection (e.g., explanation).
With respect to the objectives, the integration of LLMs into vulnerability detection introduces unique advantages over traditional approaches.
Unlike rule-based approaches, which require definitions of vulnerable patterns, the language understanding capabilities of LLMs allow for highly flexible goal setting; detection objectives can be refined or expanded, e.g., adding reasoning or vulnerability localization, simply by adjusting natural language prompts.
LLMs further have extensive pre-trained world knowledge, such as software architecture patterns, programming language-specific constructs, and broad cybersecurity concepts. Such domain knowledge is difficult to encode explicitly in traditional tools or classical ML approaches, but serves as a versatile starting point for LLM-based detection.
In the following, we outline and discuss the most common detection task formulations used with LLMs in the surveyed studies: classification and generation tasks.

\subsubsection{Classification (F1)}
Vulnerability detection is most commonly formulated as a classification problem, differentiating binary and multi-class forms.
    \textit{\textbf{Binary Classification (F1.1)}} determines whether a given code contains a security vulnerability or not ('Yes'/'No').
        \textit{\textbf{Vulnerability-Specific Classification (F1.1.1)}}~\cite{shereen2024sok}, as a refinement of the binary formulation, determines whether a given code contains a specific vulnerability, represented, e.g., by CWE-ID.
    \textit{\textbf{Multi-Class Classification (F1.2)}} determines which specific type of vulnerability the given code contains (type prediction), often using CWE-IDs as class labels. This formulation may use a given pre-defined list of vulnerability types in a prompt engineering setting. Similarly, \textit{\textbf{Multi-Label Classification (F1.3)}} aims to determine any vulnerability types present in the code, recognizing that a code snippet may contain multiple distinct vulnerabilities simultaneously.

Binary classification (F1.1, \numBinaryStudies studies) simplifies the detection task as it is reduced to a decision of whether a code sample is vulnerable or not. 
While this approach facilitates model training and deployment, it lacks fine-grained insight into the specific type or cause of the vulnerability, limiting its use in real-world vulnerability remediation. 

Vulnerability-specific classification (F1.1.1, \numVulSpecificStudies studies) is more targeted, which is valuable in contexts with high-impact CWEs or regulatory compliance.
However, it depends on curated datasets and may struggle to generalize to related vulnerability types not present in the training data.

Multi-class (F1.2, \numMultiClassStudies studies) and multi-label (F1.3, \numMultiLabelStudies studies) classification provide the most detailed perspective by distinguishing among various types of vulnerabilities. 
These formulations align more closely with real-world vulnerability remediation, where identifying the exact nature of a vulnerability is essential for vulnerability patching. 
However, these formulations introduce additional complexity in both modeling and evaluation. 
Their effectiveness also depends heavily on high-quality, CWE-labeled, and well-distributed datasets.

Some studies combine or cascade these formulations. 
For example, a binary classifier may act as a preliminary filter or router, forwarding likely vulnerable samples to a CWE-specific classifier for further analysis~\cite{yangMoEVDEnhancingVulnerability2025}.

\begin{figure*}[tb]
    \centering  
        \includegraphics[width=0.85\textwidth,trim={5em 0 5em 2em},clip]{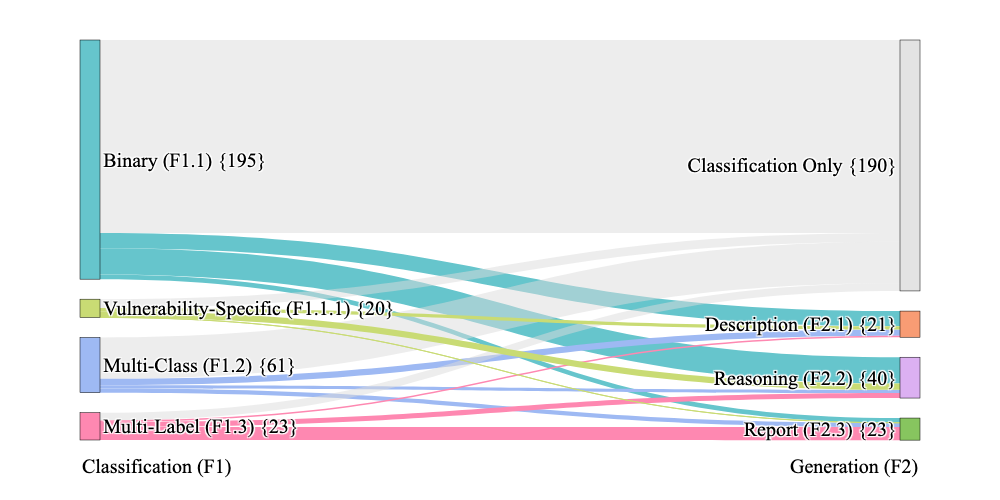} 
        \caption{Relationship between classification and generation task formulations. The minority of studies combines the classification task with a generative task.
        Numbers in braces indicate the absolute count of unique studies applying the task formulation; a single study may perform multiple tasks.}
        \Description{A Sankey diagram relating classification tasks on the left to generation tasks on the right. The largest node on the left is Binary classification. Flows connect these classification nodes to generation nodes such as Reasoning and Description.}
        \label{fig:sankey_task}
\end{figure*} 

\subsubsection{Generation (F2)}
While classification is the primary formulation, it offers limited explanations and insights into the root causes of vulnerabilities. 
Therefore, several studies extend the scope of classification and leverage the language understanding and generation capabilities of LLMs to generate additional outputs.
This generative approach takes advantage of the model's flexibility, allowing researchers to request diverse detection outputs, such as reasoning chains or fix suggestions, through simple natural language instructions.
We categorize generative tasks as (i)~vulnerability description, (ii)~root cause reasoning, and (iii)~structured vulnerability report.

\textit{\textbf{Vulnerability Description (F2.1)}} focuses on generating a high-level natural language summary or description of the specific detected vulnerability, providing developers with its context~\cite{chanTransformerbasedVulnerabilityDetection2023,cipolloneAutomatingDetectionCode2025,ezenwoyeExploringAIVulnerability2024,fuVulExplainerTransformerBasedHierarchical2023a,gajjarBridgingSemanticsStructure2025,ingemanntuffvesonjensenSoftwareVulnerabilityFunctionality2024,islamUnbiasedTransformerSource2023,kouliaridisAssessingEffectivenessLLMs2024,linEvaluatingLargeLanguage2025,liSteeringLargeLanguage2025,liVulnTeamTeamCollaboration2025,mohajerEffectivenessChatGPTStatic2024,safdarRealVulLLMLLMBased2025,shuLargeLanguageModels2025,sovranoLargeLanguageModels2025,steenhoekClosingGapUser2024,tambergHarnessingLargeLanguage2024,wangANVILAnomalybasedVulnerability2024,wangM2CVDEnhancingVulnerability2024,yildirimEvaluatingImpactConventional2024,yildizBenchmarkingLLMsLLMbased2025}.

\textit{\textbf{Root Cause Reasoning (F2.2)}} moves beyond high-level descriptions to provide a code-level justification or reasoning of the specific vulnerability and its root cause~\cite{ahmedSecVulEvalBenchmarkingLLMs2025,dolcettiHelpingLLMsImprove2024,duVulRAGEnhancingLLMbased2024,espinhagasibaMaySourceBe2024,hairanCompressingLargeLanguage2025,islamEnhancingSourceCode2024,jararwehLLaVulMultimodalLLM2025,keltekBoostingCybersecurityVulnerability2024,khareUnderstandingEffectivenessLarge2024,kholoosiQualitativeStudyUsing2024,liCryptoScopeUtilizingLarge2025,liEverythingYouWanted2025,liLLMAssistedStaticAnalysis2024,liMAVULMultiAgentVulnerability2025,liuSoftwareVulnerabilityDetection2023,liuVulDetectBenchEvaluatingDeep2024,liVulnTeamTeamCollaboration2025,maoEffectivelyDetectingExplaining2024,mathewsLLbezpekyLeveragingLarge2024,mhatreLLMGUARDLargeLanguage2025,niDistinguishingLookAlikeInnocent2023,safdarRealVulLLMLLMBased2025,sayaghThinkBroadAct2025,simoniImprovingLLMReasoning2025,steenhoekComprehensiveStudyCapabilities2024a,sunLLM4VulnUnifiedEvaluation2024,ullahLLMsCannotReliably2024,wangLeveragingLargeLanguage2025,wangSecureMindFrameworkBenchmarking2025,wangVulAgentHypothesisValidationBased2025,weiDistillingLightweightLanguage2025,wenBoostingVulnerabilityDetection2025,weyssowR2VulLearningReason2025,widyasariChatGPTEnhancingSoftware2024,yangContextEnhancedVulnerabilityDetection2025,yangDLAPDeepLearning2025,zhuSpecificationGuidedVulnerabilityDetection2025,zibaeiradReasoningLLMsZeroShot2025,chenGPTVDVulnerabilityDetection2026,nguyenHumanUnderstandableExplanationSoftware2024}.
However, the reliability of LLM-generated reasoning remains under scrutiny. 
For instance, Steenhoek et al.~\cite{steenhoekComprehensiveStudyCapabilities2024a} perform an error analysis on LLM responses, demonstrating that the models struggle to reason about the code semantics relevant to identifying vulnerabilities, especially subtle semantic differences caused by small textual changes.
Wen et al.~\cite{wenBoostingVulnerabilityDetection2025} propose forward and backward reasoning, where forward reasoning aims to deduce the causes of vulnerabilities, while backward reasoning seeks to understand the code changes in the fixes. Their results show that LLMs can reason about vulnerabilities but often lack the domain-specific knowledge required to effectively invoke such reasoning.

Other approaches request a \textit{\textbf{Structured Vulnerability Report (F2.3)}}~\cite{dubniczkyCASTLEBenchmarkingDataset2025,gajjarMalCodeAIAutonomousVulnerability2025a,gnieciakLargeLanguageModels2025,keltekBoostingCybersecurityVulnerability2024,khareUnderstandingEffectivenessLarge2024,mahmudEnsembleTransformerApproach2025,noeverCanLargeLanguage2023,pelofskeAutomatedSoftwareVulnerability2024,pengICodeReviewerImprovingSecure2025,shashwatPreliminaryStudyUsing,sunLLM4VulnUnifiedEvaluation2024,weiDistillingLightweightLanguage2025,weyssowR2VulLearningReason2025,widyasariLetTrialBegin2025,xiaStaticPatternMatching2025,yangDLAPDeepLearning2025,yuInsightSecurityCode2024,zhouSSRFSeekLLMbasedStatic2025,zhuSpecificationGuidedVulnerabilityDetection2025,zibaeiradReasoningLLMsZeroShot2025,jiaoDeepVulHunterEnhancingCode2025,tsaiLeveragingIntraInterReferences2025,nguyenHumanUnderstandableExplanationSoftware2024}: 
Studies often aggregate multiple predictive sub-tasks, such as vulnerability localization, severity estimation, type justification, and repair fixes, into a single structured output (e.g., JSON, structured text, or standardized static analysis formats~\cite{gnieciakLargeLanguageModels2025}).
For example, Zibaeirad and Vieira~\cite{zibaeiradReasoningLLMsZeroShot2025} prompt models to generate a report containing the CWE-ID, fix suggestion, vulnerability severity, and exploitability. 
Similarly, Yang et al.~\cite{yangDLAPDeepLearning2025} structure the output to include the label, confidence score, vulnerability type, and affected components.
Zhou et al.~\cite{zhouSSRFSeekLLMbasedStatic2025} produce reports detailing function names, namespaces, source/sink type, and dangerous parameter indices.

These generative tasks reflect a broader shift in the role of LLMs from simple classifiers to end-to-end vulnerability assistants that support developers interactively throughout the software development life cycle.
Figure~\ref{fig:sankey_task} maps the relationships between classification and generation tasks across the surveyed studies.
Each node is annotated with the absolute number of studies using that task formulation, providing a quantitative overview of the landscape. 
The flows further visualize studies that combine classification with generative tasks.
These combined tasks create opportunities for multitask benchmarks that evaluate detection accuracy alongside, e.g., reasoning quality, localization precision, and severity estimation, offering a more holistic perspective on vulnerability remediation.



\subsection{Input Representation (I)}
\label{sec:input_representation}
The representation of the input code and additional context plays a pivotal role in effective vulnerability detection. 
Distinct from traditional static analysis tools that require language-specific rules, LLMs offer unique capabilities in how they process input.
First, their extensive pre-trained world knowledge allows them to interpret diverse input forms, e.g., code and natural language instructions.
This flexibility enables the processing of multi-language source code with varied vulnerability types and auxiliary information within a unified framework.
Second, their semantic understanding allows LLMs to generalize; by reasoning about the input rather than matching fixed vulnerable patterns, LLMs may identify rare or novel vulnerabilities.
Considering these capabilities, we categorize input representations along the code and auxiliary information axes. 

\subsubsection{Code (I1)}
\textit{\textbf{Raw (I1.1)}} input representations of code involve feeding source code directly into the model during inference as plain text sequences (used by 139 studies).
This approach relies entirely on the model's understanding and pre-learned knowledge of programming syntax to detect vulnerable patterns.
This representation is widely used for training or fine-tuning LLMs. 
However, raw code inputs may struggle to capture the deeper semantics and context necessary for detecting subtle or complex vulnerabilities. 
As such, this approach may underperform in scenarios requiring precise reasoning about program logic, data flow, or code structure.

Combining raw code with or expanding representations to \textit{\textbf{Structured (I1.2)}} code enhances the model’s understanding of code by capturing deeper syntactic and semantic structures of code and vulnerabilities, thereby addressing the limitations of relying solely on textual patterns. 
Code is transformed into graph-based or other structured formats that expose control flow, data flow, and dependency relationships relevant to vulnerability semantics. 
Common \textit{graph representations} include Abstract Syntax Trees (ASTs)~\cite{ahmedSecVulEvalBenchmarkingLLMs2025,bahaaDBCBILDistilBertBasedTransformer2024,jianjieCodeDefectDetection2023,liRevisitingPretrainedLanguage2025,wenSCALEConstructingStructured2024,zhangVulTrLMLLMassistedVulnerability2026,yuVulnerabilityDetectionFramework2025}, specifically flattened~\cite{liRevisitingPretrainedLanguage2025} or decomposed~\cite{zhangVulTrLMLLMassistedVulnerability2026}, 
Control Flow Graphs (CFGs)~\cite{gajjarBridgingSemanticsStructure2025,liuEnhancingVulnerabilityDetection2025,luGRACEEmpoweringLLMbased2024,sunEnhancingSourceCode2024,zhangVulnerabilityDetectionLearning2023}, 
Data Flow Graphs (DFGs)~\cite{jiangDFEPTDataFlow2024,liuExplorationPromptingLLM2024,wangSCLCVDSupervisedContrastive2024,yuPATVDVulnerabilityDetection2022}, forms of
Program Dependency Graphs (PDGs)~\cite{duGeneralizationEnhancedCodeVulnerability2024,hinLineVDStatementlevelVulnerability2022,luGRACEEmpoweringLLMbased2024,sunHgtJITJustinTimeVulnerability2025,wuSoftwareVulnerabilityDetection2024}, combining data and control dependencies, and 
Code Property Graphs (CPGs)~\cite{changFineTuningPretrainedModel2024,fengCGPTuningStructureAwareSoft2025,lekssaysLLMxCPGContextAwareVulnerability2025,liCLeVeRMultimodalContrastive,liuVulLMGNNsFusingLanguage2025,nganMultimodalFusionVulnerability2025,niFunctionLevelVulnerabilityDetection2023,rangapuramGraphCodeBERTAugmentedGraphAttention2025,liangSourceCodeVulnerability2024,yuVulnerabilityDetectionFramework2025}, which integrate multiple code views (AST, CFG, PDG).
Some studies employ composite or multi-layered graph structures that integrate several of the above~\cite{jiang2022multi,liuMakingVulnerabilityPrediction2024,islamUnbiasedTransformerSource2023,islamUnintentionalSecurityFlaws2024,zhengLearningFocusContext2025,zhengSVulDetectorVulnerabilityDetection2024,liuEnhancingVulnerabilityDetection2025,luGRACEEmpoweringLLMbased2024}, which can outperform other graphs~\cite{zhangComparingPerformanceDifferent2023a}, or combine multiple modalities including text, graphs, and visual representations~\cite{niFunctionLevelVulnerabilityDetection2023}.

In addition to graph structures, other structured inputs aim to isolate semantically relevant code regions, such as \textit{program or code slices}~\cite{wu2025sparse,tianEFVDFrameworkSource2025,dingLeveragingDeepLearning2023,huangBBVDBERTbasedMethod2022,sunEnhancedVulnerabilityDetection2023,zhuBERTBasedVulnerabilityType2022}, \textit{chunks}~\cite{halderFuncVulEffectiveFunction2025},
or \textit{code gadgets}~\cite{phamDefectscannerComparativeEmpirical2024,gujarDetectBERTCodeVulnerability2024,gurfidanVULREMFineTunedBERTBased2024,kimVulDeBERTVulnerabilityDetection2022,liAutomatedSoftwareVulnerability2022,omarVulDefendNovelTechnique2023,omarVulDetectNovelTechnique2023,pengPTLVDProgramSlicing2023,purbaSoftwareVulnerabilityDetection2023,quanXGVBERTLeveragingContextualized2023,thapaTransformerBasedLanguageModels2022,zhaoAdversarialTrainingRobustness2025,zhaoHowGetBetter2023,chenGPTVDVulnerabilityDetection2026}.
Code gadgets, in particular, consist of program statements that are semantically related to each other through data or control dependencies~\cite{VulDeePecker}, isolating vulnerability-relevant code regions.
Other studies include program traces, i.e., execution-aware code structure~\cite{dingTRACEDExecutionawarePretraining2024,machtleTraceGadgetsMinimizing2025}. Mächtle et al.~\cite{machtleTraceGadgetsMinimizing2025} specifically build \textit{trace gadgets}, i.e., trace the code before slicing.
By making structural and semantic dependencies explicit, structure-aware representations enable models to reason more effectively about vulnerability-related behaviors, especially those that span across the codebase or require understanding of program logic and execution flow.

\subsubsection{Auxiliary Information (I2)}
To enhance the effectiveness of vulnerability detection, studies move beyond source code by augmenting the input with auxiliary information, particularly in prompt-engineering settings. 
This additional information provides the model with domain knowledge or code context. 
We categorize this auxiliary information into (i)~vulnerability information, (ii)~semantic artifacts, (iii)~execution artifacts, and (iv)~tool output.

\textit{\textbf{Vulnerability Information (I2.1)}} involves explicitly equipping the model with domain vulnerability knowledge.
Most commonly, studies include a pre-defined list of CWEs to the prompt to narrow the classification scope~\cite{bakhshandehUsingChatGPTStatic2023,caoLLMCloudSecLargeLanguage2024,dozonoLargeLanguageModels2024,fuChatGPTVulnerabilityDetection2023,gnieciakLargeLanguageModels2025,liEverythingYouWanted2025,liuVulDetectBenchEvaluatingDeep2024,sayaghThinkBroadAct2025, shashwatPreliminaryStudyUsing,steenhoekComprehensiveStudyCapabilities2024a,wangEnhancingLargeLanguage2023,xiaStaticPatternMatching2025,yuInsightSecurityCode2024} or provide labeled examples (few-shot settings)~\cite{chanTransformerbasedVulnerabilityDetection2023,dingVulnerabilityDetectionCode2024,dolcettiHelpingLLMsImprove2024,farrExpertintheLoopSystemsCrossDomain2025,hannanSelectingFewShotExamples2025,liRevisitingPretrainedLanguage2025,liuExplorationPromptingLLM2024,liuSoftwareVulnerabilityDetection2023,maoEffectivelyDetectingExplaining2024,niLearningbasedModelsVulnerability2024,nongChainofThoughtPromptingLarge2024,ouchebaraLlamabasedSourceCode2025,safdarRealVulLLMLLMBased2025,shuLargeLanguageModels2025,sovranoLargeLanguageModels2025,tradManualPromptEngineering2025,tsaiSequentialMultiStageApproach2025,wangSecureMindFrameworkBenchmarking2025,widyasariChatGPTEnhancingSoftware2024,yangDLAPDeepLearning2025,yildizBenchmarkingLLMsLLMbased2025,yinMultitaskBasedEvaluationOpenSource2024,zhouComparisonStaticApplication2024,zhouLargeLanguageModel2024c,chenGPTVDVulnerabilityDetection2026,jiaoDeepVulHunterEnhancingCode2025,tsaiLeveragingIntraInterReferences2025,leSoftwareVulnerabilityPrediction2024,luGRACEEmpoweringLLMbased2024,shengLProtectorLLMdrivenVulnerability2024a,zhangBenchmarkingLargeLanguage2025}.
Other studies include more detailed vulnerability (CWE) definitions and descriptions~\cite{sunLLM4VulnUnifiedEvaluation2024,shimmiSoftwareVulnerabilityDetection2024,liCryptoScopeUtilizingLarge2025,kouliaridisAssessingEffectivenessLLMs2024,ullahLLMsCannotReliably2024,zhuSpecificationGuidedVulnerabilityDetection2025,mohajerEffectivenessChatGPTStatic2024,widyasariLetTrialBegin2025,cekaCanLLMPrompting2024,pelofskeAutomatedSoftwareVulnerability2024,mathewsLLbezpekyLeveragingLarge2024,duVulRAGEnhancingLLMbased2024}, particularly as part of retrieval architectures, or full analysis reports~\cite{luoHALURustExploitingHallucinations2025}.

\textit{\textbf{Semantic Artifacts (I2.2)}} focus on the high-level intent and functionality of the code, bridging the gap between syntax and understanding. 
For example, studies provide natural language explanations of the code functionality~\cite{gajjarMalCodeAIAutonomousVulnerability2025a,torkamaniStreamliningSecurityVulnerability2025}, questions regarding the code's behavior~\cite{jararwehLLaVulMultimodalLLM2025}, explicit descriptions of data sources and sinks~\cite{liCLeVeRMultimodalContrastive}, or code comments and docstrings~\cite{huynhDetectingCodeVulnerabilities2025}.

\textit{\textbf{Execution Artifacts (I2.3)}} capture the behavior, structural dependencies, or runtime environment of the code.
Studies enrich inference input with data and control dependencies, interprocedural interactions, and environment constraints~\cite{ahmedSecVulEvalBenchmarkingLLMs2025}.
For instance, researchers include information on callee functions~\cite{sunLLM4VulnUnifiedEvaluation2024,liEverythingYouWanted2025}, dynamic execution signals such as unit test results~\cite{mhatreLLMGUARDLargeLanguage2025}, or code-context~\cite{liLLMAssistedStaticAnalysis2024,sayaghThinkBroadAct2025,liEverythingYouWanted2025}, e.g., global variables, type declarations, and library imports.
Data flow analysis techniques are also used~\cite{liRevisitingPretrainedLanguage2025,tambergHarnessingLargeLanguage2024,yangContextEnhancedVulnerabilityDetection2025,zhouSSRFSeekLLMbasedStatic2025}, e.g., to extract dependencies from call graphs~\cite{wenVulEvalRepositoryLevelEvaluation2024}.

\textit{\textbf{Tool Output (I2.4)}} augments the model context with the outputs of other tools, most commonly SAST tools, allowing the LLM to interpret the static analysis findings to refine further predictions~\cite{bakhshandehUsingChatGPTStatic2023,keltekBoostingCybersecurityVulnerability2024,munsonLittleHelpMy2025,wangVulAgentHypothesisValidationBased2025,yangDLAPDeepLearning2025}.
Yang et al.~\cite{yangDLAPDeepLearning2025} inputs the results of other deep learning models (i.e., the prediction probability) to inform the final classification.


\subsection{System Architecture (S)}
\label{sec:system_architecture}
Vulnerability detection systems using LLMs differ significantly in which and how models are integrated into the overall system. Broadly, we categorize existing approaches into two groups: LLM-centric systems, where the LLM serves as the main predictive component, and hybrid systems, which integrate an LLM with other learning-based components.

\subsubsection{LLM (S1)}
In LLM-centric approaches, the LLM performs the core predictive tasks, such as classification and reasoning. The model is typically adapted to the detection task through prompt engineering or fine-tuning.
Additional tools, such as SAST, only provide auxiliary information rather than predictive signals.
The surveyed studies employ a variety of LLMs with different architectures, pre-training domains, and scales. 
In the following, we discuss LLMs within these categories to categorize systems in more detail.  

\paragraph{Model Architecture (S1.1)} 
Most LLMs are based on the Transformer architecture, which includes encoder and decoder modules with a self-attention mechanism. 
We follow other works and categorize the LLMs used in the surveyed studies into the three architectural groups (i)~encoder-only, (ii)~encoder-decoder, and (iii)~decoder-only, as introduced by Pan et al.~\cite{panUnifyingLargeLanguage2024}.

\begin{itemize}
    \item \textit{\textbf{Encoder-Only (S1.1.1)}} LLMs, e.g., models from the BERT~\cite{BERT} family (CodeBERT~\cite{feng2020codebert}, GraphCodeBERT~\cite{guo2020graphcodebert}, RoBERTa \cite{ROBERTA}, DistilBERT~\cite{sanh2019distilbert}), use only the encoder to encode the sentence and comprehend the relationships between words~\cite{panUnifyingLargeLanguage2024}. The encoder maps the input code into a fixed-size dense vector representation (embedding) that captures semantic relationships. In vulnerability detection, these models are typically adapted via fine-tuning by adding a classification head (e.g., a multi-layer perceptron) to output a probability score or class label. Alternatively, they function as frozen feature extractors, generating embeddings that serve as input for external classifiers.    
    
    \item \textit{\textbf{Encoder-Decoder (S1.1.2)}} LLMs, e.g., T5~\cite{raffel2020T5}, CodeT5~\cite{wang2021codet5,wang2023codet5+}, or UniXcoder~\cite{guo2022unixcoder}, use both the encoder and decoder module in a sequence-to-sequence framework~\cite{panUnifyingLargeLanguage2024}. The encoder maps the input code into a latent hidden-state representation, which the decoder uses to generate a target output sequence. For vulnerability detection, this architecture allows the task to be framed as text generation, where the input is the source code and the output is a generated text string, e.g., the label "vulnerable" or a natural language explanation.
    
    \item \textit{\textbf{Decoder-Only (S1.1.3)}} LLMs use only the decoder module, using auto-regressive mechanisms to predict the next token in a sequence. The input, typically a prompt containing instructions and the code snippet, is processed sequentially, and the output is generated as a continuation of this input, enabling the generation of reasoning chains and structured vulnerability reports. Many state-of-the-art LLMs follow this architecture, e.g., the GPT family~\cite{gpt2,gpt3,gpt4} and the Llama family~\cite{touvron2023llama,touvron2023llama2,grattafiori2024llama3}. Decoder-only LLMs are the dominant architecture used for vulnerability detection in the surveyed studies.
\end{itemize}

\paragraph{Pre-Training Domain (S1.2)} 
Surveyed studies rely on LLMs from two main domains: general-purpose LLMs and code LLMs.
\textit{\textbf{General-Purpose LLMs (S1.2.1)}} are pre-trained on natural language corpora and tend to perform well on explanation-based or natural language-heavy tasks. Examples from this domain are the BERT family, the GPT family, the Llama family, the Qwen series~\cite{bai2023qwen,yang2024qwen2technicalreport,qwen2025qwen25technicalreport}, Gemma~\cite{team2024gemma,team2024gemma2}, DeepSeek~\cite{guo2025deepseekr1,liu2024deepseekv2}, Mistral~\cite{jiang2023mistral7b}, Mixtral~\cite{jiang2024mixtral}, Claude~\cite{anthropic2024claude}, Gemini~\cite{team2024gemini15}, and Phi~\cite{javaheripi2023phi2,abdin2024phi3}.
\textit{\textbf{Code LLMs (S1.2.2)}} are pre-trained specifically on large-scale code corpora for code-related tasks, enhancing performance on syntax-sensitive or program-structure-aware tasks. Examples include CodeBERT~\cite{feng2020codebert}, GraphCodeBERT~~\cite{guo2020graphcodebert}, StarCoder~\cite{li2023starcoder,lozhkov2024starcoder2}, UniXcoder~\cite{guo2022unixcoder}, the GPT-based Codex~\cite{chen2021codex}, CodeLlama~\cite{roziere2023codellama}, CodeQwen 1.5~\cite{bai2023qwen}, Qwen2.5-Coder~\cite{hui2024qwen25coder}, CodeGemma~\cite{team2024codegemma}, DeepSeek-Coder~\cite{guo2024deepseekcoder,zhu2024deepseekcoderv2}, and CodeT5~\cite{wang2021codet5,wang2023codet5+}.

\paragraph{Model Scale (S1.3)}
The parameter count of an LLM typically serves as a proxy for its capacity to learn complex patterns and encode domain knowledge. 
However, increasing scale imposes proportional demands on computational resources for training and inference.
Notably, there exists no formal consensus regarding the exact parameter threshold that constitutes a "large" language model.
Despite this ambiguity, to analyze the trade-off between performance and resource efficiency, we categorize the models used in the surveyed studies into four distinct groups:

\begin{itemize}
    \item \textit{\textbf{Tiny (S1.3.1)}}: Models with fewer than 1 billion parameters \textbf{($<$1B)}. This category primarily includes pre-trained language models such as CodeBERT or UniXcoder. Due to their lightweight nature, they are highly efficient for fine-tuning and inference, making them suitable for resource-constrained environments.
    \item \textit{\textbf{Small (S1.3.2)}}: Models with 1 to 8 billion parameters \textbf{(1B--8B)}, such as Gemma-2B, Llama2-7B, or Llama3.1-8B. This scale represents a trade-off for many researchers; these models provide a significant improvement in generative capabilities over tiny models while remaining small enough to be deployed or fine-tuned on high-end consumer-grade hardware.
    \item \textit{\textbf{Medium (S1.3.3)}}: Models between 8 and 70 billion parameters \textbf{(8B--70B)}, including CodeLlama-34B, Mixtral 8x7B, and DeepSeek-R1-32B. These models typically outperform smaller counterparts in complex reasoning tasks but exceed the memory capacity of standard consumer hardware, necessitating enterprise-grade GPUs or multi-GPU setups.
    \item \textit{\textbf{Large (S1.3.4)}}: Models exceeding 70 billion parameters \textbf{($>$70B)}, such as GPT-4, Llama3.1-405B, and DeepSeek-v3. These models achieve state-of-the-art performance on vulnerability detection benchmarks. However, due to their immense computational requirements, they are predominantly accessed via APIs or require extensive high-performance computing clusters for local deployment.
\end{itemize}

\begin{figure*}[tb]
    \centering  
        \includegraphics[width=\textwidth,trim={5em 0 5em 18em},clip]{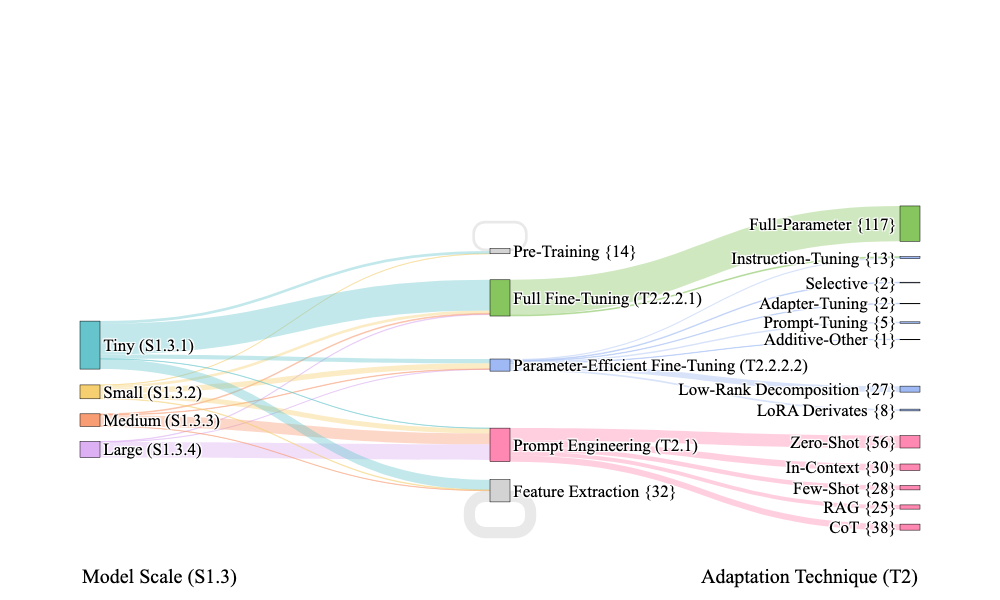} 
        \caption{Relationship between model scale and adaptation techniques across surveyed studies. \textit{Tiny} and \textit{small} models are predominantly adapted via full-parameter fine-tuning, whereas \textit{medium} and \textit{large} models increasingly utilize parameter-efficient fine-tuning and prompt engineering. Numbers in braces indicate the absolute count of unique studies applying the respective adaptation techniques; a single study may apply multiple adaptation techniques.}
        \Description{A three-stage Sankey diagram. The left stage lists model scales (Tiny, Small, Medium, Large). The middle stage lists adaptation categories (Full Fine-Tuning, PEFT, Prompt Engineering, Pre-Training). The right stage lists specific techniques. Thick flows connect Tiny and Small models to Full Fine-Tuning, while flows from Large models connect primarily to Prompt Engineering and PEFT.}
        \label{fig:sankey_model_technique}
\end{figure*}

Figure~\ref{fig:sankey_model_technique} visualizes the relationship between model scale and adaptation techniques, e.g., fine-tuning, parameter-efficient fine-tuning, and prompt engineering; highlighting that tiny and small models are often fine-tuned in a full-parameter setting while large models are mainly prompted.

\subsubsection{Hybrid (S2)}
Hybrid architectures combine LLMs with other learning-based components to leverage complementary strengths. Often, the LLM is used to produce embeddings that are subsequently processed by neural networks such as Recurrent Neural Networks (RNNs), Convolutional Neural Networks (CNNs), or Graph Neural Networks (GNNs). 

\textit{\textbf{RNNs (S2.1)}} are neural architectures designed to handle sequential data by modeling temporal and contextual dependencies between elements, e.g., syntactic, semantic, and control-flow relationships~\cite{allamanis2018survey}.
Prominent RNN variants include Long Short-Term Memory networks (LSTMs) and their bidirectional forms (BiLSTMs), which are designed to handle long-range dependencies effectively. 
In several studies, BERT embeddings are directly fed into (Bi-)LSTM networks~\cite{phamDefectscannerComparativeEmpirical2024,ziemsSecurityVulnerabilityDetection2021,chengLeveragingSelfPacedLearning2025,liAutomatedSoftwareVulnerability2022,xiongVulDCodeBERTCodeBERTBasedVulnerability2024}.

\textit{\textbf{CNNs (S2.2)}} are neural architectures specialized in identifying local spatial patterns or features within data through convolutional filters. Initially developed for image recognition tasks, CNNs have been successfully adapted to textual and sequential data analysis by treating sequences as one-dimensional spatial inputs. This capability allows CNNs to effectively capture localized syntactic and semantic patterns in code sequences, making them useful for tasks such as vulnerability detection~\cite{lin2020software}.
Studies apply CNNs directly to code or graph embeddings~\cite{zhangVulnerabilityDetectionLearning2023,phamDefectscannerComparativeEmpirical2024}. 
Others use pre-trained embeddings within TextCNN architectures~\cite{hanifVulBERTaSimplifiedSource2022}, or integrate CNNs with attention mechanisms to enhance feature extraction~\cite{wangDefectHunterNovelLLMDriven2023,wuSoftwareVulnerabilityDetection2024}.

\textit{\textbf{GNNs (S2.3)}} are specialized neural architectures designed to operate directly on graph-structured data, effectively modeling relationships between nodes and their structural dependencies.
Given structured representations of code, such as AST and CFG, several studies combine LLMs with GNN-based architectures to take advantage of both sequential and structural representations of code~\cite{lin2020software}.
A common approach is to pass code embeddings, e.g., generated by CodeBERT or GraphCodeBERT, into GNN-based classifiers~\cite{yuVulnerabilityDetectionFramework2025,tianEFVDFrameworkSource2025,kongSourceCodeVulnerability2024,liuVulLMGNNsFusingLanguage2025,zhengSVulDetectorVulnerabilityDetection2024,hinLineVDStatementlevelVulnerability2022,rangapuramGraphCodeBERTAugmentedGraphAttention2025,sunHgtJITJustinTimeVulnerability2025,sunEnhancedVulnerabilityDetection2023,nganMultimodalFusionVulnerability2025}. 
Other studies use LLMs to feed graph embeddings~\cite{islamUnbiasedTransformerSource2023,islamUnintentionalSecurityFlaws2024,jiangDFEPTDataFlow2024,quanXGVBERTLeveragingContextualized2023} or multiple modalities of a function, including text, graph, and image representations~\cite{niFunctionLevelVulnerabilityDetection2023} into Graph Convolution Networks (GCNs).
Yang et al.~\cite{yangSecurityVulnerabilityDetection2024} use a GNN as a light-weight adapter layer and concatenate its learned embedding with a fine-tuned LLM. 
Similarly, Gajjar et al.~\cite{gajjarBridgingSemanticsStructure2025} use GCN and GAT to extract structural features, subsequently fusing these graph embeddings with semantic embeddings derived from an LLM.

Several hybrid architectures combine multiple neural components, integrating RNNs, CNNs, and GNNs in sequence or parallel.
For example, Chang et al.~\cite{changFineTuningPretrainedModel2024} process CodeBERT embeddings using a two-layer GCN and a BiLSTM.
Bahaa et al.~\cite{bahaaDBCBILDistilBertBasedTransformer2024} use a multi-layer approach composed of a DistilBERT embedding layer followed by CNN and BiLSTM layers.
Mei et al.~\cite{meiDetectingVulnerabilitiesIoT2023} use CodeBERT embeddings in three hybrid models, i.e., CLSTM, CBiLSTM (sequential structure), and CNN-BiLSTM (parallel structure).
JianJie and Le~\cite{jianjieCodeDefectDetection2023} present a multi-base classifier combining BERT-encoder, GCN, CNN, and BiLSTM; the outputs are fused via stacking ensemble learning.
Sun et al.~\cite{sunEnhancedVulnerabilityDetection2023} propose a stacking ensemble strategy that integrates serialization-based and graph-based neural networks to exploit complementary code semantics.

Some studies~\cite{jiangEnhancingFineGrainedVulnerability2025a,pengPTLVDProgramSlicing2023,xueAIDetectVulSoftwareVulnerability2025,meloAreSparseAutoencoders2025a,phamDefectscannerComparativeEmpirical2024} combine LLMs with \textit{\textbf{Other (S2.4)}} learning-based components that do not rely on RNNs, CNNs, or GNNs.
For example, Melo et al.~\cite{meloAreSparseAutoencoders2025a} evaluate the effectiveness of Sparse Autoencoders (SAEs) when applied to representations from LLMs, which are then classified using traditional machine learning models such as random forests. 
Jiang et al.~\cite{jiangEnhancingFineGrainedVulnerability2025a} pass embeddings to a Policy network for prediction.
Peng et al.~\cite{pengPTLVDProgramSlicing2023} use CodeBERT embeddings as input to a Transformer-based classifier.
Similarly, Xue et al.~\cite{xueAIDetectVulSoftwareVulnerability2025} perform feature fusion from multiple pre-trained models and process the combined representation with a Transformer model.


\subsection{Technique (T)}
To use LLMs effectively for vulnerability detection, studies employ various techniques ranging from simple prompting to complex system and training designs. 
We broadly categorize techniques into three distinct strategies: (i)~feature extraction, (ii)~adaptation, and (iii)~orchestration.

\subsubsection{Feature Extraction}
In the most simple approach, language models are used in their pre-trained form without any additional training to extract features by generating input embeddings (\textit{\textbf{Feature Extraction (T1)}})~\cite{jiang2022multi,yuVulnerabilityDetectionFramework2025,tianEFVDFrameworkSource2025,phamDefectscannerComparativeEmpirical2024,wuSoftwareVulnerabilityDetection2024,haqueZeroshotFrameworkCrossproject2026,liangSourceCodeVulnerability2024,duJointGeometricalStatistical2023,gujarDetectBERTCodeVulnerability2024,hinLineVDStatementlevelVulnerability2022,islamUnbiasedTransformerSource2023,islamUnintentionalSecurityFlaws2024,jararwehLLaVulMultimodalLLM2025,jiangDFEPTDataFlow2024,jiangEnhancingFineGrainedVulnerability2025a,leFineTuningTransformerLLMs2025,mahmudEnsembleTransformerApproach2025,meloAreSparseAutoencoders2025a,nganMultimodalFusionVulnerability2025,nguyenMulVulnEnhancingPretrained2025,niFunctionLevelVulnerabilityDetection2023,pengPTLVDProgramSlicing2023,rangapuramGraphCodeBERTAugmentedGraphAttention2025,sultanEnhancedLLMBasedFramework2024,sunEnhancedVulnerabilityDetection2023,sunHgtJITJustinTimeVulnerability2025,wangANVILAnomalybasedVulnerability2024,wangDefectHunterNovelLLMDriven2023,xueAIDetectVulSoftwareVulnerability2025,zhangVulnerabilityDetectionLearning2023,zhaoHowGetBetter2023,zhuAutomatedCodeReview2025}. These high-dimensional vector representations capture the semantic and syntactic properties of the code and serve as input for downstream classifiers.

\subsubsection{Adaptation Technique (T2)}
\label{sec:adaptation_techniques}
Beyond using LLMs as static feature extractors, recent work increasingly focuses on adapting LLMs to the vulnerability detection task through various state-of-the-art techniques, which differ in how they modify the model's parameters.
The primary adaptation techniques are \textit{pre-training}, \textit{fine-tuning}, and \textit{prompt engineering}.
Each technique reflects different trade-offs between computational cost, model adaptation, and task-specific performance. 
In the following, we discuss these adaptation techniques as observed in the surveyed studies.

\paragraph{Prompt Engineering (T2.1)}
\label{sec:prompt_engineering}
Prompt engineering refers to the design and structuring of input prompts in a way that elicits desired behaviors or outputs without requiring any modifications to the model’s weights. 
There are typically two types of prompts: a \textit{system prompt} (providing general context or behavioral instructions) and a \textit{user prompt} (specifying the task and providing the code to be analyzed).
As LLMs remain sensitive to the phrasing, context, and formatting of prompts, numerous prompting strategies have emerged to improve task alignment and performance.

In \textit{\textbf{Zero-Shot Prompting (T2.1.1)}}, the model is expected to rely solely on its pre-trained knowledge to perform vulnerability detection.
The LLM is provided with the task description and input (i.e., a potentially vulnerable code snippet under detection)~\cite{ambatiNavigatingInSecurityAIGenerated2024,baeEnhancingSoftwareCode2024,bakhshandehUsingChatGPTStatic2023,chanTransformerbasedVulnerabilityDetection2023,cheshkovEvaluationChatGPTModel2023,cipolloneAutomatingDetectionCode2025,dozonoLargeLanguageModels2024,dubniczkyCASTLEBenchmarkingDataset2025,eberhardtVulnGPTEnhancingSource2024,espinhagasibaMaySourceBe2024,ezenwoyeExploringAIVulnerability2024,farrExpertintheLoopSystemsCrossDomain2025,gongHowWellLarge2024,guoOutsideComfortZone2024,hairanCompressingLargeLanguage2025,huynhDetectingCodeVulnerabilities2025,ingemanntuffvesonjensenSoftwareVulnerabilityFunctionality2024,khareUnderstandingEffectivenessLarge2024,kholoosiQualitativeStudyUsing2024,kouliaridisAssessingEffectivenessLLMs2024,linEvaluatingLargeLanguage2025,liRevisitingPretrainedLanguage2025,liuExplorationPromptingLLM2024,liuVulDetectBenchEvaluatingDeep2024,liVulnTeamTeamCollaboration2025,luGRACEEmpoweringLLMbased2024,luoHALURustExploitingHallucinations2025,maligazhdarovaComparativeStudyMachine2025,niLearningbasedModelsVulnerability2024,noeverCanLargeLanguage2023,ouchebaraLlamabasedSourceCode2025,ozturkNewTricksOld2023,purbaSoftwareVulnerabilityDetection2023,ridleyEnhancingCodeSecurity2024,safdarRealVulLLMLLMBased2025,saimbhiVulnerAIGPTBased2024,shuLargeLanguageModels2025,simoniImprovingLLMReasoning2025,steenhoekClosingGapUser2024,steenhoekComprehensiveStudyCapabilities2024a,sunLLM4VulnUnifiedEvaluation2024,tradManualPromptEngineering2025,tsaiSequentialMultiStageApproach2025,ullahLLMsCannotReliably2024,wangLeveragingLargeLanguage2025,wangM2CVDEnhancingVulnerability2024,wangSecureMindFrameworkBenchmarking2025,weiDistillingLightweightLanguage2025,wuExploringLimitsChatGPT2023,yildirimEvaluatingImpactConventional2024,yuInsightSecurityCode2024,zhangBenchmarkingLargeLanguage2025,zhouComparisonStaticApplication2024,zhouLargeLanguageModel2024c,zibaeiradReasoningLLMsZeroShot2025,zibaeiradVulnLLMEvalFrameworkEvaluating2024}.
Typical prompts include simple binary instructions such as "Is the following code vulnerable? Yes or No," targeted CWE-specific queries such as "Does the following code have a CWE-xxx vulnerability?", or openly phrased "Does this code have any vulnerabilities?"
Studies often experiment with arbitrary variations to the base prompt, such as modifying task phrasing~(e.g., "vulnerable", "buggy", "insecure"). In a role-based setting, studies include a role in the prompt ~(e.g., "Act as a vulnerability detection system"~\cite{liuExplorationPromptingLLM2024}) to guide model behavior. 
Due to its simplicity, zero-shot prompting is commonly used as a starting point and serves as a baseline for comparison with more advanced prompting techniques.

\textit{\textbf{In-Context Learning (T2.1.2)}} refers to the prompt engineering technique of embedding additional contextual information (drawing upon the auxiliary information discussed, see Section~\ref{sec:input_representation}) into the model's input to improve detection without parameter updates.
Researchers frequently provide vulnerability definitions, e.g., a candidate list of CWEs~\cite{dozonoLargeLanguageModels2024,bakhshandehUsingChatGPTStatic2023,shashwatPreliminaryStudyUsing,steenhoekComprehensiveStudyCapabilities2024a,wangEnhancingLargeLanguage2023,xiaStaticPatternMatching2025,yuInsightSecurityCode2024,fuChatGPTVulnerabilityDetection2023} or detailed vulnerability descriptions~\cite{caoLLMCloudSecLargeLanguage2024,shimmiSoftwareVulnerabilityDetection2024,gnieciakLargeLanguageModels2025,mathewsLLbezpekyLeveragingLarge2024,mohajerEffectivenessChatGPTStatic2024,ullahLLMsCannotReliably2024}.
Adopting a code-centric perspective, studies also enrich prompts with execution context~\cite{sayaghThinkBroadAct2025,sunLLM4VulnUnifiedEvaluation2024,tambergHarnessingLargeLanguage2024,zhangVulTrLMLLMassistedVulnerability2026,ahmedSecVulEvalBenchmarkingLLMs2025,mhatreLLMGUARDLargeLanguage2025,liLLMAssistedStaticAnalysis2024} or descriptions of the code's intended functionality~\cite{liuExplorationPromptingLLM2024,liuSoftwareVulnerabilityDetection2023,yangContextEnhancedVulnerabilityDetection2025,torkamaniStreamliningSecurityVulnerability2025}, supporting the vulnerability detection task.
Other approaches use a hybrid workflow by embedding prior detection results from static analysis or similar tools, asking the LLM to verify or refine these findings~\cite{keltekBoostingCybersecurityVulnerability2024,munsonLittleHelpMy2025,wangVulAgentHypothesisValidationBased2025,yangDLAPDeepLearning2025}.

\textit{\textbf{Few-Shot Prompting (T2.1.3)}}~\cite{gpt3}, a specific in-context learning technique, provides the model with one or multiple labeled input-output examples relevant to the task.
By presenting these examples in the prompt, the model is guided to infer the desired task behavior, enabling in-context learning without modifying model weights.
The examples used for few-shot prompting can be static, i.e., manually selected and reused across prompts~\cite{dingVulnerabilityDetectionCode2024,dolcettiHelpingLLMsImprove2024,leSoftwareVulnerabilityPrediction2024,liCryptoScopeUtilizingLarge2025,maoEffectivelyDetectingExplaining2024,niLearningbasedModelsVulnerability2024,sovranoLargeLanguageModels2025,tradManualPromptEngineering2025,wangSecureMindFrameworkBenchmarking2025,yildizBenchmarkingLLMsLLMbased2025,yinMultitaskBasedEvaluationOpenSource2024,chenGPTVDVulnerabilityDetection2026}, or vulnerability-specific, i.e., tailored examples targeting particular vulnerability types~\cite{farrExpertintheLoopSystemsCrossDomain2025,ouchebaraLlamabasedSourceCode2025,ullahLLMsCannotReliably2024}.
In another study, the examples are generated by the LLM  itself~\cite{chanTransformerbasedVulnerabilityDetection2023}.
Other common strategies for example selection, as discussed by Steenhoek et al.~\cite{steenhoekComprehensiveStudyCapabilities2024a}, include using random vulnerable and non-vulnerable examples from the training dataset~\cite{liRevisitingPretrainedLanguage2025,mohajerEffectivenessChatGPTStatic2024,niLearningbasedModelsVulnerability2024,ouchebaraLlamabasedSourceCode2025,steenhoekComprehensiveStudyCapabilities2024a,widyasariChatGPTEnhancingSoftware2024,yangContextEnhancedVulnerabilityDetection2025,zhouComparisonStaticApplication2024,zhouLargeLanguageModel2024c}, constructing contrastive pairs where both a vulnerable code snippet and its corresponding fixed version are provided within the same prompt~\cite{cekaCanLLMPrompting2024,nongChainofThoughtPromptingLarge2024,steenhoekComprehensiveStudyCapabilities2024a,yangContextEnhancedVulnerabilityDetection2025,yildizBenchmarkingLLMsLLMbased2025}, or selecting examples from the training dataset based on similarity to the input code~\cite{hannanSelectingFewShotExamples2025,niLearningbasedModelsVulnerability2024,shuLargeLanguageModels2025,steenhoekComprehensiveStudyCapabilities2024a,widyasariChatGPTEnhancingSoftware2024,zhouLargeLanguageModel2024c,tsaiLeveragingIntraInterReferences2025}.

\textit{\textbf{Retrieval-Augmented Generation (T2.1.4)}}~\cite{lewis2020retrieval} extends the few-shot paradigm by dynamically augmenting the prompt with relevant external information retrieved from a private knowledge base. 
Retrieval-Augmented Generation (RAG) systems use a retriever component to identify and fetch contextually similar entries based on embedding similarity or other relevance metrics, e.g., BM25.
The LLM then works with the provided knowledge to generate the output.
A key advantage of RAG over few-shot prompting is its ability to dynamically incorporate domain-specific and up-to-date knowledge, making it especially useful for integrating evolving vulnerability knowledge. 
In the context of vulnerability detection, RAG-based approaches use a wide range of retrieval content.
Some studies retrieve the most similar code snippets or code dependencies along with known vulnerability information~\cite{liuExplorationPromptingLLM2024,liuSoftwareVulnerabilityDetection2023,luGRACEEmpoweringLLMbased2024,ouchebaraLlamabasedSourceCode2025,safdarRealVulLLMLLMBased2025,shengLProtectorLLMdrivenVulnerability2024a,steenhoekComprehensiveStudyCapabilities2024a,tsaiSequentialMultiStageApproach2025,wenVulEvalRepositoryLevelEvaluation2024,yangDLAPDeepLearning2025,zhangBenchmarkingLargeLanguage2025,zhouSSRFSeekLLMbasedStatic2025,zhuSpecificationGuidedVulnerabilityDetection2025,jiaoDeepVulHunterEnhancingCode2025,tsaiLeveragingIntraInterReferences2025}.
Others rely on abstract code functionality and vulnerability descriptions to better capture the vulnerability semantics and context of the input code~\cite{duVulRAGEnhancingLLMbased2024,liCryptoScopeUtilizingLarge2025,shimmiSoftwareVulnerabilityDetection2024,sunLLM4VulnUnifiedEvaluation2024}.
Additionally, vulnerability reports and CWE descriptions are retrieved to guide the model's reasoning using structured vulnerability information~\cite{caoLLMCloudSecLargeLanguage2024,keltekBoostingCybersecurityVulnerability2024,kouliaridisAssessingEffectivenessLLMs2024,mathewsLLbezpekyLeveragingLarge2024,pelofskeAutomatedSoftwareVulnerability2024,sayaghThinkBroadAct2025}.

\textit{\textbf{Chain-of-Thought (CoT) Prompting (T2.1.5)}}~\cite{wei2022chain} aims at enhancing the reasoning capabilities of LLMs by encouraging intermediate reasoning steps before arriving at a final prediction.   
Reasoning behavior is typically triggered by simple cues such as "Let's think step by step"~\cite{baeEnhancingSoftwareCode2024,dingVulnerabilityDetectionCode2024,dolcettiHelpingLLMsImprove2024,huynhDetectingCodeVulnerabilities2025,niLearningbasedModelsVulnerability2024,pelofskeAutomatedSoftwareVulnerability2024,safdarRealVulLLMLLMBased2025,shengLProtectorLLMdrivenVulnerability2024a,steenhoekComprehensiveStudyCapabilities2024a,ullahLLMsCannotReliably2024,wangSecureMindFrameworkBenchmarking2025,wenBoostingVulnerabilityDetection2025,zhouComparisonStaticApplication2024,zibaeiradReasoningLLMsZeroShot2025,yildizBenchmarkingLLMsLLMbased2025},
requests for explanation~\cite{liVulnTeamTeamCollaboration2025,yangContextEnhancedVulnerabilityDetection2025,chenGPTVDVulnerabilityDetection2026,khareUnderstandingEffectivenessLarge2024},  
or specific step-by-step instructions~\cite{cekaCanLLMPrompting2024,duGeneralizationEnhancedCodeVulnerability2024,liCryptoScopeUtilizingLarge2025,liEverythingYouWanted2025,liMAVULMultiAgentVulnerability2025,liuExplorationPromptingLLM2024,maoEffectivelyDetectingExplaining2024,pengICodeReviewerImprovingSecure2025,shengLProtectorLLMdrivenVulnerability2024a,simoniImprovingLLMReasoning2025,sunLLM4VulnUnifiedEvaluation2024,ullahLLMsCannotReliably2024,wangSecureMindFrameworkBenchmarking2025,widyasariChatGPTEnhancingSoftware2024,yangDLAPDeepLearning2025,yuInsightSecurityCode2024,zhouSSRFSeekLLMbasedStatic2025,zhuSpecificationGuidedVulnerabilityDetection2025,zibaeiradReasoningLLMsZeroShot2025}. 
Automated CoT prompt construction has also been explored~\cite{tradManualPromptEngineering2025}.
Several studies adapt CoT prompting with domain-specific techniques. 
For instance, Nong et al.~\cite{nongChainofThoughtPromptingLarge2024} propose a vulnerability-semantics-guided prompt format to guide the model through relevant data and control flow facts. 
More advanced CoT-inspired prompting techniques include self-consistency~\cite{wang2022self}, where the prompt is run multiple times, and predictions are aggregated (e.g., a file is marked vulnerable if the model returns a positive classification in two out of three runs)~\cite{tambergHarnessingLargeLanguage2024}, and Tree-of-Thought (ToT) prompting~\cite{tambergHarnessingLargeLanguage2024,munsonLittleHelpMy2025}, which branches the reasoning process by generating multiple alternative steps per stage, followed by a selection of the most promising path.


\paragraph{Training (T2.2)}
\label{sec:fine_tuning}
The adaptation of LLMs to software vulnerability detection tasks often follows the established pre-training and fine-tuning paradigm: models are \textit{trained from scratch} (\textit{\textbf{Pre-Training (T2.2.1)}}) using large code corpora to capture programming-specific syntax and semantics~\cite{alrashedyLearningDefectPrediction2024,ahmadUnifiedPretrainingProgram2021,burattiExploringSoftwareNaturalness2020,dingLeveragingDeepLearning2023,dingTRACEDExecutionawarePretraining2024,hanifVulBERTaSimplifiedSource2022,huangBBVDBERTbasedMethod2022,jararwehLLaVulMultimodalLLM2025,jiangStagedVulBERTMultigranularVulnerability2024,liCLeVeRMultimodalContrastive,liuPretrainingPredictingProgram2024,wangVuLMCBERTVulnerabilityDetection2025,xuSoftwareVulnerabilitiesDetection2023,yuPATVDVulnerabilityDetection2022} and further \textit{fine-tuned} for domain-specific applications. 
\textit{Fine-Tuning (T2.2.2)} is the most widely adopted technique.
Studies adopt a range of fine-tuning strategies that vary in their degree of parameter update and architectural design.

\textit{\textbf{Full-Parameter Fine-Tuning (T2.2.2.1)}} refers to tuning all parameters of a pre-trained language model for the downstream task of vulnerability detection. 
Models are either fine-tuned by replacing the last output layer~\cite{grishinaEarlyBIRDCatchesBug2023} with a classification head (e.g., a multi-layer perceptron), which return the probability of vulnerability; or fine-tuned as a generative model where the response is matched to keywords such as “vulnerable” or “non vulnerable”.
Full-parameter fine-tuning is widely adopted in the surveyed studies~\cite{duDMVL4AVDDeepMultiview2025,wu2025sparse,liuMakingVulnerabilityPrediction2024,alam2025improving,liuPIONEERImprovingRobustness2026,alrashedyLearningDefectPrediction2024,ahmadUnifiedPretrainingProgram2021,aslanUtilizingLargeProgramming2025,atiiqGeneralistSpecialistExploring2024,atiiqVulnerabilityDetectionPopular2024,bahaaDBCBILDistilBertBasedTransformer2024,burattiExploringSoftwareNaturalness2020,chanTransformerbasedVulnerabilityDetection2023,chenBridgeHintExtending2024,chenDiverseVulNewVulnerable2023b,chengLeveragingSelfPacedLearning2025,cipolloneAutomatingDetectionCode2025,curtoMultiVDTransformerbasedMultitask2024,daloisioCompressionLanguageModels2024,dengImprovingLongTailVulnerability2024,dingLeveragingDeepLearning2023,dingTRACEDExecutionawarePretraining2024,dingVulnerabilityDetectionCode2024,ehrenbergPythonSourceCode2024,ferragSecureFalconAreWe2024,fuAIBugHunterPracticalTool2024,fuLineVulTransformerbasedLineLevel2022,fuVulExplainerTransformerBasedHierarchical2023a,gajjarBridgingSemanticsStructure2025,gaoKeepingPaceEverIncreasing2023,grishinaEarlyBIRDCatchesBug2023,gsBiT5BidirectionalNLP2024,guoOutsideComfortZone2024,gurfidanVULREMFineTunedBERTBased2024,halderFuncVulEffectiveFunction2025,hanAutomaticClassificationModel2025,hanifVulBERTaSimplifiedSource2022,huangBBVDBERTbasedMethod2022,ibanez-lissenLPASSLinearProbes2025,jiangStagedVulBERTMultigranularVulnerability2024,jianjieCodeDefectDetection2023,jiApplyingContrastiveLearning2024a,kaananLLMBasedApproachBuffer2024,kalouptsoglouVulnerabilityClassificationSource2024,kalouptsoglouVulnerabilityPredictionUsing2024,kimVulDeBERTVulnerabilityDetection2022,kongSourceCodeVulnerability2024,lanSmartCutsEnhance2025,leSoftwareVulnerabilityPrediction2024,liAutomatedSoftwareVulnerability2022,liCleanVulAutomaticFunctionLevel2024,linDistilledContextualizedNeural2022,liOutDistributionOut2025,liRevisitingPretrainedLanguage2025,liuEnhancingVulnerabilityDetection2025,liuPretrainingPredictingProgram2024,liuVulLMGNNsFusingLanguage2025,luAssessingEffectivenessVulnerability2023,luoDetectingIntegerOverflow2021,machtleTraceGadgetsMinimizing2025,maligazhdarovaComparativeStudyMachine2025,mamedeExploringTransformersMultiLabel2022,mechriSecureQwenLeveragingLLMs2025,meiDetectingVulnerabilitiesIoT2023,mockCrossDomainEvaluationTransformerBased2025,nguyenSAFEAdvancingLarge2024,niDistinguishingLookAlikeInnocent2023,omarVulDefendNovelTechnique2023,omarVulDetectNovelTechnique2023,panichellaMetamorphicBasedManyObjectiveDistillation2025,purbaSoftwareVulnerabilityDetection2023,qiEnhancingPreTrainedLanguage2024,quanXGVBERTLeveragingContextualized2023,rahmanCausalDeepLearning2024,ridoyEnStackEnsembleStacking2024,risseUncoveringLimitsMachine2024,russoLeveragingMultitaskLearning2025,safdarDataContextMatter2025,shestovFinetuningLargeLanguage2024,shiGreeningLargeLanguage2024,shimmiSoftwareVulnerabilityDetection2024,shimmiVulSimLeveragingSimilarity,simoniImprovingLLMReasoning2025,singhCyberSecurityVulnerability2022,steenhoekClosingGapUser2024,steenhoekEmpiricalStudyDeep2023,steenhoekLanguageModelsLearn2023,sultanaCodeVulnerabilityDetection2024,sunEnhancingSourceCode2024,suOptimizingPretrainedLanguage2023,takieldeenAIPoweredVulnerabilityDetection2025,thapaTransformerBasedLanguageModels2022,tianYouOnlyTrain2025,torkamaniStreamliningSecurityVulnerability2025,wangLinelevelSemanticStructure2024,wangM2CVDEnhancingVulnerability2024,wangVuLMCBERTVulnerabilityDetection2025,wenBoostingVulnerabilityDetection2025,wenSCALEConstructingStructured2024,wenVulEvalRepositoryLevelEvaluation2024,wenWhenLessEnough2023,xiongVulDCodeBERTCodeBERTBasedVulnerability2024,xuSoftwareVulnerabilitiesDetection2023,yangMoEVDEnhancingVulnerability2025,yinMultitaskBasedEvaluationOpenSource2024,yuanDeepNeuralEmbedding2022,zengIntelligentDetectionVulnerable2023,zhangMVDMultiLingualSoftware2024,zhaoAdversarialTrainingRobustness2025,zhaoPythonSourceCode2024,zhengLearningFocusContext2025,zhengSVulDetectorVulnerabilityDetection2024,zhouComparisonStaticApplication2024,zhuAutomatedCodeReview2025,zhuBERTBasedVulnerabilityType2022,ziemsSecurityVulnerabilityDetection2021,nguyenHumanUnderstandableExplanationSoftware2024}, particularly with tiny models, see Figure~\ref{fig:sankey_model_technique}.
While full-parameter fine-tuning is computationally intensive and less reusable across tasks, it allows the model to specialize in vulnerability detection by adapting all internal representations.

A specific variant of fine-tuning is \textit{\textbf{Instruction-Tuning (T2.2.2.1.1)}}, where a language model is fine-tuned on labeled input-output pairs annotated with natural language instructions~\cite{bappyCaseStudyFinetuning2025,duGeneralizationEnhancedCodeVulnerability2024,islamEnhancingSourceCode2024,jiangInvestigatingLargeLanguage2024,liVulnTeamTeamCollaboration2025,maoEffectivelyDetectingExplaining2024,shuLargeLanguageModels2025,steenhoekComprehensiveStudyCapabilities2024a,tianSQLInjectionVulnerability2024,widyasariLetTrialBegin2025,yusufYourInstructionsAre2024,zhangBenchmarkingLargeLanguage2025,zhangMethodSQLInjection2024}, e.g., "Instruction: Detect whether the following code contains vulnerabilities"~\cite{jiangInvestigatingLargeLanguage2024}.
This method allows the model to better follow human-written prompts by learning general patterns for task completion.
In \cite{widyasariLetTrialBegin2025}, Widyasari et al. use GPT-4o for generating the instructions.
Instruction-tuning has shown promise in enhancing performance on unseen instructions and is often combined with parameter-efficient (rather than full) fine-tuning techniques to reduce training cost~\cite{duGeneralizationEnhancedCodeVulnerability2024,jiangInvestigatingLargeLanguage2024,liVulnTeamTeamCollaboration2025,maoEffectivelyDetectingExplaining2024,shuLargeLanguageModels2025,tianSQLInjectionVulnerability2024,yusufYourInstructionsAre2024,zhangBenchmarkingLargeLanguage2025,zhangMethodSQLInjection2024}. \\

Key drawbacks of full-parameter fine-tuning are its high computational cost and the potential harm to the model’s generalization ability. 
To address these drawbacks, several studies adopt \textit{Parameter-Efficient Fine-Tuning (PEFT) (T2.2.2.2)} techniques, which adapt only a small subset of model parameters or learn external modules for new tasks while keeping most of the pre-trained parameters frozen.
To categorize the PEFT methods, we adopt the taxonomy by Han et al.~\cite{hanParameterEfficientFineTuningLarge2024}, covering (i)~selective, (ii)~additive, and (iii)~reparameterized PEFT techniques.

In \textit{\textbf{Selective PEFT (T2.2.2.2.1)}}, only a small subset of model parameters is selected for fine-tuning, while keeping the other parameters frozen. 
In the surveyed studies, the final~\cite{zhangVulTrLMLLMassistedVulnerability2026} or upper model layer(s)~\cite{chenImprovingVulnerabilityType2025} were fine-tuned to generate contextual embeddings.

\textit{Additive PEFT (T2.2.2.2.2)} techniques introduce a small number of trainable parameters that are strategically positioned within the model architecture. 
During fine-tuning, only the weights of these additional modules or parameters are updated, reducing storage, memory, and computational resource requirements. 
Prominent additive PEFT techniques are adapter-tuning and prompt-tuning.

\textit{\textbf{Adapter-Tuning (T2.2.2.2.2.1)}} inserts small trainable neural modules, i.e., \textit{adapters}, into a frozen pre-trained model. 
During fine-tuning, only the adapter parameters are updated, while the model's parameters remain unchanged~\cite{akliAutoAdaptApplicationAutoML2025,liCLeVeRMultimodalContrastive}.
Rather than employing default configurations that use the same adapter settings across all layers, Akli et al.~\cite{akliAutoAdaptApplicationAutoML2025} present AutoAdapt to automatically discover task-specific, layer-wide adapter configurations, allowing each layer to adopt distinct parameters. 

\textit{\textbf{Prompt-Tuning (T2.2.2.2.2.2)}} involves learning prompts to improve model performance on a downstream task. Two main prompt forms are distinguished: hard prompts, which are manually crafted sequences of tokens appended to the input, and soft prompts, which are learned continuous embeddings that serve as virtual tokens prepended to the input.
Prompt-tuning keeps the model weights frozen while tuning only the parameters of the soft prompts~\cite{changFineTuningPretrainedModel2024,chenVulPrPromptLearningbased2025,fengCGPTuningStructureAwareSoft2025,luAssessingEffectivenessVulnerability2023,renProRLearnBoostingPrompt2024}.
Various adaptations of prompt-tuning have been investigated.
Ren et al.~\cite{renProRLearnBoostingPrompt2024} integrate prompt-tuning with a reinforcement learning framework.
Feng et al.~\cite{fengCGPTuningStructureAwareSoft2025} explore graph-enhanced prompt-tuning, which incorporates structural code information by embedding graph features into soft prompts.

Apart from the methods mentioned above, another additive approach has been used that strategically trains additional parameters during the fine-tuning process (\textit{\textbf{Additive-Other (T2.2.2.2.2.3)}}):
Li et al.~\cite{liSteeringLargeLanguage2025} introduce a vulnerability steering vector that represents the concept of vulnerability in the representation space. The vector is injected into the activation values of the corresponding layer, guiding the model’s behavior without modifying all its parameters.

\textit{Reparameterized PEFT (T2.2.2.2.3)} techniques transform a model’s architecture by modifying how its parameters are represented and trained. 
This typically involves replacing large components of the model with smaller, low-rank versions during training, reducing the number of parameters that need to be updated.
At inference time, the model can be converted to its original parameterization~\cite{hanParameterEfficientFineTuningLarge2024}.

The most widely adopted reparameterization technique in the surveyed studies is categorized as \textit{\textbf{Low-Rank Decomposition (T2.2.2.2.3.1)}}: LoRA (Low-Rank Adaptation)~\cite{hu2022lora}.
LoRA injects trainable low-rank decomposition matrices into the attention layers of a frozen language model, enabling task-specific adaptation with significantly fewer parameters~\cite{prasadSinkVulnerabilityType2025,alqarniAdvancedDetectionFramework2025,aslanUtilizingLargeProgramming2025,caoRealVulCanWe2024,chenVulPrPromptLearningbased2025,curtoCanLlamaBe2024,duGeneralizationEnhancedCodeVulnerability2024,gajjarMalCodeAIAutonomousVulnerability2025a,goncalvesEvaluatingLLaMA322025,guoOutsideComfortZone2024,jararwehLLaVulMultimodalLLM2025,jiangInvestigatingLargeLanguage2024,lekssaysLLMxCPGContextAwareVulnerability2025,liRevisitingPretrainedLanguage2025,liuEnhancingVulnerabilityDetection2025,liVulnTeamTeamCollaboration2025,luoHALURustExploitingHallucinations2025,maoEffectivelyDetectingExplaining2024,shestovFinetuningLargeLanguage2024,shuLargeLanguageModels2025,wangSCLCVDSupervisedContrastive2024,wenBoostingVulnerabilityDetection2025,weyssowR2VulLearningReason2025,yusufYourInstructionsAre2024,zhangMethodSQLInjection2024,zhouComparisonStaticApplication2024,zhuDetectingSourceCode2024}.

\textit{\textbf{LoRA Derivatives (T2.2.2.2.3.2)}} have also been explored to improve efficiency and adaptability. 
Tian et al.~\cite{tianSQLInjectionVulnerability2024} use PiSSA (Principal Singular Values and Singular Vectors Adaptation)~\cite{meng2024pissa}, which shares the same architecture as LoRA but adopts a different initialization method.
Ibanez-Lissen et al.~\cite{ibanez-lissenLPASSLinearProbes2025} use GaLore (Gradient Low-Rank Projection)~\cite{zhao2024galore}, a memory-efficient training strategy that allows full-parameter training while requiring less memory than common low-rank approaches.
Another efficient PEFT design is QLoRA (Quantized LoRA)~\cite{dettmers2023qlora}, which applies LoRA to a quantized version of the language model to further reduce memory consumption. 
QLoRA has been used in multiple surveyed studies to efficiently fine-tune large models for code vulnerability tasks~\cite{zhangBenchmarkingLargeLanguage2025,sultanaCodeVulnerabilityDetection2024,yangSecurityVulnerabilityDetection2024,ouchebaraLlamabasedSourceCode2025,sunEnsemblingLargeLanguage2025,weiDistillingLightweightLanguage2025}.


\paragraph{Learning Paradigms (T2.3)}
\label{sec:learning_paradigms}
Beyond standard supervised fine-tuning, several advanced learning paradigms have been employed to enhance the performance and generalization capabilities of vulnerability detection models, e.g., contrastive learning, causal learning, multi-task learning, knowledge distillation, continual learning, and reinforcement learning.

\textit{\textbf{Contrastive Learning (T2.3.1)}} aims to learn more discriminative representations by encouraging the model to pull semantically similar instances closer and push dissimilar ones apart in the representation space, thereby improving the model’s ability to distinguish between structurally similar but semantically different code, i.e., secure versus vulnerable code.
Several studies adapt contrastive learning techniques to the vulnerability detection domain~\cite{jiang2022multi,wu2025sparse,niDistinguishingLookAlikeInnocent2023,wangSCLCVDSupervisedContrastive2024,duJointGeometricalStatistical2023,dingVulnerabilityDetectionCode2024,jiApplyingContrastiveLearning2024a, gajjarBridgingSemanticsStructure2025,liCLeVeRMultimodalContrastive,wangVuLMCBERTVulnerabilityDetection2025}.
Notably, Du et al.~\cite{duJointGeometricalStatistical2023} propose mutual nearest neighbor contrastive learning to align the source and target domains.
Ding et al.~\cite{dingVulnerabilityDetectionCode2024} introduce class-aware contrastive learning, which minimizes similarity only between samples with different class labels to improve type-specific discrimination.
Ji et al.~\cite{jiApplyingContrastiveLearning2024a} present a hierarchical contrastive learning framework to bring vector representations of related CWEs closer together, combining supervised and self-supervised contrastive losses to promote geometric spread and improved inter-class separation.

\textit{\textbf{Causal Learning (T2.3.2)}} guides models to focus on genuine cause-and-effect relationships rather than relying on spurious correlations in the code, e.g., variable or API names that are not actual causes of vulnerabilities. 
Rahman et al.~\cite{rahmanCausalDeepLearning2024} identify such spurious features via semantic-preserving perturbations (e.g., variable renaming or dead-code injection) and then systematically remove their influence using causal inference techniques.
Specifically, they simulate interventions, i.e., asking how the model would behave if certain spurious features were actively changed, and block non-causal paths by conditioning on known spurious features.
This strategy allows the model to rely on causally relevant features and, thereby, improve robustness and generalization to unseen or perturbed code scenarios, such as different projects or naming conventions.

\textit{\textbf{Multi-Task Learning (T2.3.3)}} has emerged as an effective paradigm for improving generalization by jointly training on related tasks, thereby reducing overfitting and enhancing model robustness across a range of tasks.
Several studies adopt multi-task learning to enhance vulnerability analysis:
Fu et al.~\cite{fuAIBugHunterPracticalTool2024} fine-tune under the multi-task setting for predicting CWE-ID and CWE-type, while
Chen et al.~\cite{chenImprovingVulnerabilityType2025} combine CWE-ID classification with line-level vulnerability detection.
Curto et al.~\cite{curtoMultiVDTransformerbasedMultitask2024} perform binary vulnerability detection alongside multi-class vulnerability categorization.
Similarly, Steenhoek et al.~\cite{steenhoekClosingGapUser2024} fine-tune for simultaneous binary classification, multi-class vulnerability type prediction, and localization.
Islam et al.~\cite{islamUnbiasedTransformerSource2023} combine vulnerability classification with description.
Du et al.~\cite{duGeneralizationEnhancedCodeVulnerability2024} apply multi-task instruction-tuning for vulnerability detection, localization, and interpretation of root causes; while Yang et al.~\cite{yangSecurityVulnerabilityDetection2024} focus on multi-task fine-tuning for vulnerability detection, explanation, and repair.
Russo et al.~\cite{russoLeveragingMultitaskLearning2025} leverage shared information between self-admitted technical debt~(SATD) and vulnerabilities to jointly detect both issues, Alqarni and Azim~\cite{alqarniAdvancedDetectionFramework2025} combine the analysis of code, system calls, and bitstreams, and Ding et al.~\cite{dingTRACEDExecutionawarePretraining2024} use multi-task pre-training to jointly learn static and dynamic code properties.

\textit{\textbf{Knowledge Distillation (T2.3.4)}} is a model compression strategy in which a smaller model, i.e., the student, is trained to replicate the behavior of a larger model, i.e., the teacher. 
By learning from the teacher’s output distributions, the student achieves comparable performance while significantly reducing computational cost, making it suitable for resource-constrained settings.
Some studies directly adopt pre-trained distilled models such as DistilBERT for vulnerability detection~\cite{daloisioCompressionLanguageModels2024,liuEnhancingVulnerabilityDetection2025,linDistilledContextualizedNeural2022}, while other studies apply knowledge distillation explicitly, using teacher and student LLMs~\cite{omarVulDefendNovelTechnique2023,shiGreeningLargeLanguage2024,weiDistillingLightweightLanguage2025,weyssowR2VulLearningReason2025,panichellaMetamorphicBasedManyObjectiveDistillation2025,liuPIONEERImprovingRobustness2026}, or a setup with CNN and GNN models to distill knowledge into or from a LLM backbone~\cite{nguyenSAFEAdvancingLarge2024,liuVulLMGNNsFusingLanguage2025,fuVulExplainerTransformerBasedHierarchical2023a,hanAutomaticClassificationModel2025}. 
A more advanced approach is presented by Fu et al.~\cite{fuVulExplainerTransformerBasedHierarchical2023a}, who introduce a hierarchical knowledge distillation framework. The CWE label space is split into multiple sub-distributions based on semantic similarity, and individual CNN teacher models are trained on each subset. A student language model then learns to generalize across the teacher outputs, improving its ability to handle multi-class vulnerability detection tasks. 
Han et al.~\cite{hanAutomaticClassificationModel2025} follow this approach with an enhanced memory mechanism.
Panichella~\cite{panichellaMetamorphicBasedManyObjectiveDistillation2025} combines metamorphic testing with many-objective optimization for distillation of LLMs for code, measuring model robustness across semantically equivalent code variants.

Retraining a model on all known vulnerabilities each time a new vulnerability is disclosed would demand considerable computational resources and time.
\textit{\textbf{Continual Learning (T2.3.5)}} enables models to incrementally learn new knowledge over time, such as new programming languages or up-to-date vulnerability type data~\cite{weiDistillingLightweightLanguage2025}, without forgetting what has already been learned (\textit{Catastrophic Forgetting}).
Gao et al.~\cite{gaoKeepingPaceEverIncreasing2023} propose an adaptive regularization strategy that preserves the most representative (i.e., informative and diverse) samples in each dataset for model retraining under project-level, language-level, and time-level continual learning settings.
Similarly, Zhang et al.~\cite{zhangMVDMultiLingualSoftware2024} present an incremental learning approach that allows the model to incorporate new programming languages without degrading performance on those it has already learned.
Tian et al.~\cite{tianYouOnlyTrain2025} introduce a training-free parameter fusion method that merges independent classification heads fine-tuned on different vulnerability types, allowing the fused model to adapt to new categories while preserving multi-class classification capabilities.

\textit{\textbf{Reinforcement Learning (RL) (T2.3.6)}} introduces a reward-based feedback loop, allowing models to optimize sequential decision-making processes.
Several studies focus on refining the model's reasoning capabilities.
Weyssow et al.~\cite{weyssowR2VulLearningReason2025} employ Reinforcement Learning from AI Feedback (RLAIF) to distill structured reasoning, reinforcing informative and coherent while penalizing misleading reasoning.
Similarly, Simoni et al.~\cite{simoniImprovingLLMReasoning2025} integrate an RL-based loss function to fine-tune instruction-based LLMs.
Ren et al.~\cite{renProRLearnBoostingPrompt2024} introduce a reward mechanism with prompt tuning that encourages the model to discover prompts that maximize detection performance.
Beyond reasoning, RL is utilized to optimize input and feature extraction. 
Jiang et al.~\cite{jiangEnhancingFineGrainedVulnerability2025a} propose an RL framework to learn vulnerability-relevant structures within code snippets.
Zhao et al.~\cite{zhaoAdversarialTrainingRobustness2025} use RL to generate high-quality adversarial examples, selecting optimal perturbations that minimize the model's vulnerability prediction probability to create harder training samples.

\textit{\textbf{Other Data-Centric (T2.3.7)}} learning techniques have been explored in the surveyed studies to improve model performance, e.g., Positive and Unlabeled (PU) learning, active learning, and adversarial training.
PU learning tackles the issue of limited and noisy labels by training models using only positively (vulnerable) labeled and unlabeled data~\cite{wenWhenLessEnough2023,kongSourceCodeVulnerability2024}.
Forms of active learning aim to reduce labeling costs by strategically selecting the most informative samples for annotation~\cite{alrashedyLearningDefectPrediction2024,lanSmartCutsEnhance2025,chengLeveragingSelfPacedLearning2025}. For example, Lan et al.~\cite{lanSmartCutsEnhance2025} combine dataset maps with active learning to prioritize valuable samples while filtering out those that may harm performance. 
Similarly, in self-paced learning, the most suitable data is selected for each training epoch, from easy to hard learning samples~\cite{chengLeveragingSelfPacedLearning2025}.
Adversarial training enhances model robustness by introducing crafted perturbed code examples during training, improving robustness to noise and attacks~\cite{yuPATVDVulnerabilityDetection2022, chenImprovingVulnerabilityType2025,zhaoAdversarialTrainingRobustness2025,duGeneralizationEnhancedCodeVulnerability2024,liuPIONEERImprovingRobustness2026}. \\




\subsubsection{Orchestration Technique (T3)}
\label{sec:orchestration_techniques}
While adaptation techniques focus on optimizing the model's internal parameters or input context, orchestration techniques focus on the design of the inference process.
Rather than relying on a simple input-model-output pipeline, orchestration techniques embed the LLM into advanced system designs that utilize multiple generation steps, iterative refinement, or collaborative workflows.
We categorize orchestration techniques into \textit{Multi-Step}, \textit{Verification}, \textit{Agentic}, \textit{Ensemble}, and \textit{Controller}-based systems.

Some studies engage the model in a \textit{\textbf{Multi-Step (T3.1)}} analysis, rather than issuing a single prompt for vulnerability detection~\cite{espinhagasibaMaySourceBe2024,huynhDetectingCodeVulnerabilities2025,liuExplorationPromptingLLM2024,liuSoftwareVulnerabilityDetection2023,ullahLLMsCannotReliably2024,wangSecureMindFrameworkBenchmarking2025,yuInsightSecurityCode2024,jiaoDeepVulHunterEnhancingCode2025,yangContextEnhancedVulnerabilityDetection2025,mhatreLLMGUARDLargeLanguage2025,pengICodeReviewerImprovingSecure2025,tsaiSequentialMultiStageApproach2025,dingVulnerabilityDetectionCode2024}. 
A common strategy within this approach is to decouple comprehension from detection, prompting the model to first analyze the code's functionality or intent before querying for vulnerabilities~\cite{espinhagasibaMaySourceBe2024,huynhDetectingCodeVulnerabilities2025,liuExplorationPromptingLLM2024,liuSoftwareVulnerabilityDetection2023,jiaoDeepVulHunterEnhancingCode2025,yangContextEnhancedVulnerabilityDetection2025,pengICodeReviewerImprovingSecure2025,wangSecureMindFrameworkBenchmarking2025}.
Beyond functional analysis, Yu et al.~\cite{yuInsightSecurityCode2024} task the model to first generate supplementary code files based on commit messages, subsequently using this supplementary context for the final detection.
Mhatre et al.~\cite{mhatreLLMGUARDLargeLanguage2025} follow a debugging style, first identifying code issues, followed by a deeper, targeted analysis of specific issues.
Tsai et al.~\cite{tsaiSequentialMultiStageApproach2025} propose a sequential approach that dynamically selects the optimal strategy, choosing between direct zero-shot classification, RAG-based prediction, or multi-agent collaboration, guided by the model's internal confidence and collaborative signals.

To mitigate hallucinations and improve precision, studies increasingly adopt \textit{\textbf{Verification (T3.2)}} mechanisms where models are prompted to review and improve generated outputs.
Verification extends the standard prompt-response interaction by re-prompting the model to evaluate, validate, or refine a prediction.
In single-model setups, the model is typically prompted to assess its prior response by verifying the correctness of its output~\cite{xiaStaticPatternMatching2025} or by recursively critiquing and improving~\cite{tambergHarnessingLargeLanguage2024,duGeneralizationEnhancedCodeVulnerability2024,sayaghThinkBroadAct2025}.
In multi-model settings, verification is implemented as an interactive process between models (collaborative refinement), possibly under different roles. 
For example, using the output of one model to prompt another model with corrective cues such as "An expert has found... please recheck it"~\cite{wangM2CVDEnhancingVulnerability2024,ambatiNavigatingInSecurityAIGenerated2024,tsaiLeveragingIntraInterReferences2025,widyasariChatGPTEnhancingSoftware2024,caoLLMCloudSecLargeLanguage2024,eberhardtVulnGPTEnhancingSource2024}. 

\textit{\textbf{Agentic (T3.3)}} systems embed LLMs into autonomous, decision-making loops, following a higher-level behavioral pattern involving multi-step planning, tool use, and self-directed reasoning.
Agents may operate individually or as part of multi-agent systems~\cite{caoLLMCloudSecLargeLanguage2024,yildizBenchmarkingLLMsLLMbased2025,widyasariLetTrialBegin2025,ahmedSecVulEvalBenchmarkingLLMs2025,eberhardtVulnGPTEnhancingSource2024,liMAVULMultiAgentVulnerability2025,tsaiSequentialMultiStageApproach2025,wangVulAgentHypothesisValidationBased2025,tsaiLeveragingIntraInterReferences2025}, where agents specialize in subtasks or roles, interact with tools, and collaboratively refine outputs.
For example, Yildiz et al.~\cite{yildizBenchmarkingLLMsLLMbased2025} employ Reasoning and Acting (ReAct) agents~\cite{yao2023react}, which follow a thought-action-observation cycle, for vulnerability detection in code repositories.
Widyasari et al.~\cite{widyasariLetTrialBegin2025} present a courtroom-inspired multi-agent framework with four role-specific agents: security researcher, code author, moderator, and review board. The agents collaborate to assess code vulnerabilities through iterative analysis, argumentation, summarization, and a final verdict based on collective reasoning.
Ahmed et al.~\cite{ahmedSecVulEvalBenchmarkingLLMs2025} propose a multi-agent pipeline consisting of a normalization, planning, context, detection, and validation agent.
Sayagh and Ghafari~\cite{sayaghThinkBroadAct2025} use multiple agentic teams, with a lister searching for potential CWEs in the input code, a reviewer requesting missing CWEs, further agents identifying external code context, and a final agent deciding on each CWE based on gathered context.

\textit{\textbf{Ensemble (T3.4)}} approaches combine the representations or predictions of diverse models to mitigate individual model limitations and biases.
To leverage complementary feature representations, e.g., semantic, contextual, and syntactic features, studies fuse representations from different pre-trained models~\cite{duDMVL4AVDDeepMultiview2025,wu2025sparse,mahmudEnsembleTransformerApproach2025,ridoyEnStackEnsembleStacking2024,shimmiVulSimLeveragingSimilarity,xiongVulDCodeBERTCodeBERTBasedVulnerability2024,sunEnhancedVulnerabilityDetection2023,jianjieCodeDefectDetection2023}.
For example, Ridoy et al.~\cite{ridoyEnStackEnsembleStacking2024} employ an ensemble stacking approach to synthesize the semantic understanding of CodeBERT, the structural representations of GraphCodeBERT, and the cross-modal capabilities of UniXcoder.
Distinct from standard feature fusion, Tian et al.~\cite{tianYouOnlyTrain2025} introduce a parameter fusion method that merges independent classification heads trained on distinct vulnerability datasets, allowing the model to adapt to new vulnerability types.
Ensemble approaches further include decision-level aggregation via voting mechanisms to combine predictions from multiple LLMs~\cite{widyasariChatGPTEnhancingSoftware2024,zhangBenchmarkingLargeLanguage2025} or ensembles of LLMs and other tools~\cite{zhangBenchmarkingLargeLanguage2025,zhouComparisonStaticApplication2024}.

Moving beyond static inference pipelines, some studies use LLMs in a \textit{\textbf{Controller (T3.5)}} or collaboration setting. 
A central model or routing mechanism analyzes the input code and coordinates downstream resources, selecting, e.g., the most appropriate expert models or prompts.
Yang et al.~\cite{yangMoEVDEnhancingVulnerability2025} and Wu and Xiao~\cite{wu2025sparse} implement a Mixture-of-Experts (MoE) framework where a router dynamically assigns input code to the most appropriate vulnerability experts.
Similarly, Wang et al.~\cite{wangVulAgentHypothesisValidationBased2025} employ a meta-agent to route code to specialized analyzers.
Peng et al.~\cite{pengICodeReviewerImprovingSecure2025} introduce a Mixture-of-Prompts architecture that activates specific prompt experts (dynamic prompt pipelines) based on code features of the code under detection.
Li et al.~\cite{liVulnTeamTeamCollaboration2025} simulate team collaboration involving a leader and various expert models specialized on syntactic features; the leader orchestrates the process and derives the final detection from the experts' analyses.
Farr et al.~\cite{farrExpertintheLoopSystemsCrossDomain2025} extend the routing logic to human collaborators. They flag code as automatic quarantine, cleared for deployment, or human analyst evaluation, where code with high model uncertainty is routed to human experts.


\section{Vulnerability Datasets}
\label{sec:datasets}
In this SLR, we set an emphasis on investigating the datasets used in the adaptation and evaluation of LLM-based vulnerability detection approaches.
Datasets play a central role: Their quality, structure, and coverage of diverse vulnerability types directly affect model performance and generalization.
Further, the choice of dataset does not merely benchmark performance.
A critical challenge in this domain is \textit{data leakage}, where data samples overlap with the corpora used to pre-train the model, potentially leading to inflated performance.

Furthermore, the use of datasets has evolved beyond standard training.
With the rise of advanced prompting techniques, datasets now serve as a dynamic input context, providing exemplars for in-context prompting or knowledge retrieval.
This shift also changes which parts of the data are used. While deep learning-based models often relied solely on code syntax, LLMs actively leverage the natural language components of datasets, such as commit messages, vulnerability descriptions, and code comments, to support high-level explanation and reasoning tasks.
Therefore, systematically comparing datasets is essential to understand their suitability for LLM-based detection, assess their limitations, and ensure meaningful evaluations.

\begin{figure}[b!]
    \centering 
    \includegraphics[width=0.8\linewidth]{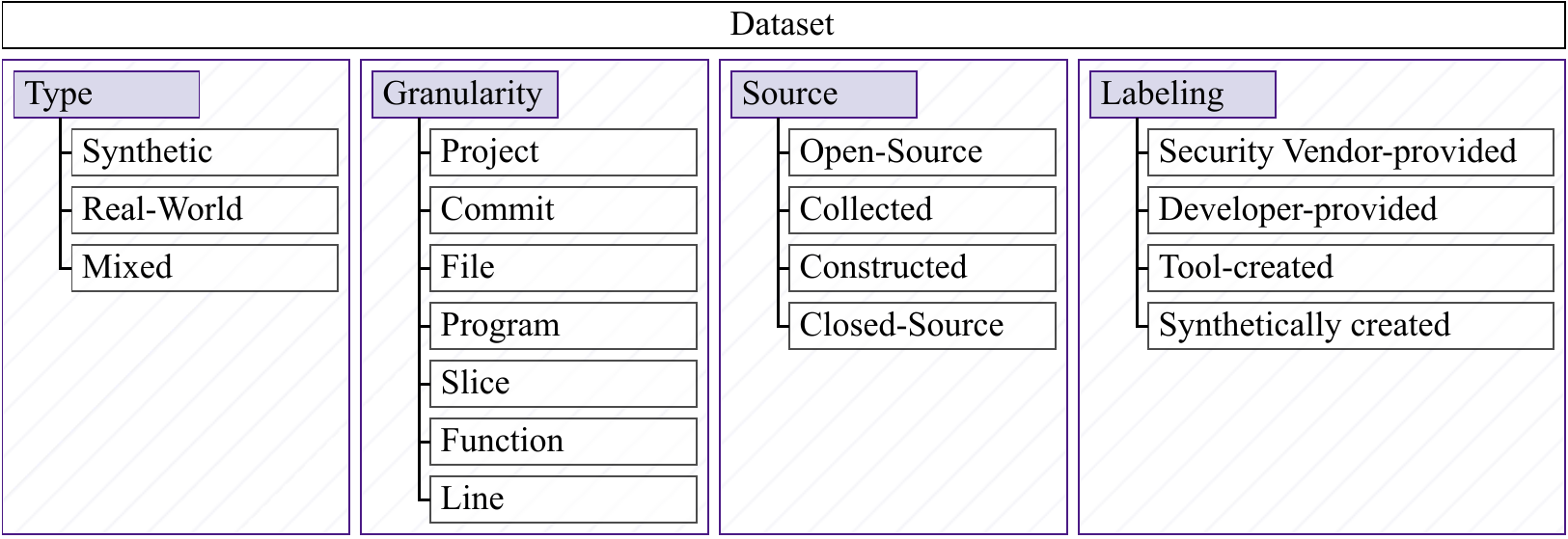}
    \caption{Taxonomy of datasets. A dataset may be associated with multiple values (i.e., white boxes) per category, e.g., providing different granularities.}
    \Description{Tree structure showing categories and attribute values for vulnerability detection datasets used.}
    \label{fig:taxonomy}
\end{figure} 

To systematically analyze the datasets used in the surveyed studies, we extend the presented taxonomy to cover datasets.
We outline the taxonomy based on the datasets used most commonly for training or evaluation in the surveyed studies, discussing type, granularity, source, and labeling.
Further, we investigate the CWE coverage and diversity of selected datasets, discussing limitations such as long-tail distributions and a lack of representative vulnerability coverage. 
Finally, we discuss trends in dataset usage, emphasizing the comparability of evaluations across different use cases.


\subsection{Taxonomy and Commonly Used Datasets}
\label{sec:taxonomy_datasets}
We categorize vulnerability datasets according to type, granularity, source, and labeling, as shown in Figure~\ref{fig:taxonomy}.
Similar to related surveys \cite{shereen2024sok, ferrag2024generative, taghavi2025large, sheng2025large, guo2024comprehensive, guo2023investigation}, we provide a comprehensive overview of commonly used datasets in the surveyed studies as well as selected, more recent datasets suited to address open challenges of LLM-based approaches, refer to Table~\ref{tab:datasets1}.
We briefly introduce these datasets along with the taxonomy and discuss the resulting practical challenges and biases in dataset construction.
In addition, we compare the selected datasets based on further characteristics such as their programming languages covered, as well as quantitative characteristics, i.e., dataset size, the share of samples representing vulnerabilities, and the number of distinct CWEs covered (cf. Table~\ref{tab:datasets2}) as a proxy for label diversity and comprehensiveness.

\subsubsection{Dataset Type}
We distinguish three dataset types: (i) synthetic, (ii) real-world, and (iii) mixed. 

\textit{\textbf{Synthetic}} datasets are artificially generated, often by injecting predefined vulnerability patterns into source code. 
The synthetic datasets Juliet \ccpp~\cite{JulietC} and Juliet Java~\cite{JulietJava} (subsets of the Software Assurance Reference Dataset~(SARD)~\cite{SARD}) provide comprehensive sets of test cases and entire test suites. These test suites are widely used to benchmark both traditional static analysis tools and deep learning-based vulnerability detectors.
A key strength of synthetic datasets lies in their reliable labels, as they inject known vulnerable patterns into code, thereby ensuring clear ground truth annotations. 
However, this injection-based approach often results in limited code diversity and reduced realism, as the code samples tend to be simple and may not reflect the complexity of real-world source code. Further, the manual effort required to represent a diverse set of vulnerabilities can be labor-intensive.

\textit{\textbf{Real-World}} datasets are sourced from open-source software projects, e.g., Devign~\cite{Devign}, or curated vulnerability databases, e.g., Big-Vul~\cite{BigVul} and MegaVul~\cite{MegaVul}.
These datasets are more suited for a realistic evaluation, capturing the challenges posed by detecting software vulnerabilities in real-world projects. 
However, due to the efforts required for crawling and parsing software projects, real-world datasets often suffer from label noise.

\textit{\textbf{Mixed}} datasets combine real-world code with synthetic data, leveraging augmentation to enrich or balance the dataset.
For example, the SARD~\cite{SARD} dataset contains a mix of production, synthetic, and academic vulnerable samples.
VulDeePecker~\cite{VulDeePecker} combines samples from SARD and the National Vulnerability Database (NVD)~\cite{NVD}.
Draper~\cite{Draper} combines synthetic functions from the Juliet test suite with open-source code from software packages and public repositories. 
Mixed datasets offer a practical trade-off between realism and control, making them particularly valuable for covering underrepresented or rare vulnerability types.

\subsubsection{Dataset Granularity}
Vulnerabilities can occur at various structural levels within the software, from single lines to entire files.
Dataset granularity refers to the level at which code is labeled or segmented within the dataset.
Following Shereen et al.~\cite{shereen2024sok}, we differentiate coarse, medium, and fine-grained levels.

\textit{Coarse-grained} datasets operate at the \textit{\textbf{Project}}, \textit{\textbf{Commit}}, or \textit{\textbf{File}} level. 
ReposVul~\cite{wangReposVulRepositoryLevelHighQuality2024} focuses on extracting dependencies on multiple granularity levels, including repository-level and file-level.
Similarly, CrossVul~\cite{CrossVul} labels samples at the file level, thereby aligning closely with practical use cases in vulnerability detection.
While this level of granularity offers realistic scenarios for analysis, it also introduces a high risk of noise due to the large code context surrounding the vulnerability.

\textit{Medium-grained} datasets segment code at the \textit{\textbf{Program}}, \textit{\textbf{Slice}} or \textit{\textbf{Function}} level. 
The largest share of commonly used vulnerability datasets is labeled at the function level~\cite{SARD,Draper,Devign,BigVul,D2A,Reveal,CVEfixes,chenDiverseVulNewVulnerable2023b,PrimeVul,duVulRAGEnhancingLLMbased2024,MegaVul,liCleanVulAutomaticFunctionLevel2024}, cf. Table~\ref{tab:datasets1}.
The VulDeePecker dataset, also known as the Code Gadget Database (CGD), represents program slices as code gadgets composed of a number of program statements that are semantically related to each other through data or control dependencies.
While this granularity supports a more focused analysis, it may miss critical contextual information when vulnerabilities span multiple functions or require interprocedural analysis.

\textit{Fine-grained} datasets label vulnerabilities at the \textit{\textbf{Line}} level, enabling precise vulnerability localization. 
Big-Vul, D2A~\cite{D2A}, FormAI, and ReposVul provide annotations at the line-level.
Such a narrow scope, however, omits the broader context required to detect certain vulnerabilities, especially those involving complex data flows or control dependencies.

\begin{table}[tb]
\fontsize{8pt}{8pt}\selectfont
\renewcommand{\arraystretch}{1.25}
\caption{Top 15 most commonly used datasets and selected datasets (*) sorted by year of publication. Categorization by type, granularity, source, and labeling. The top 3 most commonly used datasets are in bold. \label{tab:datasets1}}
\resizebox{\textwidth}{!}{%
	\begin{tabular}{@{} p{2.85cm}p{0.4cm}lp{3.1cm}ll >{\raggedleft}p{0.6cm} r @{}}
			\toprule
              \textbf{Dataset} & \textbf{Year} & \textbf{Type} & \textbf{Granularity} & \textbf{Source} & \textbf{Labeling} & \textbf{\#Used} & \textbf{Resource} 
              \\
              \midrule
              SARD \cite{SARD} & 2006 & Mixed & File & Open-Source & Synthetically & 32 & \cite{SARD} \\ 
              Juliet \ccpp \cite{JulietC} & 2017 & Synthetic & File & Open-Source & Synthetically & 9 & \cite{JulietC} \\ 
              Juliet Java \cite{JulietJava} & 2017 & Synthetic & File & Open-Source & Synthetically & 6 & \cite{JulietJava} \\ 
              VulDeePecker \cite{VulDeePecker} & 2018 & Mixed & Slice & Constructed & Security Vendor & 11 & \cite{VulDeePecker_git} \\ 
              Draper \cite{Draper} & 2018 & Mixed & Function & Constructed & Tool, Synthetically & 11 & \cite{DraperVDISC_osf} \\      
              \textbf{Devign} \cite{Devign} & 2019 & Real & Function & Open-Source & Developer & \textbf{69} & \cite{Devign_git} \\ 
              \textbf{Big-Vul} \cite{BigVul} & 2020 & Real & Function, Line & Collected & Security Vendor & \textbf{63} & \cite{BigVul_git} \\ 
              D2A \cite{D2A} & 2021 & Real & Function, Line & Open-Source & Tool & 9 & \cite{D2A_git} \\
              \textbf{ReVeal} \cite{Reveal} & 2021 & Real & Function & Collected & Developer & \textbf{40} & \cite{ReVeal_git} \\ 
              CVEfixes (v1.0.8) \cite{CVEfixes} & 2021 & Real & Function, Commit, File & Collected & Security Vendor & 19 & \cite{CVEfixes_zenodo} \\ 
              CrossVul \cite{CrossVul} & 2021 & Real & File & Collected & Security Vendor & 4 & \cite{CrossVul_zenodo} \\
              SecurityEval (v2.1) \cite{SecurityEval} & 2022 & Mixed & Program & Constructed & Synthetically, Tool & 4 & \cite{SecurityEval_git} \\
              DiverseVul \cite{chenDiverseVulNewVulnerable2023b} & 2023 & Real & Function & Collected & Developer & 27 & \cite{DiverseVul_git} \\
              SVEN \cite{SVEN} & 2023 & Real & Program & Constructed & Developer & 8 & \cite{SVEN_git} \\
              FormAI* \cite{FormAI} & 2023 & Synthetic & Program, Line & Constructed & Tool & 2 & \cite{FormAI_git} \\
              ReposVul* \cite{wangReposVulRepositoryLevelHighQuality2024} & 2024 & Real & Project, File, Function, Line & Collected & Tool & 3 & \cite{ReposVul_git} \\
              PrimeVul \cite{dingVulnerabilityDetectionCode2024} & 2024 & Real & Function & Constructed & Security Vendor & 27 & \cite{PrimeVul_git} \\ 
              PairVul*~\cite{duVulRAGEnhancingLLMbased2024} & 2024 & Real & Function & Collected & Security Vendor & 1 & \cite{PairVul_git} \\
              MegaVul* (2024/04) \cite{MegaVul} & 2024 & Real & Function & Collected & Security Vendor & 2  & \cite{MegaVul_git} \\ 
              CleanVul*~\cite{liCleanVulAutomaticFunctionLevel2024} & 2024 & Real & Function & Open-Source & Developer, Tool & 3 & \cite{CleanVul_git} \\ 
            \bottomrule 
	\end{tabular}
}%
\end{table}

\subsubsection{Dataset Source}
To categorize the source of the dataset, we adapt the four categories defined by Hou et al.~\cite{hou2023large}, i.e., (i) open-source, (ii) collected, (iii) constructed, and (iv) closed-source.

\textit{\textbf{Open-Source}} datasets are derived from public data collections accessible through open-source platforms or repositories, e.g., Github repositories. Examples for open-source datasets are SARD, Devign, D2A, and CleanVul~\cite{liCleanVulAutomaticFunctionLevel2024}.
Their accessibility ensures transparency and reproducibility, making them widely used in academic research. 
However, datasets sourced directly from GitHub repositories often lack reliable ground-truth labels and are sensitive to repository selection and curation quality, i.e., the specific projects chosen and how samples are filtered, labeled, and balanced, leading to biased, noisy, or unrepresentative data.

\textit{\textbf{Collected}} datasets are those scraped, mined, and extracted by researchers from various sources such as security trackers or related databases (e.g., NVD~\cite{NVD}, CVE~\cite{CVECommonVulnerabilities}, or CWE database~\cite{CWECommonWeakness}). 
MegaVul, for example, is collected by crawling the CVE database along with CVE-related open-source projects hosted across Git-based platforms. 
While authoritative vulnerability records provide structured metadata and standardized vulnerability classifications, the mapping of records to source code can be non-trivial and may miss undocumented or undisclosed vulnerabilities.

\textit{\textbf{Constructed}} datasets are created by modifying or augmenting one or multiple other (collected) datasets, either manually or using semi-automatic methods, to better align with domain-specific objectives, e.g., focusing on specific CWEs or programming languages.
Constructed datasets include the mixed datasets VulDeePecker and Draper.
Further, SecurityEval~\cite{SecurityEval} is constructed from sources such as GitHub’s CodeQL documentation, the CWE database, Sonar static analyzer rules, and prior Copilot-based studies. Additional samples were crafted by the authors.
SVEN~\cite{SVEN} refines and validates samples from Big-Vul, CrossVul, and VUDENC~\cite{VUDENC}, focusing on nine common CWE types.
The PrimeVul dataset~\cite{PrimeVul} merges and de-duplicates data from Big-Vul, CrossVul, CVEfixes~\cite{CVEfixes}, and DiverseVul~\cite{chenDiverseVulNewVulnerable2023b}.

\textit{\textbf{Closed-Source}} datasets are obtained from commercial or industrial entities.
Closed-sourced datasets are absent from the listed commonly used datasets. 
Notably, only few surveyed studies~\cite{dingLeveragingDeepLearning2023,mockCrossDomainEvaluationTransformerBased2025,pengICodeReviewerImprovingSecure2025,xueAIDetectVulSoftwareVulnerability2025} use an industrial, closed-sourced dataset. 
Although such datasets offer significant potential for research targeting real-world deployment scenarios, they are often subject to restrictions that limit the publication and sharing of company-internal data. Such restricted access limits reproducibility, public benchmarking, and broader adoption within the research community.

\subsubsection{Dataset Labeling}
Accurate labeling is essential for constructing high-quality vulnerability datasets and reliable evaluations.
To categorize the labeling of datasets, we adopt the four main label origin categories by Croft et al.~\cite{croft2023data}: (i)~security vendor-provided, (ii)~developer-provided, (iii)~tool-created, and (iv)~synthetically created.

\textit{\textbf{Security Vendor-provided}} labels are derived from curated vulnerability databases maintained by security vendors, i.e., organizations that collect and standardize information on disclosed vulnerabilities from various advisories.
An example database is the NVD. 
Many entries include links to corresponding patches on security trackers or GitHub sites, enabling the mapping of vulnerabilities to real-world source code.
For example, the Big-Vul dataset was collected by crawling the public CVE database and linking the CVE entries to related source code changes in GitHub projects.
This labeling strategy is reliable and consistent with real-world vulnerability disclosures. However, it is limited to publicly disclosed cases and includes historical data with labels that may not align with up-to-date vulnerability mappings~\cite{wangReposVulRepositoryLevelHighQuality2024}.

\textit{\textbf{Developer-provided}} labels are extracted directly from a project's development history or issue tracker systems on platforms such as GitHub. 
A commonly used strategy involves \textit{vulnerability fixing commits (VFCs)}, identifying vulnerable code before a fix and secure code after a fix. 
VFCs are often identified via project references by the NVD or keyword search.
The Devign dataset (also referred to as FFmpeg and QEMU) consists of vulnerable functions extracted from VFCs identified via keyword-based filtering and manually labeled by security researchers. Devign is also part of the CodeXGLUE benchmark~\cite{luCodeXGLUEMachineLearning}.
Contrary to manually labeling VFCs, relying solely on VFCs often results in noisy labels due to the simplifying assumption that all pre-commit code is vulnerable and all post-commit code is secure~\cite{croft2023data}.
In practice, VFCs may contain unrelated changes such as refactoring or test updates, and fixes are sometimes distributed across multiple commits. 
Further, labeling based solely on modified lines may overlook vulnerable context and cross-line dependencies.
To address these challenges, newer datasets propose refined VFC labeling strategies. 
The PrimeVul dataset~\cite{PrimeVul} uses two labeling strategies: \textit{PRIMEVUL-ONEFUNC}, which marks a function as vulnerable only if it is the only function modified by a VFC, and \textit{PRIMEVUL-NVDCHECK}, which cross-references CVE descriptions with modified functions. 
CleanVul addresses the label noise using an LLM, which assigns a confidence score from 0 (no vulnerability detected) to 4 (very high likelihood of vulnerability) to identify vulnerability-fixing changes and filter out unrelated modifications such as test updates or refactorings.

\textit{\textbf{Tool-created}} labels are automatically generated using, e.g., static analyzers or formal verification tools. 
For instance, in the Draper dataset, labels are generated using static analyzers and then manually mapped to binary vulnerability labels and CWEs by a team of security experts. 
Similarly, the D2A dataset applies differential analysis to VFC version pairs from several open-source projects using static analysis tools, identifying disappearing bugs likely to present true vulnerabilities. 
Notably, the FormAI dataset~\cite{FormAI} consists of compilable and self-contained C programs generated using GPT-3.5-Turbo, incorporating varying levels of complexity to simulate a broad spectrum of coding patterns. Labels are derived using formal verification.
ReposVul jointly employs LLMs for evaluating the relevance between code changes and vulnerability fixes and static analysis tools for checking the vulnerabilities in the changes.
While tools enable large-scale dataset creation, they are often criticized for generating a high volume of false positives.
Static analysis tools, in particular, produce numerous warnings, many of which originate from low-confidence detections~\cite{croft2023data}.
Formal verification offers higher detection reliability but at higher computational costs.

\textit{\textbf{Synthetically created}} labels result from injecting known vulnerable patterns into code, generating controlled artificial samples, e.g., as used by the synthetic Juliet test suites. This method is useful for training on rare vulnerability types and balancing datasets. However, such patterns may oversimplify the complexity of real-world vulnerabilities.

\begin{table}[tb]
\fontsize{8pt}{8pt}\selectfont
\renewcommand{\arraystretch}{1.25}
\setlength{\tabcolsep}{4.5pt}
\caption{Top 15 most commonly used datasets and selected datasets (*) sorted by year of publication. Overview of supported programming languages, size (corresponding to the granularity first in line, e.g., number of functions), share of vulnerable samples, support of multi-class classification, and number of CWEs covered. The numbers stated are reproduced with the officially published datasets, cf. Table~\ref{tab:datasets1}. The top 3 most commonly used datasets are in bold. \label{tab:datasets2}}
\resizebox{\textwidth}{!}{%
	\begin{tabular}{@{} p{2.85cm}p{4cm}p{3.05cm}rrcr @{}}
			\toprule
              \textbf{Dataset} & \textbf{Programming Language} & \textbf{Granularity} & \textbf{Size} & \textbf{\%Vuln} & \shortstack[l]{\textbf{Multi-}\\\textbf{Class}}  & \textbf{\#CWEs} \\ \midrule
              SARD \cite{SARD} & \ccpp, C\#, Java, PHP & File &  450,000+ & n/s & \cmark & 150+ \\ 
              Juliet \ccpp \cite{JulietC} & \ccpp & File &  64,099 & n/s & \cmark & 118 \\ 
              Juliet Java \cite{JulietJava} & Java & File & 28,881 & n/s & \cmark & 112 \\ 
              VulDeePecker \cite{VulDeePecker} & \ccpp & Slice & 61,638 & 28.7 &  \cmark & 2 \\ 
              Draper \cite{Draper} & \ccpp & Function & 1,274,366 & 6.5 & \cmark & 4  \\      
              \textbf{Devign} \cite{Devign} & \ccpp & Function & 48,687 & 48.0 & \xmark & n/a  \\ 
              \textbf{Big-Vul} \cite{BigVul} & \ccpp & Function, Line & 264,919 & 4.5 & \cmark & 91 \\ 
              D2A \cite{D2A} & \ccpp & Function, Line & 1,295,623 & 1.4 & \cmark & n/a \\
              \textbf{ReVeal} \cite{Reveal} & \ccpp & Function & 18,169 & 9.2 & \xmark & n/a \\ 
              CVEfixes (v1.0.8) \cite{CVEfixes} & 27 languages & Function, Commit, File & 277,948 & 45.5 & \cmark & 272  \\ 
              CrossVul \cite{CrossVul} & 40+ languages & File & 27,476 & 50.0 & \cmark & 168 \\
              SecurityEval (v2.1) \cite{SecurityEval} & Python & Program & 121 & 100.0 & \cmark & 69 \\
              DiverseVul \cite{chenDiverseVulNewVulnerable2023b} & \ccpp & Function & 330,492 & 5.7 & \cmark & 150  \\
              SVEN \cite{SVEN} & \ccpp, Python & Program & 1,606 & 50.0 & \cmark & 9 \\
              FormAI* \cite{FormAI} & C & Program, Line & 246,549 & 80.2 & (\cmark) & (41) \\
              ReposVul* \cite{wangReposVulRepositoryLevelHighQuality2024} & \ccpp, Java, Python & Project, File, Function, Line & 232,465 & 0.7 & \cmark & 236 \\
              PrimeVul \cite{dingVulnerabilityDetectionCode2024} & \ccpp & Function & 235,768 & 2.9 &  \cmark & 140  \\ 
              PairVul*~\cite{duVulRAGEnhancingLLMbased2024} & \ccpp & Function & 8,628 & 50.0 & \cmark & 95 \\
              MegaVul* (2024/04) \cite{MegaVul} & \ccpp & Function & 353,873 & 5.1 & \cmark & 176 \\ 
              CleanVul*~\cite{liCleanVulAutomaticFunctionLevel2024} & \ccpp, C\#, Java, JavaScript, Python & Function & 11,632 & 50.0 & \xmark & n/a \\ %
              \bottomrule 
		\end{tabular}
}%
\end{table}

\subsubsection{Dataset Programming Language}
\label{sec:dataset_prog_lan}
Commonly used vulnerability datasets are heavily biased towards \ccpp, as shown in Table~\ref{tab:datasets2}. 
While this bias reflects the prevalence of these languages in technological infrastructure, it creates a significant imbalance, as vulnerability patterns are intrinsically linked to the programming language.
Qian et al.~\cite{qianSoftwareVulnerabilityAnalysis2025a} highlight that unmanaged languages such as \ccpp\, are prone to memory safety violations (e.g., buffer overflows), whereas managed languages such as JavaScript or Python frequently suffer from logic bugs.
In a comparative analysis of interpreted languages (Java, JavaScript, Python, and Ruby), Ruohonen~\cite{ruohonenSimilaritiesSoftwareVulnerabilities2021} observes that while certain weaknesses such as cross-site scripting and input validation errors are ubiquitous across all ecosystems, language-specific trends emerge.
For instance, cryptographic issues were found to be prevalent in JavaScript but significantly less common in Java or Python~\cite{ruohonenSimilaritiesSoftwareVulnerabilities2021}.
Consequently, the predominance of \ccpp\, risks over-optimizing models for memory safety violations while potentially neglecting the distinct vulnerability landscapes of modern web languages.

Fewer datasets cover multiple programming languages, i.e., SARD, CVEfixes, CrossVul, and CleanVul; however, they are not represented in the same share.
Recent studies have begun constructing datasets for less covered languages, e.g., Python~\cite{ozturkNewTricksOld2023} and Rust~\cite{luoHALURustExploitingHallucinations2025}, contributing to broader language diversity in vulnerability research.

\begin{figure}[b]
    \centering  
    \includegraphics[width=0.8\linewidth]{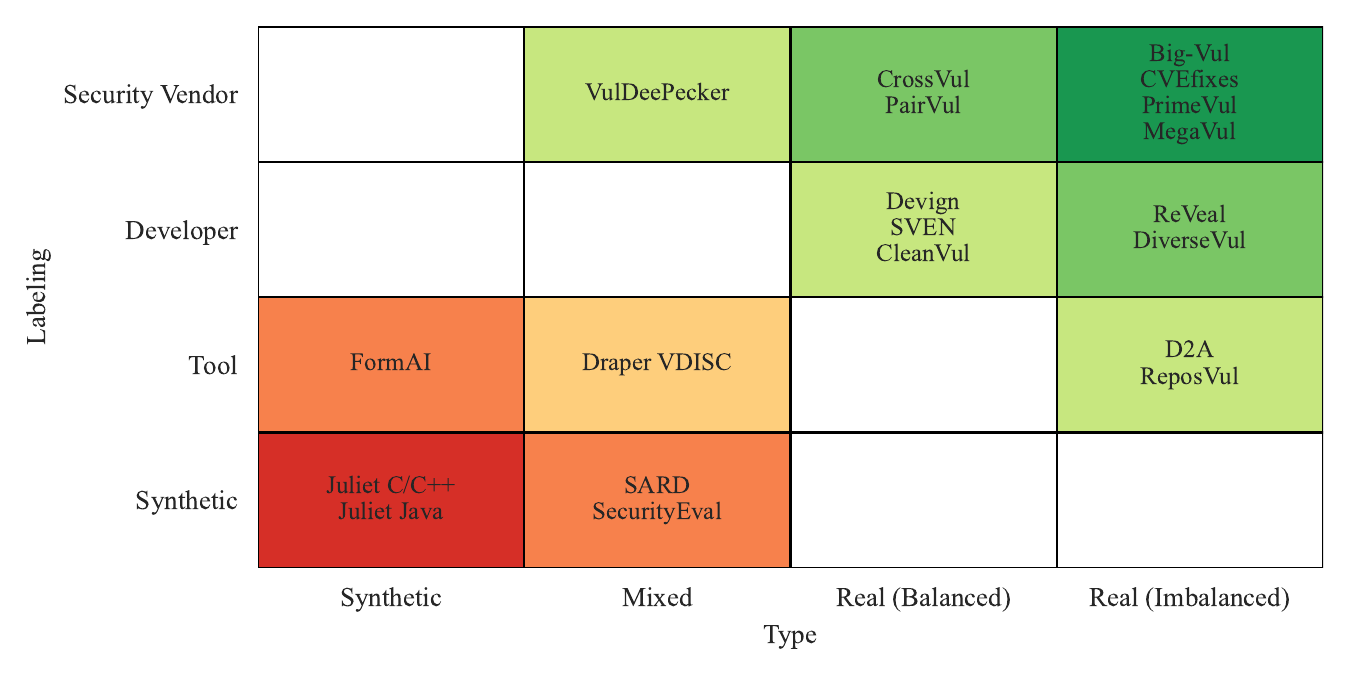}
    \caption{Datasets and their realistic nature. Colors range from red to green, symbolizing increasing realistic nature. Red is the most synthetic, and green is the most realistic. Figure adapted from \cite{Reveal}, extended.}
    \Description{Matrix symbolizing the datasets' realistic nature (type vs. vulnerability diversity and balance).}
    \label{fig:datasets_tax_map}
\end{figure} 

\subsubsection{Dataset Size and Share of Vulnerable Samples}
The size of a dataset is strongly influenced by the dataset source and its labeling method. 
Smaller datasets typically involve manual curation, enabling high label quality but limited scalability.
SecurityEval, for example, focuses on evaluating secure code generation in Python. It consists of 121 manually curated prompt-based samples across 75 CWE types, resulting in a dataset with 100\% vulnerable samples.
Similarly, SVEN focuses on evaluating secure code generation, offering improved label reliability with a 50\% share of vulnerable samples, but is limited in scope due to its focus on a fixed subset of CWEs.
In contrast, larger datasets rely on automated pipelines for scalability.
DiverseVul~\cite{chenDiverseVulNewVulnerable2023b} scales up the volume and variety of functions compared to earlier VFC-based datasets such as Devign, ReVeal~\cite{Reveal}, Big-Vul, CrossVul, and CVEfixes, with a share of 5\% vulnerable samples. 

The dataset type and the labeling method used in dataset construction influence both the quality and realism, cf. Figure~\ref{fig:datasets_tax_map}. 
Synthetic datasets, such as Juliet or FormAI, offer high label accuracy due to controlled injection of known vulnerability patterns but often lack the structural complexity and semantic diversity found in real-world code.
In contrast, datasets derived from real-world sources provide greater realism, better reflecting practical vulnerability scenarios. 
However, they often suffer from label noise due to assumptions inherent in VFC-based labeling.
These datasets also vary in the proportion of vulnerable samples. 
Highly imbalanced datasets, i.e., where vulnerabilities are underrepresented, reflect the natural rarity of vulnerabilities in real code. 
While realistic, such an imbalance poses challenges for training models, which may become biased towards predicting the majority class. 
To mitigate this challenge, preprocessing techniques are often required, e.g., resampling: oversampling the minority (vulnerable) class or undersampling the majority (non-vulnerable) class to achieve a more balanced class distribution.
Figure~\ref{fig:datasets_tax_map} visualizes this trade-off between labeling accuracy and realism, providing a practical framework for selecting datasets for different research objectives.

\subsubsection{Dataset Multi-Class Labeling}
Vulnerability detection is often framed as a binary classification task, i.e., distinguishing between vulnerable and non-vulnerable code. 
For example, the ReVeal dataset provides security-related patches collected from bug tracking systems, offering paired vulnerable and fixed code samples labeled in binary form. 

Many datasets go beyond binary classification by providing labels for specific vulnerability types, enabling multi-class and multi-label classification. 
CVEfixes is a comprehensive and automatically curated dataset collected from CVE entries in the NVD. 
It identifies linked open-source projects and extracts function-level vulnerable and fixed code samples from VFCs. 
The latest release at the time of writing (v1.0.8) includes all CVEs published up to July 23, 2024, covering 272 distinct CWEs. 
The PairVul dataset~\cite{duVulRAGEnhancingLLMbased2024} takes a more controlled approach by focusing on pairs of vulnerable and non-vulnerable code with high lexical similarity. 
These pairs are extracted from Linux kernel CVEs, labeled using VFCs, and filtered to ensure that the associated fixes are not reverted or modified in subsequent commits. 
While the dataset offers high-quality and balanced samples, its scope is currently limited to selected Linux kernel vulnerabilities.

A common limitation across vulnerability datasets is the lack of coverage of recently disclosed vulnerabilities. 
Static datasets may quickly become outdated, reducing their relevance for real-time or forward-looking research. 
In addition, this temporal lag introduces a critical validity threat for LLMs in the form of data leakage.
Since widely used static datasets are likely contained within the pre-training corpora of LLMs and new models are continuously emerging, high performance on these benchmarks may indicate memorization rather than genuine generalization.
This challenge highlights the need for continuously maintained and extensible datasets, e.g., CVEfixes, which is supported by an automated pipeline, to ensure evaluation against novel, unseen vulnerabilities.

Datasets such as Big-Vul, CrossVul, CVEfixes, and ReposVul further offer rich \textit{metadata}, including CWE descriptions and severity scores. 
Such metadata supports diverse generative tasks, for example, root-cause reasoning and context-aware vulnerability analysis, broadening their use beyond classification.


\subsection{CWE Coverage and Diversity}
\label{sec:datasets_cwe_analysis}
To assess the real-world applicability and limitations of LLM-based vulnerability detection approaches, it is crucial to use datasets that capture a diverse range of vulnerability types.
Such diversity enables more realistic evaluations, as it reflects the variety and complexity of vulnerabilities encountered in practical software development.
As shown in Table~\ref{tab:datasets2}, the number and variety of vulnerability types (i.e., number of CWE classes) represented vary considerably across the selected datasets. 
Some datasets focus narrowly on specific vulnerability types and CWE classes, while others aim for broader CWE diversity.
To assess the diversity and composition of CWEs, we analyze selected real-world and mixed datasets providing multi-class labels (cf. Table~\ref{tab:datasets1}) in more depth, focusing on their CWE coverage and distribution. 
Specifically, we analyze VulDeePecker, Draper, Big-Vul, CVEfixes, CrossVul, SecurityEval, DiverseVul, SVEN, ReposVul, PrimeVul, PairVul, and MegaVul.
D2A and FormAI are excluded from this analysis.
D2A is not considered, as it provides only static analyzer outputs without CWE annotations. 
Similarly, FormAI, while performing manual CWE mapping, does not include these mappings as part of the dataset but only error types provided by the formal verification tool. 

\begin{figure}[b]
    \centering  
    \includegraphics[width=\columnwidth]{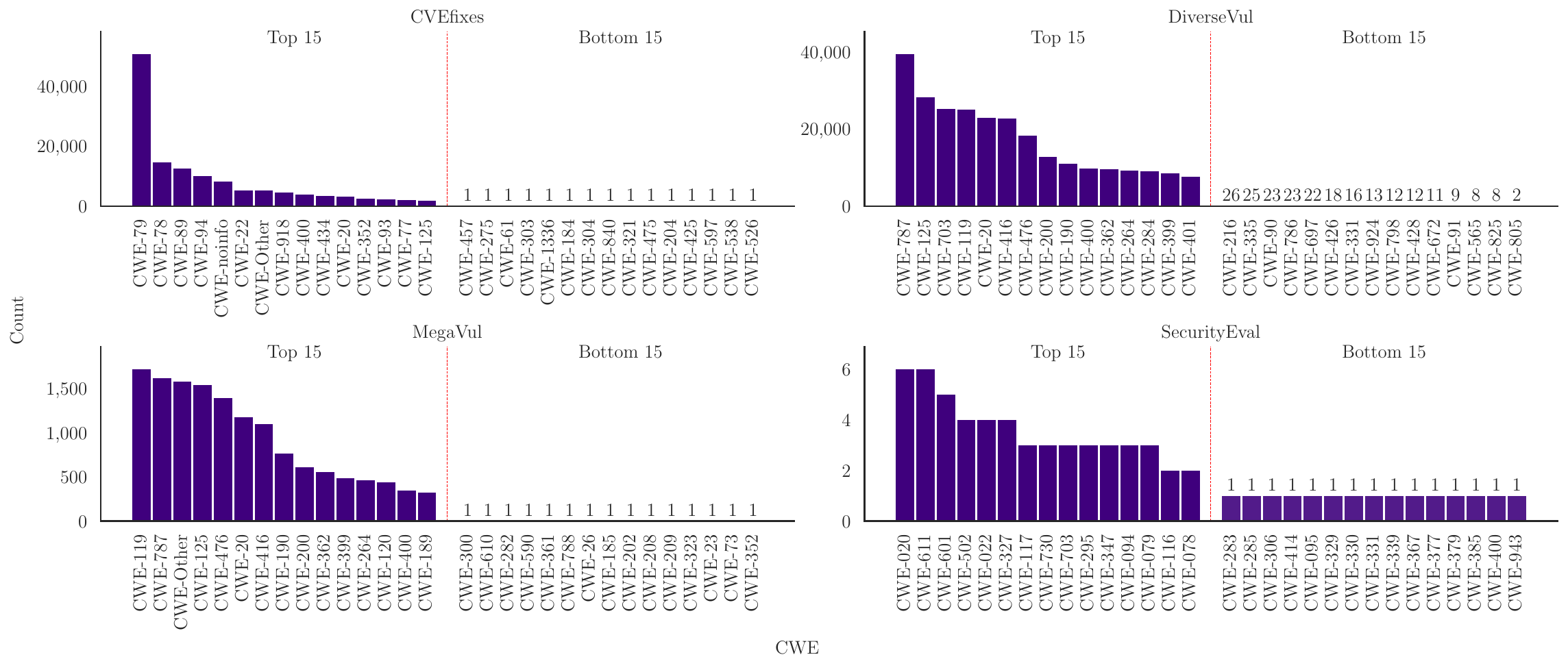}
    \caption{Long-tail distribution of CWEs in selected datasets: A small number of CWE types are heavily represented, while the majority of CWE types appear only rarely.}
    \Description{Skewed long-tail distribution of CWEs in multi-class datasets: A small number of CWE types are heavily represented, while the majority of CWE types appear only rarely.}
    \label{fig:long_tail}
\end{figure} 

\begin{table}[b]
\centering
\caption{CWE-1000~\cite{cwe1000} research view pillars with the number of hierarchically subordinated CWEs.}
\label{tab:pillars}
\fontsize{8pt}{8pt}\selectfont
\renewcommand{\arraystretch}{1.15}
\resizebox{\textwidth}{!}{%
    \begin{tabular}{@{}l p{4.75cm} p{6.6cm} r@{}}
    \toprule
    \textbf{CWE-ID} & \textbf{Pillar Name} & \textbf{Description} & \textbf{\#CWEs} \\
    \midrule
    CWE-284 & Improper Access Control & Weaknesses related to protection mechanisms such as authentication, authorization, and accountability & 166\\
    CWE-435 & Improper Interaction Between Multiple Correctly-Behaving Entities & Weaknesses that arise due to unexpected consequences when multiple entities interact, even though each entity behaves correctly in isolation & 16\\
    CWE-664 & Improper Control of a Resource Through its Lifetime & Resource management weaknesses such as improper initialization, reuse, or cleanup; includes cases where explicit instructions for resource creation, usage, or destruction are not properly followed & 367\\
    CWE-682 & Incorrect Calculation & Calculations that generate incorrect or unintended results later used in security-critical decisions or resource management & 14\\
    CWE-691 & Insufficient Control Flow Management & Logic or flow vulnerabilities such as incorrect conditions or decision-making & 84\\
    CWE-693 & Protection Mechanism Failure & Weaknesses caused by bypassing, misconfiguring, or incorrectly using security mechanisms & 100\\
    CWE-697 & Incorrect Comparison & Weaknesses in comparison logic between variables, objects, or values & 22\\
    CWE-703 & Improper Check or Handling of Exceptional Conditions & Weaknesses where exceptional conditions that rarely occur during normal operation are not properly anticipated or handled & 59\\
    CWE-707 & Improper Neutralization & Inadequate sanitation or escaping of input/output data where the application fails to ensure that data conforms to expected formats and is safe & 144 \\
    CWE-710 & Improper Adherence to Coding Standards & Violations of safe and established programming practices & 195\\
    \bottomrule
    \end{tabular}
}%
\end{table}

As discussed in prior works~\cite{dengImprovingLongTailVulnerability2024,jiangInvestigatingLargeLanguage2024,MegaVul}, real-world vulnerability datasets exhibit a characteristic \textit{long-tail distribution}: 
A small number of CWE types are heavily represented, while the majority of CWE types appear only rarely, often with just one or two examples. This characteristic is visualized in Figure~\ref{fig:long_tail}, which shows the count of CWEs for the 15 most and least represented CWE classes for selected datasets.
For example, while CVEfixes covers 272 distinct CWE types, the majority are sparsely represented. A similar pattern is observed for DiverseVul, which covers 150 CWEs. Such skewed distributions lead to biased learning, where models disproportionately focus on a small subset of vulnerability types.
MegaVul and SecurityEval represent efforts to encourage a more balanced representation.
SecurityEval, in particular, provides one to six manually curated examples for 69 CWEs. These efforts represent a step towards reducing long-tail effects and developing more generalizable vulnerability detection models.

For a more structured comparison of the CWE distributions across the datasets, we adopt the \textbf{CWE-1000 Research View}~\cite{cwe1000}.
This view provides a hierarchical grouping of CWEs intended for research and analysis.
It groups CWEs into broad research concepts to help identify inter-dependencies, shared characteristics, and root causes of vulnerabilities. 
The CWE-1000 view spans multiple levels of abstraction:
At the highest level are \textit{pillars} and \textit{categories}, which provide high-level groupings but are not intended for direct vulnerability mapping.
Beneath these are the \textit{class}, \textit{base}, and \textit{variant} levels, with variant representing the most specific level, i.e., language- or technology-specific weaknesses.
By design, the CWE-1000 view organizes every weakness in the CWE catalog into one of ten pillars, cf. Table~\ref{tab:pillars}.

We consider the mapping of CWE-1000 to cluster the CWEs represented in the individual datasets. 
This clustering enables a structured comparison beyond CWE counts by highlighting which vulnerability types are emphasized or underrepresented, assessing the thematic diversity and coverage of vulnerability datasets.
Notably, many datasets were built using the NVD, which also includes deprecated or discouraged CWE entries not mapped by CWE-1000, as well as general categories such as \texttt{CWE-OTHER} or \texttt{CWE-NOINFO}, used when insufficient information is available. 
To account for these cases, we introduce an additional group labeled \textit{not mapped}, capturing all CWE entries that fall outside the scope of the CWE-1000 view. 
For each dataset, we compute the relative share of each CWE based on the vulnerable functions (or files, depending on the granularity provided).
A special case is the CVEfixes dataset, where VFCs are associated with CVE identifiers, and each CVE may be linked to one or more CWEs. 
Some commits address multiple CVEs, and the associated CWE mappings may span different levels of specificity. 
To normalize the distributions and ensure a consistent comparison, we compute the share of each CWE based on the total number of CWE labels assigned to all vulnerable samples, rather than the number of vulnerable functions.

\begin{figure*}[tb]
    \centering  
        \includegraphics[width=\textwidth]{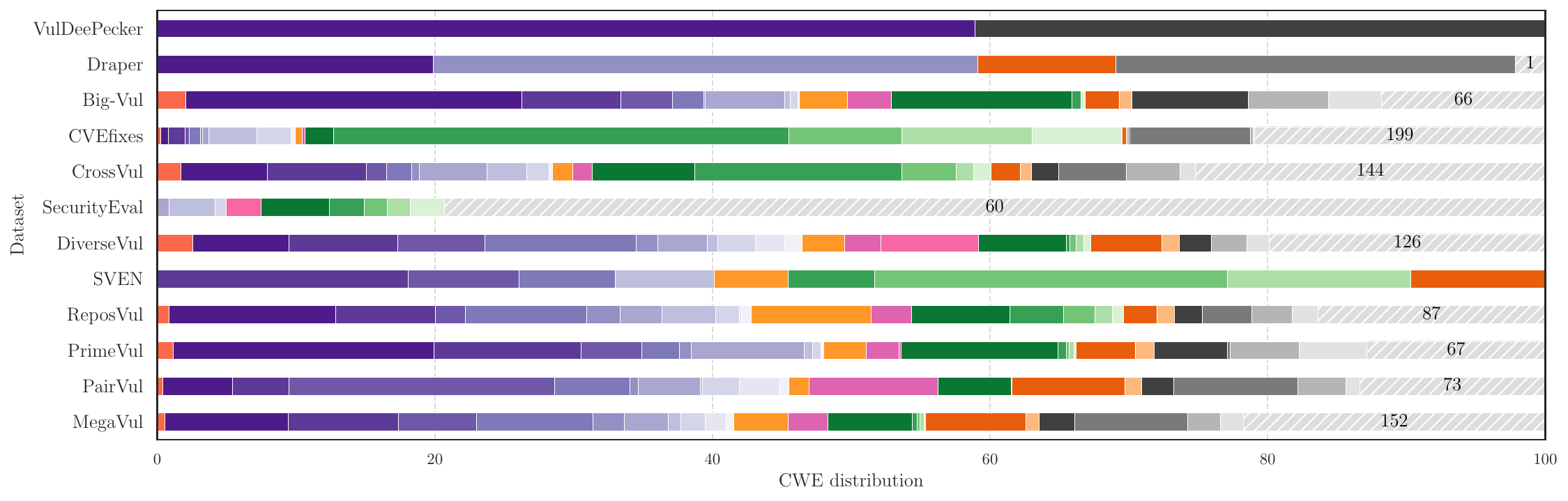}
        \hspace*{1cm} 
        \includegraphics[height=2.62cm]{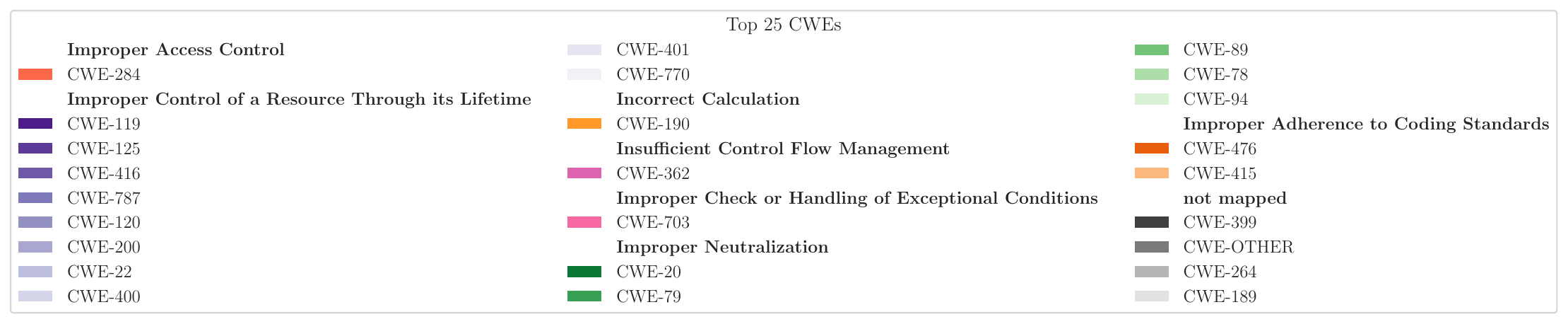} 
        \caption{Top 25 CWEs across selected datasets. Relative share of each CWE within the dataset's total labeled vulnerable samples. Hatched segments indicate the number of additional CWEs in the dataset that are not part of the Top 25 CWEs.}
        \Description{Relative CWE share of each Top 25 CWE within the selected dataset's total labeled vulnerable samples.}
        \label{fig:cwe_top25}
\end{figure*}

Figure~\ref{fig:cwe_top25} visualizes the distribution of the Top 25 CWE types covered by vulnerable samples across the selected datasets. 
Among the mixed datasets and those with a CWE-specific focus, VulDeePecker, Draper, and SVEN include only a small number of CWEs, i.e., two, four, and nine, respectively. 
SecurityEval features the widest variety of CWEs among mixed datasets, with a similar distribution across its covered types. The CWEs represented in the Top 25 across all datasets account for only around 20\% of its total CWE labels.
In comparison, the real-world datasets exhibit a more pronounced long-tail distribution: the Top 25 CWEs typically account for around 80\% of all vulnerability labels. This highlights a strong skew towards a small set of frequently occurring vulnerability types.

\begin{figure*}[tb]
        \includegraphics[width=\textwidth, trim=0 0 95 0, clip]{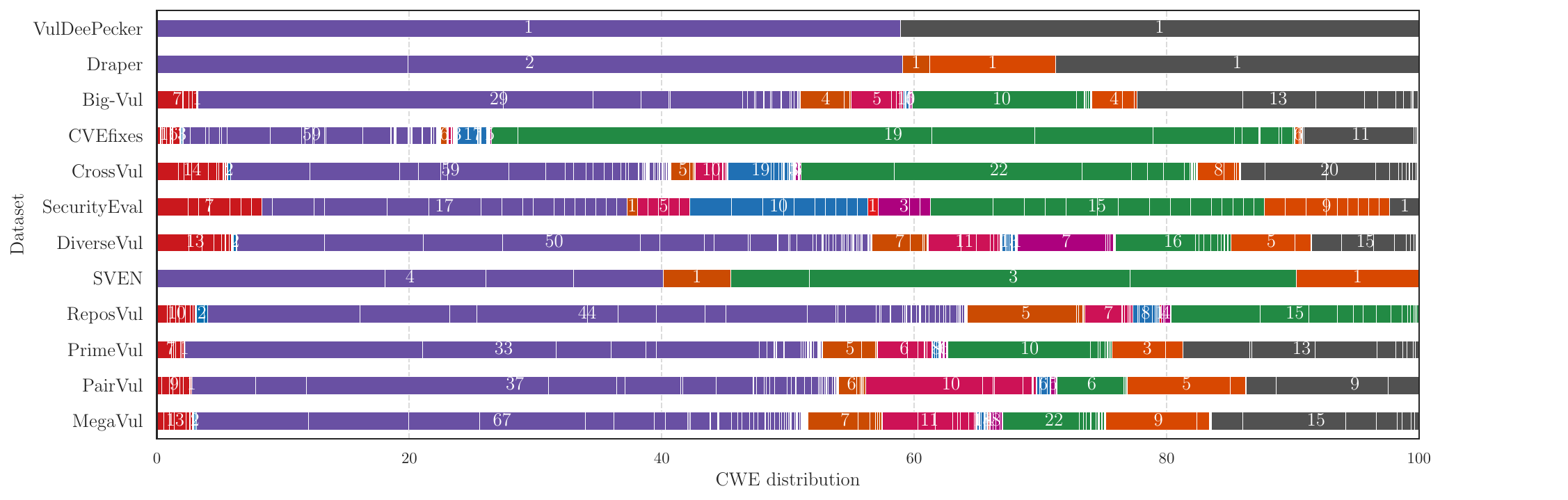}
        \hspace*{1.1cm} 
        \includegraphics[width=0.91\textwidth]{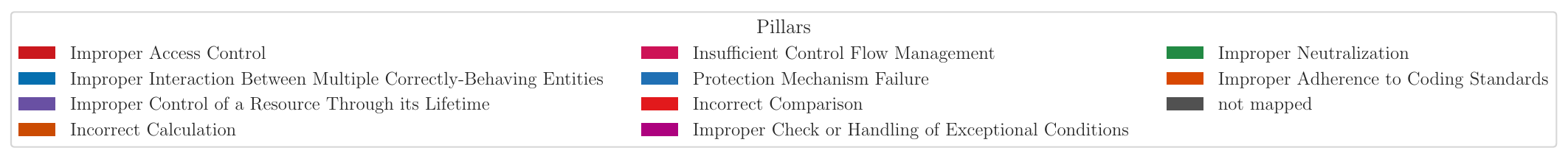}
        \caption{Number of all covered CWEs across selected datasets. CWEs grouped by pillar.}
        \Description{Relative CWE share of all CWEs grouped by pillar within the selected dataset's total labeled vulnerable samples.}
        \label{fig:cwe_all_app}
\end{figure*}

The most widely represented CWEs across datasets are those associated with the pillar \textit{Improper Control of a Resource Through its Lifetime}, which includes memory and resource management vulnerabilities that are commonly found in \ccpp, the most dominant language across the studied datasets (cf. discussion in Section~\ref{sec:dataset_prog_lan}). 
This pillar is also the largest in terms of CWE count (cf. Table~\ref{tab:pillars}), followed by \textit{Improper Neutralization}, encompassing issues related to input validation and sanitization.
Other pillars are underrepresented in comparison to their relative size in the CWE view. These pillars include \textit{Improper Adherence to Coding Standards} and \textit{Improper Access Control}. 
Notably, some pillars, such as \textit{Improper Interaction Between Multiple Correctly-Behaving Entities}, \textit{Protection Mechanism Failure}, and \textit{Incorrect Comparison}, are not represented at all among the Top 25 CWEs, and are less represented in the overall CWEs covered, cf. Figure~\ref{fig:cwe_all_app}. 
These categories are inherently more difficult to capture in real-world datasets, particularly at the function level, as they often involve complex inter-component dependencies, system-level interactions, or contextual analysis that is not easily observable from isolated code snippets.


A considerable share of CWEs in the analyzed datasets fall into the \textit{not mapped} category. 
This is often the result of outdated or abstract vulnerability labels sourced from the NVD. 
For instance, VulDeePecker covers two high-level CWEs, i.e., the class \textit{CWE-119 Improper Restriction of Operations within the Bounds of a Memory Buffer}~\cite{cwe119} and the category \textit{CWE-399 Resource Management Errors}~\cite{cwe399}, flattening the existing CWE hierarchy and obscuring more specific vulnerability types. 
Importantly, the use of abstract categories such as classes for mapping has been \textbf{discouraged} since 2019, as they are frequently misapplied in low-information vulnerability reports where more specific child CWEs would be more appropriate~\cite{cwe119}. The use of categories for vulnerability mapping is explicitly \textbf{prohibited}~\cite{cwe399}.
Several other CWEs commonly found in the selected datasets are also discouraged for use in vulnerability mappings, e.g., \textit{CWE-20 Improper Input Validation}, \textit{CWE-200 Exposure of Sensitive Information to an Unauthorized Actor}, and \textit{CWE-400 Uncontrolled Resource Consumption}. 
In addition, pillars such as \textit{CWE-284 Improper Access Control} and \textit{CWE-703 Improper Check or Handling of Exceptional Conditions} are often used as placeholders in vulnerability reports lacking detailed analysis, leading to imprecise dataset annotations.

Some CWEs are marked as allowed with review, meaning they can be used in vulnerability mapping but should be applied carefully and only when more specific alternatives are not available. 
For example, \textit{CWE-120 Buffer Copy without Checking Size of Input} is frequently selected based on keyword presence ('Classic Buffer Overflow'), but is only appropriate when the vulnerability involves unchecked buffer copy operations~\cite{cwe120}. 
Similar caution applies to CWEs such as \textit{CWE-362 Concurrent Execution using Shared Resource with Improper Synchronization} ('Race Condition') and \textit{CWE-94 Improper Control of Generation of Code} ('Code Injection').

The use of general or deprecated CWE labels presents challenges to dataset quality and limits the effectiveness of CWE-based classification and reasoning. 
Notably, there are 555 CWEs in the CWE-1000 view that are allowed for vulnerability mapping but are not represented as vulnerable samples in any of the analyzed datasets.
This indicates that current datasets cover only a subset of CWE types, leaving many mappable software weaknesses absent from training and benchmarking.
As a result, tools trained on these datasets may lack generalization capabilities when faced with previously unseen vulnerabilities.
Further, evaluation results appear artificially strong, as they are often biased towards commonly occurring CWEs while excluding rare and long-tail vulnerability types. 
To address these limitations, more effort is needed to incorporate the CWE hierarchy into training and evaluation protocols, moving beyond binary labels to consider hierarchically related CWEs. 
For example, Tamberg and Bahsi~\cite{tambergHarnessingLargeLanguage2024} follow an evaluation method in which both parent and child CWEs are considered valid classifications using MITRE’s Research Concepts view.


\subsection{Trends and Use Cases in Dataset Utilization}
\label{sec:datasets_trends_comparability}

Understanding how vulnerability datasets are used over time provides insight into community practices, widely used benchmark datasets, and emerging research directions. 
Analyzing temporal trends also reveals how quickly new datasets are adopted, how long benchmark datasets remain influential, and where gaps or biases may persist in evaluation setups. 
Such an analysis is particularly important in this rapidly evolving field to guide future dataset development and advance performance through more representative and robust evaluations.

\begin{figure*}[b!]
    \centering  
    \includegraphics[width=\linewidth]{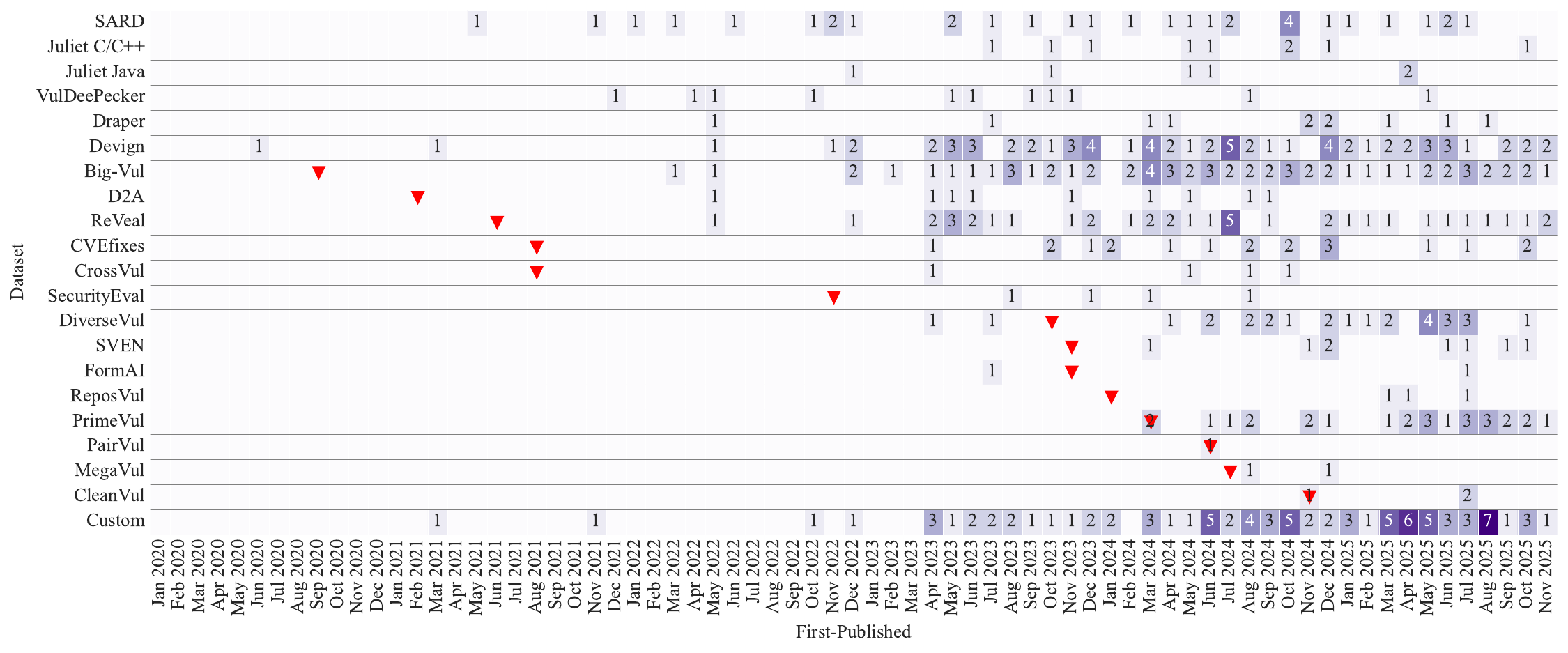}
    \caption{Use of popular datasets in studies from Jan 2020 to Mar 2025. Markers (red triangles) indicate the publication month of the paper that introduced the dataset.}
    \Description{Heatmap showing the use of popular datasets in studies from Jan 2020 to Mar 2025.}
    \label{fig:datasets_heatmap}
\end{figure*}

Figure~\ref{fig:datasets_heatmap} visualizes the temporal usage and usage frequency of the selected vulnerability datasets from Jan-2020 to Nov-2025 by the surveyed studies.
Red markers indicate the publication dates of the datasets. 
The Figure shows that even today, early deep learning datasets such as Devign, which serves as a foundational dataset for graph-based and learning-based vulnerability detection models, and  Big-Vul remain in widespread use. 
Classic test suites such as SARD and Juliet are also frequently used, cf. Figure~\ref{fig:datasets_heatmap}, useful for controlled evaluations due to their synthetic nature.
However, these earlier datasets often lack current CWE coverage, may reinforce outdated vulnerability mappings, and may already be part of training corpora.

\begin{figure}[!b]
    \centering  
    \includegraphics[width=0.6\linewidth]{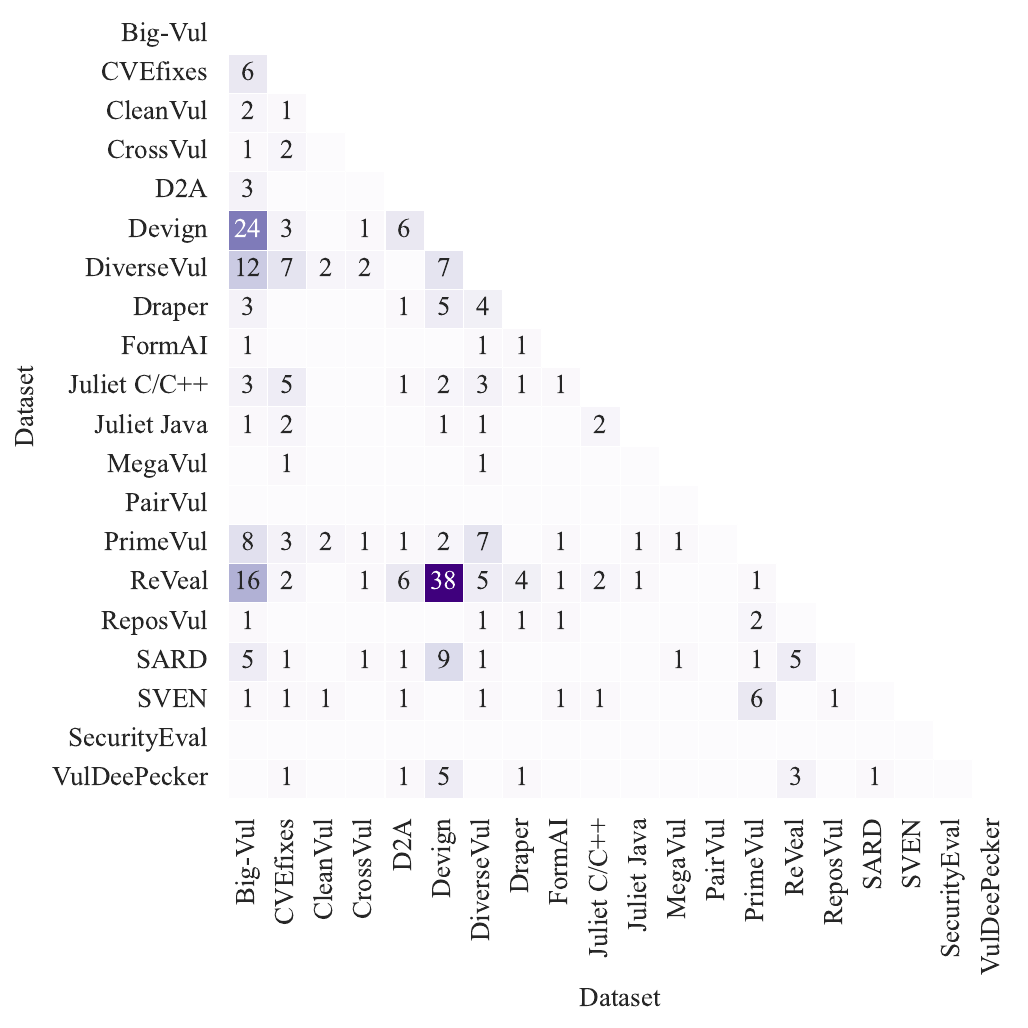}
    \caption{Common combination pairs of selected datasets.}
    \Description{Heatmap showing common combination pairs of selected datasets.}
    \label{fig:datasetxdataset_all}
\end{figure}

Figure~\ref{fig:datasets_heatmap} also shows that since 2021, there has been a clear trend towards real-world vulnerability datasets, e.g., CVEfixes, CrossVul, and DiverseVul. 
Several more recent datasets, such as FormAI, PrimeVul, PairVul, and CleanVul, are explicitly designed for evaluating LLM-based vulnerability detection approaches.
For these datasets, adoption lags of approximately one to six months can be observed between a dataset's release and adoption in published research. 
This lag reflects the time needed for several practical steps, including dataset validation, integration into existing toolchains and model pipelines, and broader community uptake. 
In many cases, adoption may also depend on the availability of documentation and preprocessing scripts.
Further, early adopters often play a key role in demonstrating a dataset's utility, which can influence its acceptance and reuse.

Another notable trend is the increasing use of newly created or custom datasets, see row "Custom" in Figure~\ref{fig:datasets_heatmap}. 
A total of \numOwnDatasets studies either use or introduce custom datasets, often to address perceived limitations in existing datasets. 
Common motivations include correcting unrealistic data distributions, expanding project and vulnerability diversity, including more recent disclosures, or improving overall dataset quality (e.g., fixing incomplete or incorrectly merged functions)~\cite{MegaVul}. 
This trend is also driven by domain-specific or task-aligned needs. 
While such efforts encourage innovation and enable tailored evaluations, they also complicate comparability and may introduce implicit biases tied to individual evaluation design choices and construction assumptions.

\begin{figure}[!b]
  \centering
  \begin{subfigure}[b]{0.48\textwidth}
    \includegraphics[width=\textwidth]{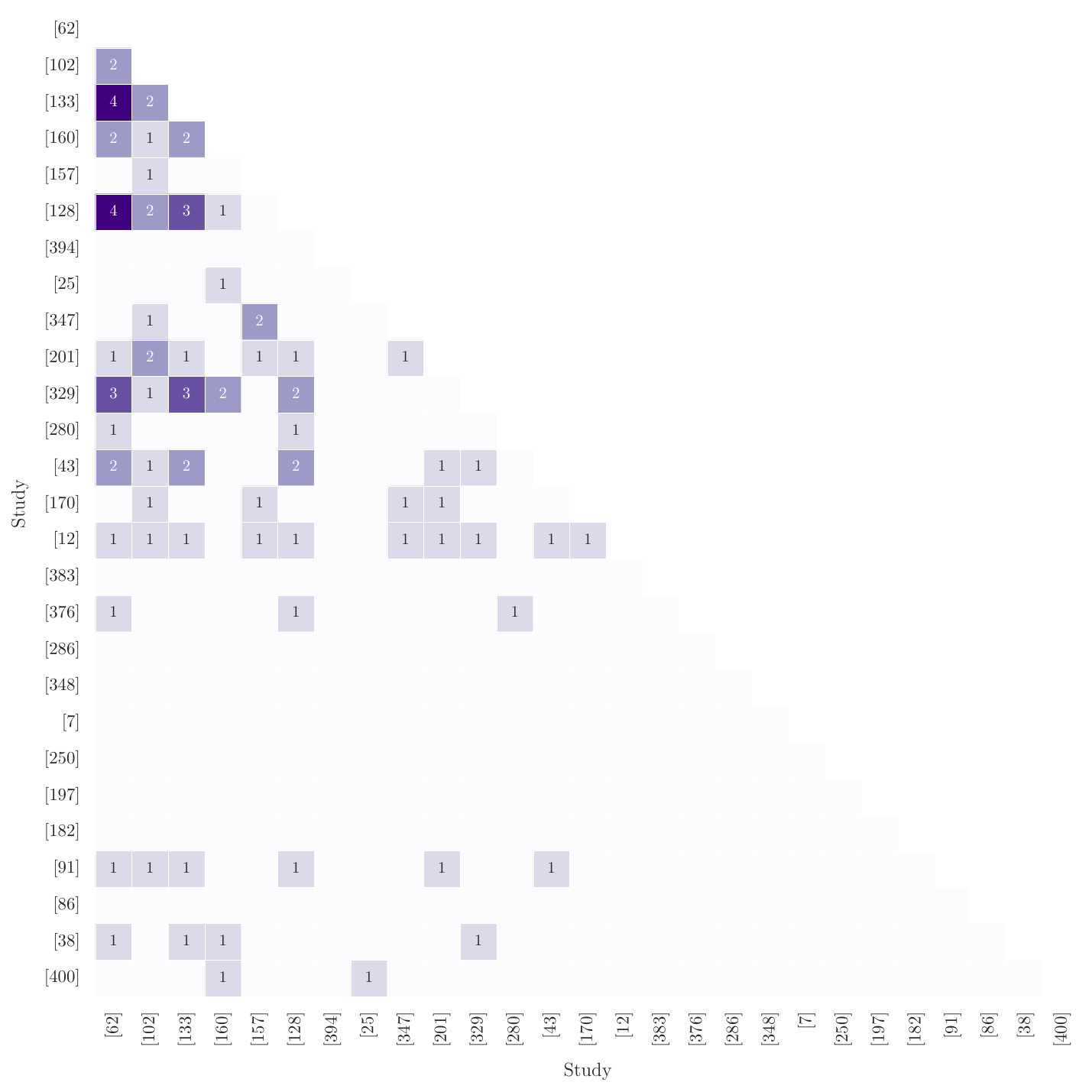}
    \caption{Low-Rank Decomposition}
    \label{fig:usecaseAdapter}
  \end{subfigure}
  \hfill
  \begin{subfigure}[b]{0.48\textwidth}
    \includegraphics[width=0.75\textwidth]{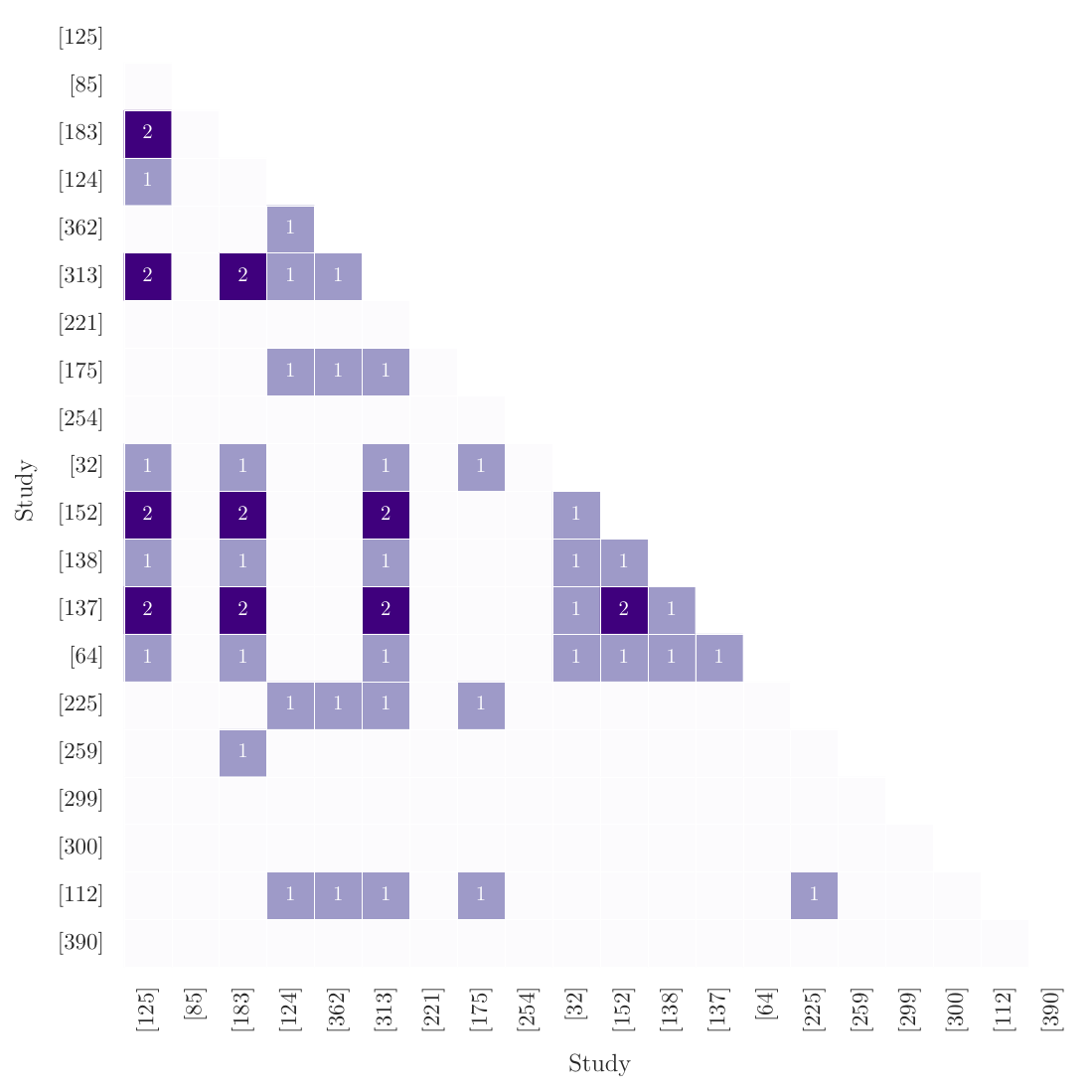}
    \caption{Hybrid - GNN}
    \label{fig:usecaseReason}
  \end{subfigure}

  \begin{subfigure}[b]{0.48\textwidth}
    \includegraphics[width=0.94\textwidth]{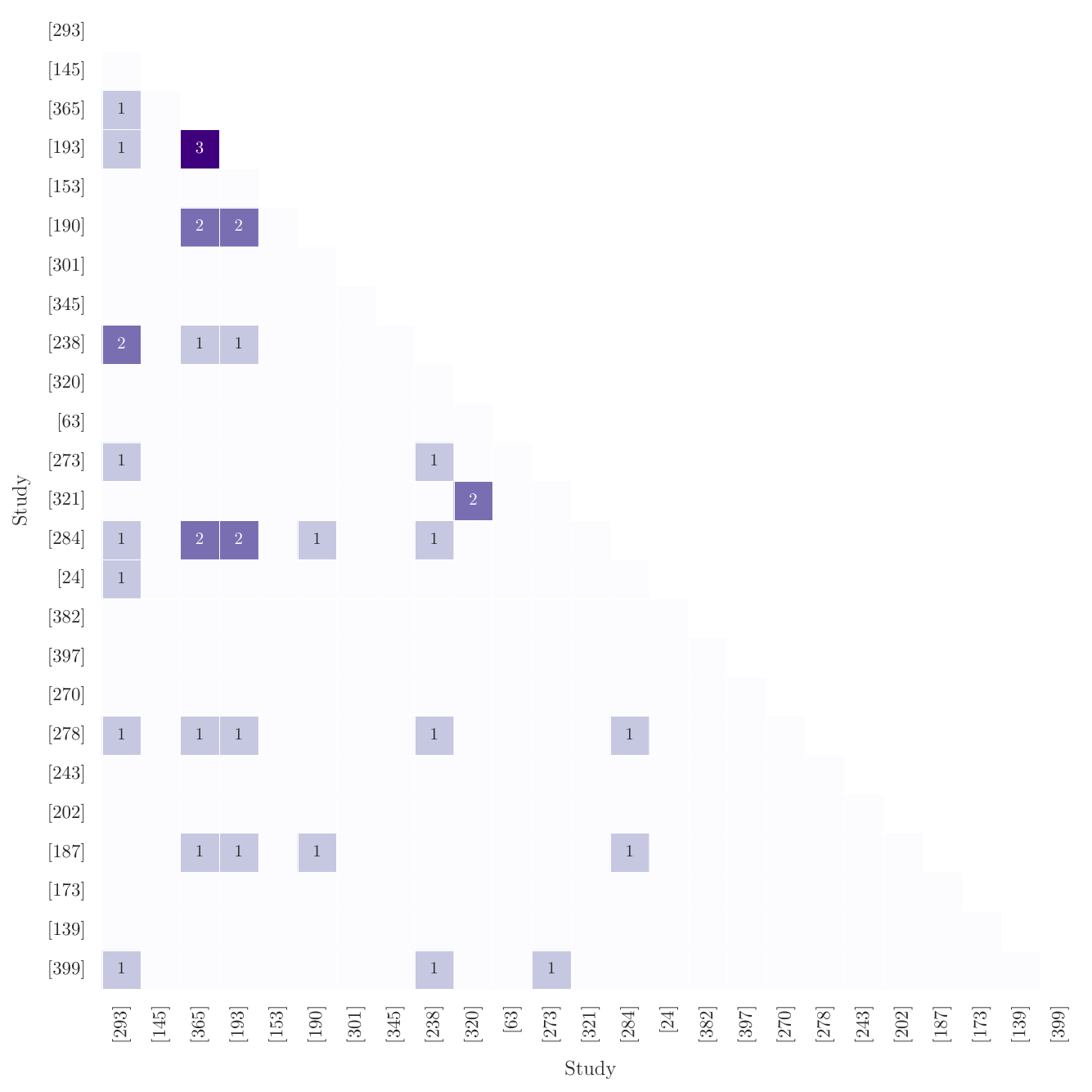}
    \caption{RAG}
    \label{fig:usecaseRAG}
  \end{subfigure}
  \hfill
  \begin{subfigure}[b]{0.48\textwidth}
    \includegraphics[width=0.45\textwidth]{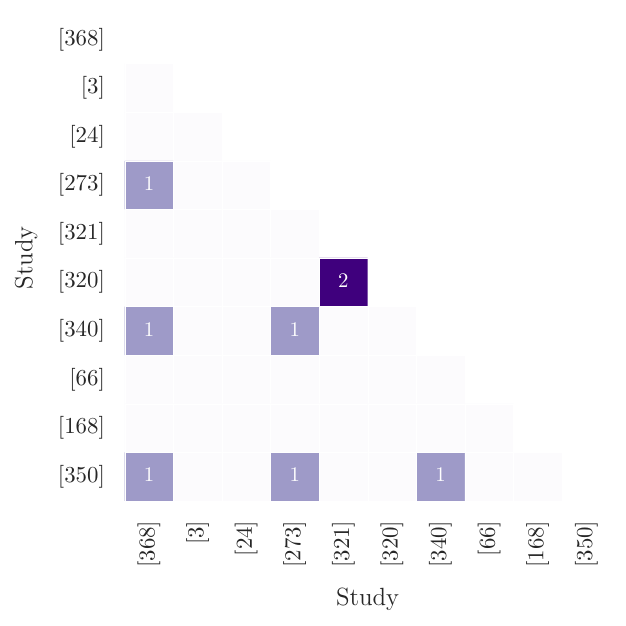}
    \caption{Agentic}
    \label{fig:usecaseAgentic}
  \end{subfigure}

  \caption{Dataset overlap heatmap: Use of \# same datasets between studies across four example use cases (wrt. adaptation techniques, orchestration techniques, and system design).}
  \Description{Matrix showing dataset overlap (same datasets used in studies) between studies across four example use cases.}
  \label{fig:dataset_overlap}
\end{figure}

The choice of suitable vulnerability datasets reflects a combination of common benchmark datasets, task alignment, and system design. 
However, the use of various datasets across the surveyed studies illustrates a high degree of fragmentation. 
Most studies either use or introduce custom datasets.
Only nine datasets are reused in more than ten studies, highlighting the lack of a standardized suite of benchmark datasets and consistency in dataset selection.
Many studies use more than one dataset, often combining datasets to increase training data diversity or to evaluate model generalization, e.g., vulnerability types and programming languages.
Figure~\ref{fig:datasetxdataset_all} highlights the most frequently used dataset pairs across all surveyed studies. 
Among these, the most common combinations are Devign and ReVeal, Devign and Big-Vul, as well as Big-Vul and ReVeal. 
These pairings often complement each other in terms of data characteristics, such as vulnerable percentage or combining datasets with broad versus narrow CWE coverage.

The use of many different datasets and combinations, however, hampers direct comparability of evaluation results, particularly when studies apply different splits for training, validation, and testing, or pre-process the same datasets differently. Figure~\ref{fig:dataset_overlap} offers insight into overlaps and divergences in dataset usage across specific use cases.
When fine-tuning LLMs for vulnerability detection, e.g., using LoRA (Low-Rank Decomposition, 27 studies), researchers often resort to large and diverse datasets, most commonly DiverseVul, Big-Vul, PrimeVul, and Devign.
However, many studies make use of additional, lesser-used datasets or introduce custom datasets (ten studies).
This practice reduces comparability and makes it difficult to replicate results or isolate model performance.
RAG studies (25 studies) most commonly use Big-Vul and Devign, appearing in six and five studies, respectively.
Big-Vul, in particular, is often used for knowledge base construction of CWE entries due to the provided metadata, e.g., CWE descriptions.
Eight RAG studies introduce custom datasets, which hinder comparison, especially concerning the semantics of the retrieved context.
Agentic approaches (10 studies) show even greater fragmentation, with only four studies aligning on PrimeVul.
GNN-based hybrid approaches (20 studies) most frequently rely on Devign, followed by custom datasets, Big-Vul, and ReVeal. Devign is particularly suited for GNNs due to its composite code representation, encoding program control and data dependencies as heterogeneous edges within a unified graph structure.  
The absence of shared benchmarks within the discussed use cases highlights the need for controlled and standardized evaluation protocols.


\section{Current Limitations and the Road Ahead}
\label{sec:discussion}
Despite progress in leveraging LLMs for software vulnerability detection, we identified several remaining limitations in this SLR.
In the following, we outline the key limitations derived from the surveyed studies and propose actionable research directions to address them. 
Table~\ref{tab:limitations} summarizes the discussed limitations and future research opportunities.

\subsection{Limitations}
The limitations identified in this review emerge from recurring issues observed across the surveyed studies and span key aspects such as dataset realism and representativeness, vulnerability diversity, methodological rigor, and evaluation practices. 
We structure these findings into seven themes \textbf{L1-L7} that outline the most pressing limitations.

\begin{table}[tb]
\centering
\caption{Limitations in surveyed LLM-based vulnerability detection studies and future research directions.}
\label{tab:limitations}
\fontsize{8pt}{8pt}\selectfont
\setlength{\tabcolsep}{5pt}  
\renewcommand{\arraystretch}{1.3} 
\resizebox{\textwidth}{!}{%
    \begin{tabular}{@{}p{7.5cm} p{7.5cm}@{}}
    \toprule
     \textbf{Limitations} & \textbf{Future Research Opportunities} \\ \midrule

    \rowcolor{gray!08} \multicolumn{2}{c}{\hspace{-0.5em}\textbf{L1 Dataset and Detection Granularity}} \\
        \textbullet\ Focus on function-level inputs & \textbullet\ Explore coarser detection granularity (project- and repo-level) \\
        \textbullet\ Failure to capture complex vulnerabilities with dependencies & \textbullet\ Augment code context \\
        \textbullet\ Focus on C/C++ code & \textbullet\ Expand to other languages \\
         & \textbullet\ Develop multilingual and cross-project benchmarks \\\midrule

    \rowcolor{gray!08} \multicolumn{2}{c}{\hspace{-0.5em}\textbf{L2 Dataset Labeling and Quality}} \\
        \textbullet\ Synthetic datasets oversimplify real-world complexity & \textbullet\ Build new datasets focusing on vulnerability diversity \\
        \textbullet\ Automated labeling introduces noise and redundancy & \textbullet\  Improve label quality (manual verification, improved VFC tools) \\
        \textbullet\ Missing reasoning ground truth & \textbullet\ Reasoning-oriented datasets and benchmarks \\
        \textbullet\ Class imbalance and long-tail distribution of CWE types & \textbullet\ Reevaluate approaches on diverse and representative datasets \\
        \textbullet\ Data leakage & \textbullet\ Build new datasets considering knowledge cutoff \\ \midrule

    \rowcolor{gray!08} \multicolumn{2}{c}{\hspace{-0.5em}\textbf{L3 Study Evaluation and Comparability}} \\
        \textbullet\ Inconsistent use of datasets & \textbullet\ Align dataset use across similar use cases for comparability\\
        \textbullet\ Lack of predefined train/validate/test splits & \textbullet\ Use datasets with documented splits \\
        \textbullet\ Diverse modifications to datasets & \textbullet\ Evaluate on multiple datasets for comparability \\
         & \textbullet\ Document all preprocessing and filtering steps \\
         & \textbullet\ Encourage open science \\ \midrule

    \rowcolor{gray!08} \multicolumn{2}{c}{\hspace{-0.5em}\textbf{L4 Up-to-Date Vulnerability Knowledge}} \\
        \textbullet\ Static datasets quickly become outdated & \textbullet\ Use automatically updated datasets (e.g., CVEfixes) \\
        \textbullet\ Retraining is expensive and often infeasible & \textbullet\ Further explore RAG and continual learning \\ \midrule

    \rowcolor{gray!08} \multicolumn{2}{c}{\hspace{-0.5em}\textbf{L5 Code and Vulnerability Representation}} \\
        \textbullet\ Raw code often used without structure or abstraction  & \textbullet\ Investigate structure-aware representations \\
        \textbullet\ Reliance on superficial text patterns & \textbullet\ Combine LLMs with structural models (e.g., GNNs) \\
        \textbullet\ Sensitivity to semantic-preserving transformations & \textbullet\ Reevaluate robustness on perturbed or transformed code \\ \midrule

    \rowcolor{gray!08} \multicolumn{2}{c}{\hspace{-0.5em}\textbf{L6 Model Interpretability and Explainability}} \\
        \textbullet\ Limited insight into predictions & \textbullet\ Develop metrics for explanation trustworthiness  \\
        \textbullet\ Hallucinated or misleading justifications, self-contradictions & \textbullet\ Use structured explanation formats \\
        & \textbullet\ Leverage external knowledge to ground reasoning \\ \midrule

    \rowcolor{gray!08} \multicolumn{2}{c}{\hspace{-0.5em}\textbf{L7 Integration into Pipelines and Workflows}} \\
        \textbullet\ Few evaluations in realistic development settings & \textbullet\ Integrate tools into IDEs or pipelines with developer feedback \\
        \textbullet\ Poor generalization to production codebases & \textbullet\ Evaluate models on proprietary or closed-source datasets \\
        \textbullet\ High computational cost of tuning and inference & \textbullet\ Develop CWE-type classifiers for finer-grained prediction \\
         & \textbullet\ Explore agentic systems for reasoning or self-assessment \\
         & \textbullet\ Explore model compression for lightweight deployment \\ \bottomrule
    \end{tabular}
}%
\end{table}

\subsubsection{\textbf{L1 Dataset and Detection Granularity}}
In the surveyed studies, vulnerability detection is typically applied at the level of individual functions or small program slices.
This focus on function-level granularity has been influenced by the dominance of function-level vulnerability datasets (cf. Table~\ref{tab:datasets1}) and partly by architectural limitations, particularly the restricted input lengths of earlier LLMs~\cite{steenhoekComprehensiveStudyCapabilities2024a}.
However, methods have been proposed to enhance context, such as the work by Chen et al.~\cite{chenBridgeHintExtending2024}, and LLMs now support significantly longer input contexts, e.g., 128,000 tokens for GPT4~\cite{gpt4}.
Despite these capabilities, the continued focus on function-level datasets fails to capture the context necessary to generalize to complex vulnerabilities that span multiple functions or files.
For example, Huynh et al.~\cite{huynhDetectingCodeVulnerabilities2025} investigate how varying levels of code context influence detection performance, comparing snippets enriched with comments and docstrings against full-file inputs. The findings suggest that even with extended context windows, current models still perform poorly on distributed vulnerabilities. 

Further, the surveyed studies focus predominantly on detecting vulnerabilities in \ccpp code (cf. Table~\ref{tab:datasets2}), with limited coverage of other widely used programming languages, such as Java or Python. 
This narrow programming language coverage limits the applicability of current models in diverse, multilingual software environments.

\subsubsection{\textbf{L2 Dataset Labeling and Quality}}
Synthetic datasets offer high label-precision, as vulnerabilities are injected and, thus, well-defined. 
However, synthetic datasets may oversimplify real-world scenarios and fail to capture the complexity and diversity of real-world vulnerabilities, limiting their effectiveness in preparing models for deployment in realistic environments.
Similarly, the common formulation as a binary classification task (cf. Figure~\ref{fig:sankey_task}) lacks the granularity needed for practical application, where identifying the specific vulnerability type is essential for severity assessment, prioritization, and appropriate repair.
Real-world datasets are typically labeled using automated methods such as VFCs or static analysis tools.
While these approaches enable large-scale dataset creation, they are associated with label noise and redundancy, which reduce the uniqueness and representativeness of the dataset~\cite{croft2023data}.
Although manual labeling or verification can improve precision, it remains a resource-intensive process, especially when scaling to larger datasets.

Moreover, as research shifts toward generative tasks, a critical gap of missing reasoning ground truth emerges.
While existing datasets typically provide the label and possibly a description of the vulnerability, they rarely provide a verified chain of thought reasoning the root cause.
This limitation makes it difficult to quantitatively benchmark the quality of LLM-generated reasoning without expensive human evaluation.

Further, most datasets exhibit a strong class imbalance, where vulnerable samples are heavily outnumbered by non-vulnerable ones (cf. Table~\ref{tab:datasets2}), and a long-tail distribution of vulnerability types, with few CWEs dominating and many others being severely underrepresented (cf. Figure~\ref{fig:long_tail} and Figure~\ref{fig:cwe_top25}).
These imbalances pose significant challenges for both training balance and evaluation reliability, causing models to overfit to frequent vulnerabilities while failing on rare ones~\cite{huynhDetectingCodeVulnerabilities2025}.

Beyond these traditional labeling and quality issues, LLMs introduce the unique challenge of data leakage.
Unlike models trained on closed datasets, LLMs are pre-trained on vast open-source corpora. Consequently, popular benchmarks (e.g., Devign, Big-Vul) are likely contained within the pre-training corpus, rendering standard train/test splits unreliable.
This challenge is compounded by the knowledge cut-off; without continuously updated datasets that post-date the model's training window, it is difficult to evaluate whether an LLM is detecting a vulnerability via reasoning or simply recalling historical data it has already seen.

\subsubsection{\textbf{L3 Study Evaluation and Comparability}}
The lack of consistency and transparency in dataset use and evaluation protocols complicates meaningful comparison across studies.
Even within the same use case, studies are often not comparable due to the use of different datasets (cf. Figure~\ref{fig:dataset_overlap}).
Further, most datasets do not provide predefined train/validate/test splits, which are essential for ensuring a consistent and fair evaluation~\cite{shereen2024sok}. 
Only a few studied datasets offer standardized splits, e.g., Draper~\cite{Draper}, CodeXGLUE~\cite{luCodeXGLUEMachineLearning} (a subset of Devign~\cite{Devign}), Big-Vul~\cite{BigVul}, and PrimeVul~\cite{PrimeVul}. 
Without fixed splits, results can vary significantly, even when using the same dataset. 
For example, if subsequent studies fine-tune on the same dataset but apply different splits, direct performance comparisons become unreliable.
In addition, several studies further modify or refine existing datasets ~\cite{zhangMVDMultiLingualSoftware2024, yinMultitaskBasedEvaluationOpenSource2024,hortSemanticPreservingTransformationsMutation2025,goncalvesEvaluatingLLaMA322025,wangLinelevelSemanticStructure2024}, or augment them with custom data \cite{qiEnhancingPreTrainedLanguage2024}. 
While such adaptations are often motivated by practical needs, e.g., improving data quality or evaluating generalization, they introduce additional variability and complicate comparability with prior and future works.

\subsubsection{\textbf{L4 Up-to-Date Vulnerability Knowledge}}
The number of disclosed software vulnerabilities continues to grow rapidly~\cite{CVEMetrics}.
As a result, static datasets may quickly become outdated.
Models trained on such outdated datasets may lack awareness of newly discovered vulnerabilities and emerging vulnerability patterns, limiting their practical relevance and effectiveness in real-time scenarios and for the detection of zero-day vulnerabilities~\cite{nongChainofThoughtPromptingLarge2024}. 
Another challenge lies in how to efficiently expose models to this growing volume of vulnerability information. 
Since retraining models from scratch is computationally expensive and impractical at high frequency, there is a critical need for scalable methods to update or augment model knowledge.

\subsubsection{\textbf{L5 Code and Vulnerability Representation}}
In most surveyed studies, the code to be analyzed is provided as raw code or embedded directly within the prompt during inference.
However, models frequently rely on superficial textual patterns rather than capturing the underlying semantic structures of vulnerabilities~\cite{dingVulnerabilityDetectionCode2024, zhaoHowGetBetter2023,liSVTrustEvalCEvaluatingStructure2025a}.
As a result, model performance degrades under semantic-preserving code transformations, such as variable renaming, code reordering, or formatting changes, highlighting a lack of robustness and true semantic understanding~\cite{niLearningbasedModelsVulnerability2024, risseUncoveringLimitsMachine2024, qiEnhancingPreTrainedLanguage2024, hortSemanticPreservingTransformationsMutation2025, panichellaMetamorphicBasedManyObjectiveDistillation2025,wangSecureMindFrameworkBenchmarking2025}.
While contrastive learning is designed to produce more discriminative embeddings, models still often fail to draw the classification boundary correctly and identify the vulnerable patterns~\cite{dingVulnerabilityDetectionCode2024}.

\subsubsection{\textbf{L6 Model Interpretability and Explainability}}
Current LLM-based vulnerability detection approaches offer limited insight into why a model classifies the input code as vulnerable or non-vulnerable.
Even in studies that use CoT prompting or explicitly investigate reasoning, issues with hallucinated or misleading justifications are commonly reported~\cite{espinhagasibaMaySourceBe2024,noeverCanLargeLanguage2023}.
Similarly, studies implementing verification report instances of inaccurate corrections, self-contradictions, and hallucinated justifications~\cite{yuFightFireFire2024}.
These issues undermine the trustworthiness of model outputs and hinder their integration in practical development workflows where interpretability and reliability are critical.

\subsubsection{\textbf{L7 Integration into Pipelines and Workflows}}
Despite growing interest, the integration of LLM-based vulnerability detection into practical development workflows remains underexplored. 
Only a few studies investigate real-world applicability through prototype tools or user studies (e.g.,~\cite{fuAIBugHunterPracticalTool2024,steenhoekClosingGapUser2024}), though such efforts are essential for assessing feasibility and trust in practical settings.
Generalization to realistic codebases also remains challenging, with models often underperforming outside curated benchmarks~\cite{chakrabortyRevisitingPerformanceDeep2024}. 

Resource efficiency is a further limiting factor for real-world deployment, as fine-tuning and inference with LLMs remain computationally expensive.
While model compression techniques such as knowledge distillation reduce inference latency and memory usage~\cite{daloisioCompressionLanguageModels2024,shiGreeningLargeLanguage2024,ibanez-lissenLPASSLinearProbes2025}, their practical application for deployment on developer machines remains unexplored. 
This gap is particularly relevant when considering the privacy constraints of proprietary codebases, which may require on-premise or offline analysis~\cite{bappyCaseStudyFinetuning2025}.


\subsection{Actionable Insights and Future Research Opportunities}
Building on the identified limitations, each theme reflects core challenges in current research and highlights open questions that future work must address to enable more robust, generalizable, and reproducible progress in LLM-based software vulnerability detection. 
In this section, we provide interpretations or promising research directions, offering actionable insights to guide future work towards addressing these challenges.

\subsubsection{\textbf{L1 Dataset and Detection Granularity}}
To move beyond the limitations of function-level detection, future work should focus on context-enhanced vulnerability detection, including project- and repository-level analysis~\cite{zhouComparisonStaticApplication2024,wenVulEvalRepositoryLevelEvaluation2024,liLLMAssistedStaticAnalysis2024,guoRepoAuditAutonomousLLMAgent2025}.
Initial steps in this direction include augmenting function-level samples with additional context, such as surrounding code lines \cite{halderFuncVulEffectiveFunction2025}, function arguments, external functions, type definitions, global variables, environmental constraints~\cite{risse2024top}, as well as dependency and execution flows~\cite{zhengLearningFocusContext2025}.
For example, Ahmed et al.~\cite{ahmedSecVulEvalBenchmarkingLLMs2025} use GPT-4 to identify the required contexts for a given vulnerability in a function, but still find limitations in pinpointing vulnerable statements and their root causes in complex real-world code.
Li et al.~\cite{liEverythingYouWanted2025} demonstrate that incorporating execution and data context improves model performance and reasoning quality, emphasizing that the challenge lies in precise, context-aware vulnerability reasoning. 
Similarly, Yang et al.~\cite{yangContextEnhancedVulnerabilityDetection2025} propose a program analysis-based approach that abstracts complex function calls into primitive API representations to enrich contextual understanding.
Further investigating such context-enhancements is essential for capturing dependencies that span multiple functions or files, enabling the detection of more complex vulnerability types that are otherwise missed in isolated, function-level analysis.

The current language focus on \ccpp\ should be expanded to include other widely used languages such as Python, Java, and Rust~\cite{PYPLPopularitYProgramming}.
Establishing multilingual and cross-project benchmarks will be essential for assessing robustness and model generalization.

\subsubsection{\textbf{L2 Dataset Labeling and Quality}}
To address current limitations in dataset construction, future work should focus on building real-world vulnerability datasets with improved label quality and balanced coverage across CWE types.
Label quality can be improved through two main strategies: (1) refining existing datasets via targeted manual labeling to correct and verify automatically generated labels~\cite{SVEN}, and (2) advancing automated labeling methods, such as VFCs, to reduce noise and improve labeling reliability~\cite{PrimeVul,liCleanVulAutomaticFunctionLevel2024,ahmedSecVulEvalBenchmarkingLLMs2025,gaoMonoYourClean2025,luICVulWelllabeledVulnerability2025a}.
Balanced coverage across frequent and less common vulnerability types can be achieved, e.g., by using LLMs to generate vulnerable samples for specific vulnerability types~\cite{yongGVIGuidedVulnerability2025,bappyCaseStudyFinetuning2025,shahzadTheoryPracticeCode2025}.

Complementary to improved context, there is a critical need for further reasoning-oriented benchmarks~\cite{sunLLM4VulnUnifiedEvaluation2024} to close the gap in ground truth.
Future dataset creation should prioritize pairing vulnerable code with verified, step-by-step explanations of the root cause and exploitation logic.
Establishing such reasoning ground truths is a prerequisite for advancing from simple pattern matching to genuine semantic understanding.

In addition, new datasets should be constructed with minimal overlap to existing corpora to prevent data leakage and inflated performance due to prior exposure during model training. 
These datasets should set a focus on programming language and vulnerability diversity, capturing variations in human-written and LLM-generated code, complex multi-line vulnerabilities, and less frequently represented CWE types~\cite{steenhoekEmpiricalStudyDeep2023}.
A practical direction for dataset expansion is to use the set of mappable-but-unused CWEs identified in Section~\ref{sec:datasets_cwe_analysis}, i.e., CWE classes that are currently not represented in the discussed datasets.
Revisiting existing vulnerability detection approaches on more representative datasets may yield new insights into their robustness and effectiveness in real-world vulnerability detection.

\subsubsection{\textbf{L3 Study Evaluation and Comparability}}
To enhance reproducibility and comparability across studies, future research should prioritize standardized evaluation protocols. 
Specifically, we recommend the use of datasets with documented train/validation/test splits (or even closed evaluation of test splits) and careful alignment with datasets used in related studies to ensure meaningful comparisons within the same use case. We provide the mappings of datasets to surveyed use cases in the artifacts~\cite{replicationpackage}.

To assess generalization and maximize comparability with prior works, models should be evaluated on multiple datasets, combining recent datasets such as PrimeVul~\cite{PrimeVul} or MegaVul~\cite{MegaVul} with established benchmark datasets such as Devign~\cite{Devign} or Big-Vul~\cite{BigVul}.
All pre-processing and filtering steps must be documented to enable replication.
In support of open science~\cite{nong2022open}, sharing of code, evaluation scripts, and model checkpoints is encouraged.

\subsubsection{\textbf{L4 Up-to-Date Vulnerability Knowledge}}
Given the broad and continuously growing number of CVEs, ensuring that models remain informed about newly discovered vulnerabilities is crucial for maintaining relevance in practical scenarios.
To address the limitation of static datasets quickly becoming outdated, future work should focus on continuously maintained and automatically extensible datasets, such as CVEfixes~\cite{CVEfixes}, which employs an automated pipeline for updates.
Advanced adaptation techniques also offer promising pathways to bridge this knowledge gap. 
RAG enables LLMs to access external, up-to-date knowledge at inference time.
Continual learning can further support adaptability by integrating new vulnerability knowledge while preserving performance on previously seen vulnerability types.
However, both techniques have seen limited adoption in the context of software vulnerability detection. 
Further research should explore their effectiveness, especially in handling large and heterogeneous vulnerability information. 
For RAG, this includes investigating how different representations, e.g., structured CWE hierarchy graphs or abstract function descriptions, affect retrieval quality and downstream detection accuracy.

\subsubsection{\textbf{L5 Code and Vulnerability Representation}}
To address the need for robust representations that capture vulnerability semantics, future work should further focus on structure-aware input representations.
Directions include graph-based representations, such as control and data flow graphs, (LLM-generated) abstract descriptions that capture semantics and code behavior~\cite{zhangVulTrLMLLMassistedVulnerability2026,wuEnhancingVulnerabilityDetection2025,duVulRAGEnhancingLLMbased2024}, and multi-modal inputs, e.g., combining API abstractions, data flow graphs, and natural language documentation~\cite{yangContextEnhancedVulnerabilityDetection2025}.
Understanding how different representations complement each other can enhance vulnerability detection capabilities.
Ideally, such representations should be language-agnostic to support generalization across diverse programming languages (cf. multi-language studies~\cite{dozonoLargeLanguageModels2024,zhangBenchmarkingLargeLanguage2025}).
In addition, hybrid architectures should be further explored.
GNNs, in particular, have demonstrated potential for capturing code semantics and structural dependencies~\cite{changFineTuningPretrainedModel2024,islamUnbiasedTransformerSource2023,islamUnintentionalSecurityFlaws2024,jiangDFEPTDataFlow2024,jianjieCodeDefectDetection2023,kongSourceCodeVulnerability2024,liuVulLMGNNsFusingLanguage2025,niFunctionLevelVulnerabilityDetection2023,yangSecurityVulnerabilityDetection2024,zhengSVulDetectorVulnerabilityDetection2024}.
Future work should also evaluate the robustness of different representation strategies against syntactic perturbations and domain shifts, such as LLM-generated or refactored code, to better assess real-world generalization capabilities.

\subsubsection{\textbf{L6 Model Interpretability and Explainability}}
To build trust in LLM-based vulnerability detection systems, future research should go beyond standard evaluation metrics (e.g., accuracy, recall, precision, F1) and incorporate dedicated metrics that assess the trustworthiness and consistency of model-generated explanations. 
As LLMs increasingly take on roles in security workflows, their predictions must not only be correct but also justifiable and transparent to developers. 
Justification includes distinguishing between correct predictions made for the wrong reasons (spurious correlations) and those grounded in valid vulnerability semantics~\cite{risse2024top}.
In this matter, studies should increasingly look into \textit{Explainable AI (XAI) }techniques~\cite{takieldeenAIPoweredVulnerabilityDetection2025,nguyenHumanUnderstandableExplanationSoftware2024,alam2025improving}.

Interpretability may further be enhanced through structured explanation formats, such as vulnerability propagation paths or flow annotations.
These structured rationales offer greater reliability than natural language justifications and can be cross-checked against program logic or expert knowledge.
Further, integrating RAG techniques with structured vulnerability knowledge, such as CWE hierarchies or exploit chains, could improve reasoning by anchoring the model's output in verifiable context. 
This approach may not only strengthen the grounding of explanations but also align with the information needs of developers and auditors in practice.

\subsubsection{\textbf{L7 Integration into Pipelines and Workflows}}
To ensure the practical applicability of LLM-based vulnerability detection, future work should focus on the integration into real-world development workflows.
Such works should, e.g., integrate solutions into IDEs and CI/CD pipelines, conduct field studies, and collect developer feedback. 
Closed-source or proprietary codebases can serve as valuable testbeds for evaluating real-world performance, robustness, and usability.

Agentic systems and reasoning-driven workflows offer promising pathways to autonomously assist developers (e.g., expert-in-the-loop system~\cite{farrExpertintheLoopSystemsCrossDomain2025}) through tasks such as self-assessment, explanation generation, or iterative refinement.

To further improve generalization to practical vulnerability remediation workflows, LLM-based approaches should move towards CWE-type classification and report generation. 
Promising directions include the development of type-specific classifiers that align with the CWE hierarchy and research views, cf.~\cite{atiiqGeneralistSpecialistExploring2024,caoRealVulCanWe2024,dengImprovingLongTailVulnerability2024}, enabling more actionable insights and effective prioritization.

A current pathway for industrial adoption lies in synergizing LLMs with traditional program analysis~\cite{liEnhancingStaticAnalysis2024}.
While LLMs excel at semantic reasoning, they lack the formal correctness guarantees of, e.g., formal verification tools.
Pipelines where SAST tools detect candidates and LLMs filter false positives can significantly reduce alert fatigue by applying semantic understanding to noisy static reports.
Conversely, workflows where LLMs propose vulnerabilities and static analyzers verify feasibility offer a mechanism to ground probabilistic outputs.

Finally, trade-offs between performance and resource efficiency must be considered for real-world deployment. 
Fine-tuning smaller models for on-premise use is both practical and privacy-preserving~\cite{bappyCaseStudyFinetuning2025}. 
Alternatively, model compression techniques such as knowledge distillation and quantization offer promising pathways for enabling lightweight deployment on developer machines.


\section{Threats to Validity}
\label{sec:limitations}
To ensure transparency of our findings, we acknowledge potential threats to the validity of this SLR along with the mitigation strategies employed.

\textbf{Paper retrieval omissions.} 
A primary threat to any SLR is the omission of relevant studies due to studies with incomplete titles, abstracts, or varied keyword terminologies.
To address this threat, we constructed a broad search string based on manually selected keywords from primary studies as well as keywords from prior SLRs.
We combined manual search, automated search, and snowballing to minimize overlooking relevant papers.
Despite these efforts, due to the living nature of the field, a static snapshot may miss very recent developments. 
To address this threat, we employ a continuous search strategy using alerts and maintain a repository to update the study corpus.

\textbf{Study selection bias.} 
The selection of studies involves subjective judgment, which introduces potential bias. 
To minimize this threat, we conducted a revision of all crawled papers, strictly applying predefined inclusion and exclusion criteria (cf. Table~\ref{tab:inclusion_exclusion}).
A specific threat regarding quality arises from our decision not to filter exclusively for high-ranked conferences or journals. 
We included a wide range of studies to provide a holistic picture of the landscape and identify gaps that might be missing in a narrower scope. 
To address the quality implications of this decision, we provide an analysis of the distribution of studies across conference and journal ranks as a quality indicator.
Additionally, we include preprints from arXiv.
While necessary to reflect the latest advancements, preprints have not undergone rigorous peer review.
To mitigate this threat, we conducted a manual quality assessment, excluding studies that lacked clear evaluation setups or coherent contributions.

\textbf{Data extraction and classification.} 
The manual extraction of information and the construction of the taxonomy are subject to researcher bias and interpretation errors. 
For instance, determining whether a study performs "reasoning" versus "description" can be subjective. 
To mitigate this risk, we aligned our taxonomy with existing definitions where possible (e.g., dataset granularity, PEFT techniques) and refined our categories iteratively during the full-text review. 
We further support the reliability of our extraction by publicly releasing the full list of categorized studies and their attributes in our repository~\cite{replicationpackage}.


\section{Conclusion}
\label{sec:conclusion}
In this SLR, we analyzed \numStudies studies on LLM-based software vulnerability detection. 
To structure this rapidly evolving field, we introduced a comprehensive taxonomy that covers detection task formulation, i.e., the classification and generation tasks presented to the LLM; input representation, i.e., how code and auxiliary information are provided to the LLM; system architecture, differentiating between LLM-centric and hybrid approaches; as well as model adaptation (prompt engineering, fine-tuning, and learning paradigms) and orchestration techniques, e.g., agentic and ensemble systems.
The analysis shows that most studies perform binary code classification (vulnerable/non-vulnerable).
Common adaptation techniques include full-parameter fine-tuning and zero-shot prompting.
More advanced adaptation techniques, learning paradigms, and orchestration designs, such as parameter-efficient fine-tuning, retrieval-augmented generation, continual learning, and agentic workflows, are only beginning to emerge.
We further analyzed common datasets used, investigating their realistic nature, diversity in vulnerability types with respect to CWE-1000 pillars, and usage trends.
Despite notable progress in the field, we identified several key limitations that hinder practical adoption of LLMs in software vulnerability detection, including limited robustness in code representations, vulnerability data leakage, and limited interpretability and explainability.
To address the identified limitations, we outlined actionable research directions, such as advancing structure-aware and language-agnostic input representations, aligning datasets across use cases to enable cross-study comparison and standardized evaluation protocols, adopting retrieval and continual learning techniques to improve adaptability, and fitting approaches into practical development workflow requirements.
By mapping existing studies, identifying open challenges, and proposing future research directions, this review aims to guide researchers and practitioners in advancing the development of reliable, generalizable, and practically applicable LLM-based vulnerability detection systems.

\begin{acks}
This work has been supported in part by the Deutsche Forschungsgemeinschaft (DFG, German Research Foundation) Project-ID 528745080 - FIP 68. The authors alone are responsible for the content of the paper.
\end{acks}

\bibliographystyle{ACM-Reference-Format}
\bibliography{literature_SLR}

\end{document}